\documentclass[11pt]{article}
\usepackage[british,UKenglish,USenglish,english,american]{babel}
\usepackage[utf8]{inputenc}
\usepackage{latexsym}
\usepackage{graphicx}
\usepackage{amsfonts}
\usepackage{amssymb}
\usepackage{amsmath}
\usepackage{mathrsfs}
\usepackage{dsfont}
\usepackage{amsthm}
\usepackage{mdframed}
\usepackage{lipsum}
\usepackage{bbm}
\usepackage{color}
\usepackage{capt-of}
\usepackage{booktabs}
\usepackage{xcolor} 
\usepackage{hyperref}

\usepackage{subcaption}

\numberwithin{equation}{section}

\makeatletter
\newcommand*{\rom}[1]{\expandafter\@slowromancap\romannumeral #1@}
\makeatother
\title{On Extensions of the Barone-Adesi \& Whaley Method to Price American-Type Options}
\author{\sc \Large Ludovic Mathys\footnote{Email: ludovic.mathys@bf.uzh.ch} \vspace{0.5em} \\
       {\it Department of Banking and Finance, University of Zurich, Switzerland.}}
\date{}
\providecommand{\keywords}[1]{\textbf{Keywords:} #1}
\providecommand{\mscclass}[2]{\textbf{MSC (2010) Classification:} #1}
\providecommand{\jelclass}[3]{\textbf{JEL Classification:} #1}
\providecommand{\acknow}[4]{\textbf{Acknowledgements:} #1}

\begin{document}

\maketitle

\thispagestyle{empty}

\begin{abstract}
\noindent The present article provides an efficient and accurate hybrid method to price American standard options in certain jump-diffusion models as well as American barrier-type options under the Black \& Scholes framework. Our method generalizes the quadratic approximation scheme of Barone-Adesi \& Whaley (cf.~\cite{ba87}) and several of its extensions. Using perturbative arguments, we decompose the early exercise pricing problem into sub-problems of different orders and solve these sub-problems successively. The obtained solutions are combined to recover approximations to the original pricing problem of multiple orders, with the $0$-th order version matching the general Barone-Adesi \& Whaley ansatz. We test the accuracy and efficiency of the approximations via numerical simulations. The results show a clear dominance of higher order approximations over their respective $0$-th order version and reveal that significantly more pricing accuracy can be obtained by relying on approximations of the first few orders. Additionally, they suggest that increasing the order of any approximation by one generally refines the pricing precision, however that this happens at the expense of greater computational~costs.  
\end{abstract}
$\;$ \vspace{2em} \\
\noindent \keywords{American-Type Options, Exotic Options, Jump-Diffusion Models, Barone-Adesi \& Whaley Approximation, Perturbation Expansion.} \vspace{0.5em} \\
\noindent \mscclass{91-08, 91B25, 91B70, 91G20, 91G60, 91G80.}{} \vspace{0.5em}\\
\noindent \jelclass{C32, C63, G12, G13.} \vspace{-0.5em} \\
\section{Introduction}
\noindent For now more than fifty years, academics have been working on the problem of pricing American-type options. Compared to their European counterparts, these options have an early exercise feature that substantially complicates their structure and their pricing problem. Indeed, valuing American-type options is directly linked to certain types of free-boundary problems. Solving these problems is not an easy task so that analytical results are only known in few special cases. For this reason, pricing American-type options is still done in most cases via numerical approximations. Among many of the proposed approximations, hybrid approximations build a class of popular methods. These methods are based on combinations of analytical and numerical techniques and often lead to very efficient results. Examples in the case of standard American options include the class of integral representations initiated in the paper of Kim (cf.~\cite{ki90} and \cite{ca92} for two prominent examples of this class) as well as the approximations proposed by MacMillan (cf.~\cite{mc86}) and by Barone-Adesi \& Whaley (cf.~\cite{ba87}) together with their extensions (cf.~\cite{ba91}, \cite{jz99}, \cite{kw04}, \cite{gh09}, \cite{cs14}). An interesting survey of the main methods used for pricing standard American options (and that were developed prior to 2005) can be found in \cite{ba05}. \vspace{1em} \\
\noindent More recently, continuous growth in the trading of exotic options has incentivized the development of new pricing methods for American (single) barrier options and corresponding hybrid methods have been proposed: While Guo et al. (cf.~\cite{gh00}) and AitSahlia et al. (cf.~\cite{ai03}) developed an approximation based on the integral representation method offered in \cite{ki90} and \cite{ca92}, AitSahlia et al. (cf.~\cite{ai03}) and Chang et al. (cf.~\cite{ch07}) extended the quadratic approximation of Barone-Adesi \& Whaley to price American (single) barrier options. In addition to these developments in the pricing of exotics, Fatone et al. (cf.~\cite{fa15}) lately proposed a novel hybrid method to price standard American options under the Black \& Scholes framework (cf.~\cite{bs73}). Using perturbative arguments these authors provided a decomposition of the early exercise pricing problem into sub-problems of different orders that generalizes the Barone-Adesi \& Whaley ansatz (cf.~\cite{ba87}). The present paper combines these two paths of development to offer an accurate pricing method for certain American-type options. \vspace{1em} \\
\noindent Our paper extends the current literature on pricing American-type options in two directions: First, we consider the problem of pricing standard American options in jump-diffusion models. Here, the ansatz introduced in \cite{fa15} is extended to a model of constant jumps as well as to Merton's jump-diffusion model (cf.~\cite{me76}). Compared to the Black \& Scholes model investigated in \cite{fa15}, adding jumps to the dynamics of the asset substantially complicates the pricing attempt. In this case, early exercise of the American option may be additionally triggered by jumps and applying the Barone-Adesi \& Whaley ansatz only leads to an ordinary integro-differential equation whose solution is not known in general. We solve the problem approximately by relying on similar ideas to the ones introduced in \cite{ba91} and provide this way a generalization of Bates' method (cf.~\cite{ba91}). When compared to the latter method, the resulting approximations allow for a substantial increase in accuracy. In particular, our ansatz offers great performances for a very large range of parameters, including long times to maturity,\footnote{We provide numerical results for times to maturity of up to 10 years. The results are are in line with the findings obtained in \cite{fa15}.}~as well as for in-the-money options, for which Bates' method is known to fail. Secondly, we consider the problem of pricing American (single) barrier options in the model of Black \& Scholes (cf.~\cite{bs73}). Here, the techniques developed in the context of standard American options are applied to extend the methods proposed in \cite{ai03} and \cite{ch07}. On the theoretical side, our extension is substantially more challenging than these methods. This is due to the form of our expansion that, in particular, increases the complexity of the resulting equations. Here again, we provide (semi-)analytical solutions to these equations and recover approximations to the original pricing problem of multiple orders. These approximations are still very efficient in practical applications and exhibit a similar performance to the one obtained for standard American options: Compared to the simple (and modified) quadratic versions of \cite{ai03} and \cite{ch07}, our ansatz allows for a considerable increase in accuracy and the difference in accuracy between both methods is accentuated, when in-the-money options are considered. As for standard American options, this is due to the fact that the Barone-Adesi \& Whaley scheme looses in accuracy when pricing in-the-money options, while this does not affect the pricing quality of our higher order versions. \vspace{1em} \\
\noindent Finally, we note that the same techniques can be used to extend the method proposed by \cite{ch07} to price floating strike lookback options. However, since the main idea does not substantially differ from the one presented in this paper, we refrain from detailing it here. Additionally, we believe that the general idea underlying our ansatz can be combined with the results obtained in \cite{kw04} and \cite{cs14} to derive higher order approximations for the pricing of American-type options within the class of hyper-exponential jump-diffusions. This could be part of future work. \vspace{1em} \\
\noindent The remaining of this paper is structured as follows: In Section \ref{GENNOT}, we introduce the general framework as well as the notation used in the rest of the paper. Section~\ref{secapprox} deals with the pricing problem for standard American options in jump-diffusion models: While our ansatz is first presented under general jump-diffusion assumptions, solutions to the sub-problems are subsequently derived under a model of constant jumps as well as under Merton's jump-diffusion model. The techniques developed here are then extended in Section~\ref{BARRR} to deal with American (single) barrier options. All methods are finally tested in Section~\ref{numSEC} and the paper concludes with Section~\ref{SEC6}. Complementary results are presented in the Appendices (Appendix A, B and C).

\section{General Setting and Notation}
\label{GENNOT}
\noindent We start by introducing the general framework as well as the notation used in the rest of the paper. We consider a filtered probability space $(\Omega, \mathcal{F}, \mathbf{F}, \mathbb{Q})$, whose filtration $\mathbf{F} = \left( \mathcal{F}_{t} \right)_{t \geq 0}$ satisfies the usual conditions and let $(W_{t})_{t \geq 0}$ denote an $\mathbf{F}$-Brownian motion. Additionally, we let $(N_{t})_{t \geq 0}$ be an $\mathbf{F}$-Poisson process with constant intensity $\lambda>0$ and consider a financial market consisting of two assets, a deterministic savings account $(B_{t}(r))_{t \geq 0}$, with
\begin{equation}
B_{t}(r) = e^{r t}, \hspace{1.5em} r \geq 0, \, t \geq 0,
\label{market1}
\end{equation}
and a risky stock $(S_{t})_{t \geq 0}$, whose dynamics, under a (chosen) pricing measure $\mathbb{Q}$, are described by the following stochastic differential equation (SDE)
\begin{equation}
dS_{t} = S_{t-} \bigg( \big(r-\delta - \lambda \zeta \big) dt + \sigma dW_{t} + d \bigg( \sum \limits_{i=1}^{N_{t}} ( e^{J_{i}} -1 ) \bigg) \bigg), \hspace{1.5em} S_{0} > 0,
\label{market2}
\end{equation}
\noindent with $\zeta := \mathbb{E}^{\mathbb{Q}} \left[e^{J_{1}} -1 \right]$. Here, the constant parameters $\delta \in \mathbb{R}$ and $\sigma>0$ denote the dividend yield and the volatility level respectively and $\mathbb{E}^{\mathbb{Q}}[ \cdot ]$ refers to expectation with respect to the pricing measure~$\mathbb{Q}$. Furthermore, we assume that the jump sizes $(J_{i})_{i \in \mathbb{N}}$ form a sequence of independent and identically distributed random variables that are also independent of $(N_{t})_{t \geq 0}$ and will denote by $f_{J_{1}}(\cdot)$ the density associated with the distribution of $J_{1}$. Numerous models in the financial literature belong to this framework. Important examples include the standard model of Black \& Scholes (cf.~\cite{bs73}), Merton's jump-diffusion model (cf.~\cite{me76}) as well as Kou's double exponential jump-diffusion model (cf.~\cite{ko02}).\vspace{1em} \\
\noindent It is well-known that Equation (\ref{market2}) has a unique solution of the form
\begin{equation}
S_{t} = S_{0} e^{X_{t}},  \hspace{1.5em}  X_{t} := \Big(r-\delta - \lambda \zeta - \frac{1}{2}\sigma^{2} \Big)t + \sigma W_{t} + \sum \limits_{i=1}^{N_{t}} J_{i}, \hspace{1.5em} t \geq 0. \label{SusuOL}
\end{equation}
\noindent Hence, Model (\ref{market2}) is of (ordinary) exponential Lévy type with drift $b_{X} := r-\delta - \lambda \zeta - \frac{1}{2}\sigma^{2} + \int_{\{ |y|\leq 1 \}} y \, \Pi_{X}(dy)$, volatility $\sigma_{X}^{2} : = \sigma^{2}$ and jump measure given by $\Pi_{X}(dy) := \lambda f_{J_{1}}(y) dy$.
\noindent We define the Lévy exponent of $(X_{t})_{t \geq 0}$, $\Psi_{X}(\cdot)$, in the usual way and obtain that it is given, for any $\theta \in \mathbb{R}$, by
\begin{align}
\Psi_{X}(\theta) := - \log \left( \mathbb{E}^{\mathbb{Q}} \left[ e^{i\theta X_{1}} \right] \right) & = -i b_{X}\theta  + \frac{1}{2}\sigma_{X}^{2} \theta^2 + \int \limits_{ \mathbb{R}} \big(1 - e^{i\theta y} + i \theta y \mathds{1}_{\{ | y | \leq 1\}} \big) \Pi_{X}(dy) \nonumber \\
&= -i\Big(r-\delta - \lambda \zeta - \frac{1}{2}\sigma^{2} \Big) \theta + \frac{1}{2} \sigma^2 \theta^2  + \lambda \int \limits_{\mathbb{R}} \big( 1-e^{i\theta y} \big) f_{J_{1}}(y) dy .
\end{align}
\noindent Similarly, its Laplace exponent, $\Phi_{X}(\cdot)$, is well-defined for any $\theta \in \mathbb{R}$ satisfying $\mathbb{E}^{\mathbb{Q}} \left[ e^{\theta J_{1}} \right] < \infty$ and is recovered from $\Psi_{X}(\cdot)$ via the following relation:
\begin{equation}
\Phi_{X}(\theta) := -\Psi_{X}(-i\theta) = \Big(r-\delta - \lambda \zeta - \frac{1}{2}\sigma^{2} \Big) \theta + \frac{1}{2} \sigma^2 \theta^2  - \lambda \int \limits_{\mathbb{R}} \big( 1-e^{\theta y} \big) f_{J_{1}}(y) dy.
\end{equation}
\noindent Finally, it should be noticed that $(S_{t})_{t \geq 0}$ has a Markovian structure. Following standard theory for Markov processes, we therefore obtain that its infinitesimal generator is a partial integro-differential operator given, for sufficiently smooth $V: [0,\infty) \times \mathbb{R} \rightarrow \mathbb{R}$, by
\begin{align}
\mathcal{A}_{S} V(\mathcal{T},x) & :=  \lim \limits_{t \downarrow 0} \; \frac{\mathbb{E}^{\mathbb{Q}}_{x} \big[ V(\mathcal{T},S_{t})\big] - V(\mathcal{T},x)}{t} \nonumber \\
& = \frac{1}{2} \sigma^{2} x^{2} \partial_{x}^{2} V(\mathcal{T},x) + \big(r-\delta - \lambda \zeta \big) x \partial_{x} V(\mathcal{T},x) +  \lambda \int \limits_{\mathbb{R}}  \big( V(\mathcal{T},xe^{y}) - V(\mathcal{T},x) \big) f_{J_{1}}(y)dy,
\end{align}
\noindent where $\mathbb{E}_{x}^{\mathbb{Q}}[ \cdot]$ denotes expectation under $\mathbb{Q}_{x}$, the pricing measure having initial distribution $S_{0}=x$. We will extensively make use of these notations in the upcoming sections.
 \section{Approximation of Standard American Options}
 \label{secapprox}
\noindent We first consider the problem of pricing standard American options and derive an approximation that generalizes the ansatz adopted by Barone-Adesi \& Whaley in the standard Black \& Scholes model (cf.~\cite{ba87}) and extended by Bates to Merton's jump-diffusion model (cf.~\cite{ba91}). Our derivations focus on the standard American call. However, we note that the case of a standard American put can be treated analogously and only requires few obvious adjustements. For this reason, our numerical discussion in Section \ref{numSEC} also provides simulation results for American put options.
\subsection{Pricing Problem and Perturbation Expansion}
\label{PricProbl}
We start by reviewing few well-known facts on pricing standard American (call) options in models of the type of (\ref{market1}), (\ref{market2}). First, we recall that the value of a standard American call option on $(S_{t})_{t \geq 0}$ having maturity $\mathcal{T} \geq 0$, initial value $S_{0}=x \geq 0$ and strike price $K \geq 0$ has the following representation
\begin{equation}
\mathcal{C}_{A}(\mathcal{T},x;K) := \sup \limits_{ \tau \in \mathfrak{T}_{[0,\mathcal{T}]} }\mathbb{E}^{\mathbb{Q}}_{x} \left[ B_{\tau}(r)^{-1} \left(S_{\tau} -K \right)^{+} \right],
\label{AMERO}
\end{equation}
\noindent where $\mathfrak{T}_{[0,\mathcal{T}]}$ denotes the set of stopping times that take values in the interval $[0,\mathcal{T}]$, and that its European counterpart is obtained via
\begin{equation}
\mathcal{C}_{E}( \mathcal{T},x;K) := \mathbb{E}^{\mathbb{Q}}_{x} \left[ B_{\mathcal{T}}(r)^{-1} \left(S_{\mathcal{T}} -K \right)^{+} \right].
\label{EUROPO}
\end{equation}
\noindent Although European Option (\ref{EUROPO}) has, for many densities $f_{J_{1}}(\cdot)$, a closed form representation, pricing American-style derivatives is not an easy task and is usually done via numerical approximations. One popular way to derive such approximations consists in decomposing Option (\ref{AMERO}) into two components: Its European counterpart (\ref{EUROPO}) and an early exercise premium $\mathcal{E}(\cdot)$, obtained via
\begin{equation}
\mathcal{E}(\mathcal{T}, x;K) := \mathcal{C}_{A}(\mathcal{T},x;K) - \mathcal{C}_{E}(\mathcal{T},x;K).
\end{equation}
\noindent This decomposition is of great practical interest, since it usually reduces the pricing problem to the valuation of the early exercise premium $\mathcal{E}(\cdot)$ and, therefore, leads for a particular approximation method to a higher pricing accuracy when compared with a direct application of the same method to (\ref{AMERO}) instead. We will also adopt this approach and derive an approximation of the early exercise premium $\mathcal{E}(\cdot)$ for finite maturities, i.e.~we fix a final maturity $T >0$ and will focus on the valuation of $\mathcal{E}(\mathcal{T},x;K)$ for $\mathcal{T} \in [0,T]$.\footnote{We understand this value as the time-$t$ value of the early exercise premium by having in mind that $\mathcal{T} = T-t$.}~To this end, we note by the same arguments as the ones provided in \cite{ma18} that the early exercise premium $\mathcal{E}(\cdot)$ is linked to a partial integro-differential equation (PIDE) and has the following properties:
\begin{itemize} \setlength \itemsep{-0.5em}
\item[1.] If  $\delta \leq 0$, the early exercise premium $\mathcal{E}(\cdot)$ satisfies
$$ \mathcal{E}(\mathcal{T},x;K) = 0, \hspace{1.5em} \forall (\mathcal{T},x) \in [0,T] \times [0,\infty) .$$
\item[2.] If $ \delta > 0$, the pair $\big(\mathcal{E}(\cdot), \mathfrak{b}(\cdot)\big)$, where $\mathfrak{b}(\cdot)$ denotes the (corresponding) early exercise boundary, is a solution of the following free-boundary problem:
\begin{equation}
-\partial_{\mathcal{T}} \mathcal{E}(\mathcal{T},x;K) + \mathcal{A}_{S} \mathcal{E} (\mathcal{T},x;K) - r \mathcal{E}(\mathcal{T},x;K)  =  0 , \; \; \; \; x \in (0,\mathfrak{b}(\mathcal{T})), \; \mathcal{T} \in (0,T], \label{P1}
\end{equation}
\noindent subject to the boundary conditions
\begin{align}
\mathcal{E}(\mathcal{T}, \mathfrak{b}(\mathcal{T});K) & =  \mathfrak{b}(\mathcal{T}) -K - \mathcal{C}_{E}(\mathcal{T}, \mathfrak{b}(\mathcal{T});K), \; \; \; \; \; \mathcal{T} \in (0,T], \label{P2} \\
\partial_{x} \mathcal{E}( \mathcal{T}, \mathfrak{b}(\mathcal{T});K) & =  1 - \partial_{x} \mathcal{C}_{E} (\mathcal{T}, \mathfrak{b}(\mathcal{T});K), \; \; \; \; \; \mathcal{T} \in (0,T], \label{P3} \\
\mathcal{E}(\mathcal{T},0;K) & =  0, \; \; \; \; \; \mathcal{T} \in (0,T], \label{P4}
\end{align}
\noindent and initial condition
\begin{equation}
\mathcal{E}(0, x;K) = 0, \; \; \; \; \; x \in (0,\mathfrak{b}(\mathcal{T})).
\label{P5}
\end{equation}
\end{itemize}
\noindent Therefore, we focus from now on on the non-trivial case 2.~and derive a solution to the above free-boundary characterization.
\subsubsection{The Barone-Adesi \& Whaley Ansatz}
\label{secBAW}
\noindent As done in \cite{ba87}, we next rewrite the early exercise premium $\mathcal{E}(\cdot)$ in the following form
\begin{equation}
\mathcal{E}(\mathcal{T},x;K) = h(\mathcal{T})F(h(\mathcal{T}),x;K),
\label{REPR}
\end{equation}
\noindent where $h(\mathcal{T}) := 1-e^{-r \mathcal{T}}$ and $F(\cdot)$ is an auxiliary ``well-behaved'' function that will be determined later. Under this representation, straightforward computations transform Equations (\ref{P1})-(\ref{P4}) into a new problem:
\begin{equation}
- \frac{r}{h(\mathcal{T})} F(h(\mathcal{T}),x;K) + \mathcal{A}_{S} F(h(\mathcal{T}),x;K) - r\big(1-h(\mathcal{T})\big) \partial_{h}F(h(\mathcal{T}),x;K) =0, \; \; \; \; x \in (0,\mathfrak{b}(\mathcal{T})), \; \mathcal{T} \in (0,T], \label{P12}
\end{equation}
\noindent with boundary conditions
\begin{align}
F(h(\mathcal{T}), \mathfrak{b}(\mathcal{T});K) & = \frac{1}{h(\mathcal{T})} \Big( \mathfrak{b}(\mathcal{T}) -K - \mathcal{C}_{E}(\mathcal{T}, \mathfrak{b}(\mathcal{T});K) \Big), \; \; \; \; \;  \mathcal{T} \in (0,T], \label{P22} \\
\partial_{x} F( h(\mathcal{T}), \mathfrak{b}(\mathcal{T});K) & =  \frac{1}{h(\mathcal{T})} \Big( 1 - \partial_{x} \mathcal{C}_{E} (\mathcal{T}, \mathfrak{b}(\mathcal{T});K) \Big), \; \; \; \; \;  \mathcal{T} \in (0,T], \label{P32} \\
F(h(\mathcal{T}),0;K) & =  0, \; \; \; \; \;  \mathcal{T} \in (0,T]. \label{P42}
\end{align}
\noindent Although Initial Condition (\ref{P5}) is not anymore required to hold, it will be naturally satisfied whenever $F(\cdot)$ has ``good properties''.\footnote{The ansatz we will follow consists in representing $F(\cdot)$ by a series of products of logarithms and power functions. Consequently, $F(\cdot)$ will have sufficiently good properties.}~This follows, from (\ref{REPR}), since $h(0)=0$ clearly holds. \vspace{1em} \\
\noindent Starting from a problem that corresponds to (\ref{P12})-(\ref{P42}) in the Black \& Scholes model, the authors in \cite{ba87} had the brilliant idea to drop the last term in their partial differential equation (PDE) corresponding to (\ref{P12}). This allowed them to convert the initial problem into a much more manageable one\footnote{Under the Black \& Scholes model the resulting equation simplifies to an ordinary differential equation (ODE).}, for which an ``analytical solution''\footnote{This solution still depends on the free-boundary $\mathfrak{b}(\cdot)$. Finding this boundary level requires however the use of numerical methods.} can be easily derived (cf.~\cite{ba87}). Also in our case this approach can be taken. However, omitting the last term $r\big(1-h(\mathcal{T})\big) \partial_{h}F(h(\mathcal{T}),x;K)$ in Equation (\ref{P12}) now transforms it into an ordinary integro-differential equation (OIDE) whose solution is not known in general. Nevertheless, approximate solutions have proven to be effective in some types of model. Such an approximation was first introduced under Merton's jump-diffusion model in \cite{ba91}. An application of the same ansatz under a model of constant jumps (cf.~Equation (\ref{mixed})) is also presented in \cite{jc02}, \cite{jy06}.
\subsubsection{Generalization of the Barone-Adesi \& Whaley Ansatz}
\label{GenBAW}
\noindent Instead of relying on the well-known Barone-Adesi \& Whaley ansatz, we follow an extended approach to Problem (\ref{P12})-(\ref{P42}) that was proposed in the classical Black \& Scholes model in \cite{fa15}. To this end, we introduce a new parameter, a perturbation parameter $\epsilon \in [0,1]$, in (\ref{P12})-(\ref{P42}) and consider the following modified problem:
\begin{equation}
- \frac{r}{h(\mathcal{T})} F^{\epsilon}(h(\mathcal{T}),x;K) + \mathcal{A}_{S} F^{\epsilon}(h(\mathcal{T}),x;K) - \epsilon r\big(1-h(\mathcal{T})\big) \partial_{h}F^{\epsilon}(h(\mathcal{T}),x;K) =0, \; \; \;  x \in (0,\mathfrak{b}^{\epsilon}(\mathcal{T})), \mathcal{T} \in (0,T], \label{P13}
\end{equation}
\noindent with boundary conditions
\begin{align}
F^{\epsilon}(h(\mathcal{T}), \mathfrak{b}^{\epsilon}(\mathcal{T});K) & =  \frac{1}{h(\mathcal{T})} \Big( \mathfrak{b}^{\epsilon}(\mathcal{T}) -K - \mathcal{C}_{E}(\mathcal{T}, \mathfrak{b}^{\epsilon}(\mathcal{T});K) \Big), \; \; \; \; \; \mathcal{T} \in (0,T], \label{P23} \\
\partial_{x} F^{\epsilon}( h(\mathcal{T}), \mathfrak{b}^{\epsilon}(\mathcal{T});K) & =  \frac{1}{h(\mathcal{T})} \Big( 1 - \partial_{x} \mathcal{C}_{E} (\mathcal{T}, \mathfrak{b}^{\epsilon}(\mathcal{T});K) \Big), \; \; \; \; \; \mathcal{T} \in (0,T], \label{P33} \\
F^{\epsilon}(h(\mathcal{T}),0;K) & =  0, \; \; \; \; \; \mathcal{T} \in (0,T]. \label{P43}
\end{align}
\noindent Switching from (\ref{P12})-(\ref{P42}) to the new problem (\ref{P13})-(\ref{P43}) clearly allows for a more general treatment of the pricing attempt. Indeed, for $\epsilon=1$, PIDEs (\ref{P13}) and (\ref{P12}) are identical and the perturbative approach reduces to the original problem. Additionally, solving the modified problem while taking $\epsilon =0$ allows to recover the classical Barone-Adesi \& Whaley ansatz. \vspace{1em} \\
\noindent In order to solve Problem (\ref{P13})-(\ref{P43}), we make use of a typical perturbative ansatz (cf.~\cite{ve05}) and assume that the solution pair $\big(F^{\epsilon}(\cdot),\mathfrak{b}^{\epsilon}(\cdot)\big)$ to Equations (\ref{P13})-(\ref{P43}) has, for any $\epsilon \in [0,1]$, a representation as ``well-behaved''\footnote{In particular, we will assume that any derivative of $F^{\epsilon}(\cdot)$ can be obtained by differentiating inside the sum.} series expansion of the form
\begin{align}
 F^{\epsilon}(h(\mathcal{T}),x;K) & = \sum \limits_{n=0}^{\infty} \epsilon^{n}  f_{n}(h(\mathcal{T}),x;K), \; \; \; \; \;  x \in (0,\mathfrak{b}^{\epsilon}(\mathcal{T})), \mathcal{T} \in [0,T],  \label{Sum1}\\
  \mathfrak{b}^{\epsilon}(\mathcal{T}) & = \sum \limits_{n=0}^{\infty} \epsilon^{n}  b_{n}(\mathcal{T}), \; \; \; \; \; \mathcal{T} \in [0,T], \label{Sum2}
\end{align} 
\noindent for some functions $\big(f_{n}(\cdot)\big)_{n \in \mathbb{N}_{0}}$ and $\big(b_{n}(\cdot)\big)_{n \in \mathbb{N}_{0}}$. Additionally, we define partial sums of $N$-th order via
\begin{align}
 F_{N}^{\epsilon}(h(\mathcal{T}),x;K) & =  \sum \limits_{n=0}^{N} \epsilon^{n}  f_{n}(h(\mathcal{T}),x;K), \; \; \; \; \;   x \in (0,\mathfrak{b}_{N}^{\epsilon}(\mathcal{T})), \mathcal{T} \in [0,T], \label{susu} \\
  \mathfrak{b}_{N}^{\epsilon}(\mathcal{T}) & =  \sum \limits_{n=0}^{N} \epsilon^{n}  b_{n}(\mathcal{T}), \; \; \; \; \;  \mathcal{T} \in [0,T].
\end{align}
\noindent Using Representation (\ref{Sum1}), we obtain upon setting $f_{-1}(h(\mathcal{T}),x;K) := 0$ that 
\begin{equation}
\sum \limits_{n=0}^{\infty} \epsilon^{n} \left( -\frac{r}{h(\mathcal{T})} f_{n}(h(\mathcal{T}),x;K) + \mathcal{A}_{S} f_{n}(h(\mathcal{T}),x;K) - r \big(1-h(\mathcal{T}) \big) \partial_{h}f_{n-1}(h(\mathcal{T}),x;K) \right) = 0
\end{equation}
\noindent and imposing this equation to hold order by order in the powers of $\epsilon$ leads to the following recurrent system of $n$-th order problems: For $n=0$, the $0$-th order problem reads
\begin{equation}
- \frac{r}{h(\mathcal{T})} f_{0}(h(\mathcal{T}),x;K) + \mathcal{A}_{S} f_{0}(h(\mathcal{T}),x;K) =0, \; \; \; \; \; x \in (0,\mathfrak{b}_{0}^{\epsilon}(\mathcal{T})), \mathcal{T} \in (0,T], \label{P14}
\end{equation}
\noindent with boundary conditions
\begin{align}
f_{0}(h(\mathcal{T}), \mathfrak{b}_{0}^{\epsilon}(\mathcal{T});K) & = \frac{1}{h(\mathcal{T})} \Big( \mathfrak{b}_{0}^{\epsilon}(\mathcal{T}) -K - \mathcal{C}_{E}(\mathcal{T}, \mathfrak{b}_{0}^{\epsilon}(\mathcal{T});K) \Big), \; \; \; \; \; \mathcal{T} \in (0,T], \label{P24} \\
\partial_{x} f_{0}( h(\mathcal{T}), \mathfrak{b}_{0}^{\epsilon}(\mathcal{T});K) & =  \frac{1}{h(\mathcal{T})} \Big( 1 - \partial_{x} \mathcal{C}_{E} (\mathcal{T}, \mathfrak{b}_{0}^{\epsilon}(\mathcal{T});K) \Big), \; \; \; \; \; \mathcal{T} \in (0,T], \label{P34} \\
f_{0}(h(\mathcal{T}),0;K) & =  0, \; \; \; \; \; \mathcal{T} \in (0,T]. \label{P44}
\end{align}
\noindent Additionally, the following higher order problems ($n \in \mathbb{N}$) are obtained:
 \begin{equation}
- \frac{r}{h(\mathcal{T})} f_{n}(h(\mathcal{T}),x;K) + \mathcal{A}_{S} f_{n}(h(\mathcal{T}),x;K) - r\big(1-h(\mathcal{T})\big) \partial_{h}f_{n-1}(h(\mathcal{T}),x;K) =0, \; \; \;  x \in (0,\mathfrak{b}_{n}^{\epsilon}(\mathcal{T})), \mathcal{T} \in (0,T], \label{P15}
\end{equation}
\noindent with boundary conditions, for $\mathcal{T} \in (0,T]$:
\begin{align}
f_{n}(h(\mathcal{T}), \mathfrak{b}_{n}^{\epsilon}(\mathcal{T});K) & = \frac{\epsilon^{-n}}{ h(\mathcal{T})} \Big( \mathfrak{b}_{n}^{\epsilon}(\mathcal{T}) -K - \mathcal{C}_{E}(\mathcal{T}, \mathfrak{b}_{n}^{\epsilon}(\mathcal{T});K)- h(\mathcal{T})F_{n-1}^{\epsilon}(h(\mathcal{T}),\mathfrak{b}_{n}^{\epsilon}(\mathcal{T});K)\Big), \label{P25} \\
\partial_{x} f_{n}( h(\mathcal{T}), \mathfrak{b}_{n}^{\epsilon}(\mathcal{T});K) & = \frac{\epsilon^{-n}}{h(\mathcal{T})} \Big( 1 - \partial_{x} \mathcal{C}_{E} (\mathcal{T}, \mathfrak{b}_{n}^{\epsilon}(\mathcal{T});K) - h(\mathcal{T}) \partial_{x} F_{n-1}^{\epsilon}(h(\mathcal{T}),\mathfrak{b}_{n}^{\epsilon}(\mathcal{T});K) \Big), \label{P35} \\
f_{n}(h(\mathcal{T}),0;K) & = 0. \label{P45}
\end{align}
\noindent Recall that our initial early exercise premium valuation attempt is related to the above problems via the following relation
\begin{equation}
 \mathcal{E}(\mathcal{T},x;K)  =  h(\mathcal{T}) F^{\epsilon=1}(h(\mathcal{T}),x;K) =:  \mathcal{E}^{\epsilon=1}(\mathcal{T},x;K).
 \end{equation}
 Assuming that $F^{\epsilon}(\cdot)$ has a representation of the form of (\ref{Sum1}), we therefore expect to obtain $N$-th order approximations of the early exercise premium by means of the following quantities: 
\begin{equation}
\mathcal{E}_{N}^{\epsilon=1}(\mathcal{T},x;K)  :=  h(\mathcal{T})F_{N}^{\epsilon=1}(h(\mathcal{T}),x;K)  =  h(\mathcal{T}) \sum \limits_{n=0}^{N}  f_{n}(h(\mathcal{T}),x;K), \hspace{2em} x \in [0,\mathfrak{b}_{N}^{\epsilon=1}(\mathcal{T})), \label{above}
\end{equation}
\begin{equation}
\mathcal{E}_{N}^{\epsilon=1}(\mathcal{T},x;K) := x-K - \mathcal{C}_{E}(\mathcal{T},x;K), \hspace{2em} x \in [\mathfrak{b}_{N}^{\epsilon=1}(\mathcal{T}), \infty ),
\label{above2}
\end{equation}
\noindent where $\mathcal{T} \in [0,T]$. Consequently, we focus in the sequel on the $n$-th order problems (\ref{P14})-(\ref{P44}) and (\ref{P15})-(\ref{P45}) for $\epsilon =1$ and will subsequently recover approximations of multiple orders via (\ref{above}), (\ref{above2}). \vspace{1em} \\
\noindent At this point, we should note that the boundary functions $(b_{n}(\cdot))_{n \in \mathbb{N}_{0}}$ play no role in the respective $n$-th order problems. Indeed, numerical experiments have shown that the partial sums of the first few orders computed by solving the $n$-th order problems applied directly to $(\mathfrak{b}_{n}^{\epsilon=1}(\cdot))_{n \in \mathbb{N}_{0}}$ provide better results than the corresponding partial sums obtained by solving the same problems but applied order by order to $(b_{n}(\cdot))_{n \in \mathbb{N}_{0}}$.\footnote{This is in line with the findings in \cite{fa15}.}~Therefore, solving the $n$-th order problems will always be carried out directly in terms of $(\mathfrak{b}_{n}^{\epsilon=1}(\cdot))_{n \in \mathbb{N}_{0}}$.

 \subsection{Solutions under Constant Jumps}
 \label{SEC2}
\noindent We next turn to the derivation of $N$-th order approximations under constant jumps, i.e.~we fix $\varphi \in \mathbb{R}$ and assume throughout the rest of this section that the jump measure $\Pi_{X}$ is given by
\begin{equation}
\lambda f_{J_{1}}(y) dy =\Pi_{X}(dy)  = \lambda \delta_{\varphi}(dy),
\end{equation}
\noindent where $\delta_{\varphi}(\cdot)$ denotes the Dirac measure at $\varphi$. This is equivalent to the assumption that the asset dynamics $(S_t)_{t \geq 0}$ evolve, under the pricing measure $\mathbb{Q}$, according to the following SDE
\begin{equation}
 dS_{t} = S_{t-} \Big( \big(r-\delta-\lambda(e^{\varphi}-1)\big) dt + \sigma dW_{t} + (e^{\varphi}-1) dN_{t} \Big), \hspace{1.5em} S_{0} > 0,
\label{mixed}
\end{equation}
\noindent where the processes $(W_{t})_{t \geq 0}$ and $(N_{t})_{t \geq 0}$ and the parameters $\lambda>0$, $r\geq 0$, $\delta\in \mathbb{R}$ and $\sigma>0$ have the same properties as in (\ref{market1}), (\ref{market2}). In this case, $(S_{t})_{t \geq 0}$ is recovered from (\ref{SusuOL}) with
\begin{equation*}
X_{t} := \Big(r-\delta - \lambda (e^{\varphi}-1)- \frac{1}{2}\sigma^{2} \Big)t + \sigma W_{t} + \varphi N_{t}, \hspace{1.5em} t \geq 0,
\label{mixed2}
\end{equation*}
\noindent and its infinitesimal generator takes the following simplified form
\begin{equation}
\mathcal{A}_{S} V(\mathcal{T},x) = \frac{1}{2} \sigma^{2} x^{2} \partial_{x}^{2} V(\mathcal{T},x) + \big(r-\delta - \lambda (e^{\varphi}-1)\big) x \partial_{x} V(\mathcal{T},x) + \lambda \big( V(\mathcal{T},xe^{\varphi}) - V(\mathcal{T},x)\big) .
\end{equation}
\noindent Whenever $\varphi \leq 0$, this will in particular allows us to derive a well-known solution to the OIDE arising in the $0$-th order problem, as it now simplifies in the continuation region to an homogeneous second order linear ODE that does not depend anymore on boundary terms. Analogously, deriving an exact solution of the OIDE arising in the $0$-th order problem for the American put requires that $\varphi \geq 0$. This will be outlined in the next section.
\subsubsection{Solution of the $0$-th Order Problem}
\label{Sol0thPro}
\noindent We start our derivations by noting that, under Model (\ref{mixed}), the Laplace exponent of $(X_{t})_{t \geq 0}$, $\Phi_{X}(\cdot)$, is well-defined for all $\theta \in \mathbb{R}$. Furthermore, it can be easily seen that $\theta \mapsto \Phi_{X}(\theta)$ is convex and satisfies $\Phi_{X}(0)=0$ and $\lim \limits_{ |\theta| \rightarrow \infty} \Phi_{X}(\theta)  =  \infty $. Therefore, the equation $ \Phi_{X}(\theta) = y$ has for any $y >0$ two solutions, a positive and a negative root. We will denote by $\Phi_{X}^{-1,+}(y)$ its positive root and by $\Phi_{X}^{-1,-}(y)$ its negative root. \vspace{1em}\\
\noindent We now turn to the $0$-th order problem. For $\varphi \leq 0$, Equation (\ref{P14}) is well-known and its general solution takes the simple form
\begin{equation}
f_{0}(h(\mathcal{T}),x;K)  =  c_{0,0}^{+}(h(\mathcal{T})) x^{\rho_{+}(h(\mathcal{T}))} + c_{0,0}^{-}(h(\mathcal{T})) x^{\rho_{-}(h(\mathcal{T}))}, \hspace{1.5em}  x \in (0,\mathfrak{b}_{0}^{\epsilon=1}(\mathcal{T})),  \, \mathcal{T} \in (0,T],
\label{AHAH}
\end{equation} 
\noindent where, for $\mathcal{T} \in (0,T]$,
\begin{equation}
\rho_{+}(h( \mathcal{T})) := \Phi_{X}^{-1,+}\Big(\frac{r}{h(\mathcal{T})}\Big), \hspace{2em}  \rho_{-}(h( \mathcal{T})) := \Phi_{X}^{-1,-}\Big(\frac{r}{h(\mathcal{T})}\Big),
\label{ROOT}
\end{equation}
\noindent and $c_{0,0}^{+}(h(\mathcal{T}))$ and $c_{0,0}^{-}(h(\mathcal{T}))$ are ``constants'' to be determined. Conversely, the ODE corresponding to (\ref{P14}) under this model takes a special form immediately below the exercise boundary when $\varphi > 0$. Indeed, for $x \in [\mathfrak{b}_{0}^{\epsilon=1}(\mathcal{T}) e^{-\varphi}, \mathfrak{b}_{0}^{\epsilon=1}(\mathcal{T}))$, Equation (\ref{P14}) becomes
\begin{eqnarray*}
 -\frac{r}{h(\mathcal{T})} f_{0}(h(\mathcal{T}),x;K) + \frac{1}{2} \sigma^{2} x^{2} \partial_{x}^{2} f_{0}(h(\mathcal{T}),x;K) + \big(r-\delta-\lambda(e^{\varphi}-1)\big)x \partial_{x} f_{0}(h(\mathcal{T}),x;K) \hspace{4em} \\
\hspace{15em} + \; \lambda \Big( \frac{1}{h(\mathcal{T})} \big(xe^{\varphi}-K-\mathcal{C}_{E}(\mathcal{T},xe^{\varphi};K) \big) - f_{0}(h(\mathcal{T}),x;K) \Big) =0
\end{eqnarray*}
\noindent and, unfortunately, there is no known solution to this equation.\footnote{When considering an American put option, $\varphi < 0$ transforms (\ref{P14}) for any $x \in (\mathfrak{b}_{0}^{\epsilon=1}(\mathcal{T}),\mathfrak{b}_{0}^{\epsilon=1}(\mathcal{T}) e^{-\varphi}]$ into a similar equation. Here again, there is no known solution to the resulting equation.}~Since we expect $f_{0}(\cdot)$ to be continuous in the jump parameter $\varphi$, it appears however sensible to approximate the solution for ``small'' jump sizes anyway via (\ref{AHAH}). This is in line with the approximation proposed in \cite{ba91} and with the discussion in \cite{jc02}. We will also follow this approach and provide numerical tests to the resulting $N$-th order approximations in Section \ref{numSEC}. \vspace{1em} \\
\noindent To derive an expression for $c_{0,0}^{+}(\cdot)$, $c_{0,0}^{-}(\cdot)$ and $\mathfrak{b}_{0}^{\epsilon=1}(\cdot)$, we use the complementary conditions (\ref{P24}), (\ref{P34}) and (\ref{P44}). First, we note that (\ref{P44}) implies that $c_{0,0}^{-}(h(\mathcal{T})) \equiv 0$. Secondly, substituting (\ref{AHAH}) into Condition~(\ref{P34}), allows us to express $c_{0,0}^{+}(\cdot)$ in terms of the free-boundary $\mathfrak{b}_{0}^{\epsilon=1}(\cdot)$ as
\begin{equation}
c_{0,0}^{+}(h(\mathcal{T}))  =  \frac{\Big( \mathfrak{b}_{0}^{\epsilon=1}(\mathcal{T}) \Big)^{1- \rho_{+}(h(\mathcal{T}))}}{h(\mathcal{T}) \rho_{+}(h(\mathcal{T}))}  \Big( 1 - \partial_{x} \mathcal{C}_{E} (\mathcal{T}, \mathfrak{b}_{0}^{\epsilon=1}(\mathcal{T});K) \Big), \; \; \; \; \;  \mathcal{T} \in (0,T].
\label{REL}
\end{equation}
\noindent Finally, substituting again (\ref{AHAH}) into (\ref{P24}) and inserting Representation (\ref{REL}) in the resulting equation, gives
\begin{equation}
\label{REFL}
\mathfrak{b}_{0}^{\epsilon=1}(\mathcal{T})  = K + \mathcal{C}_{E}( \mathcal{T}, \mathfrak{b}_{0}^{\epsilon=1}(\mathcal{T});K) + \frac{\mathfrak{b}_{0}^{\epsilon=1}(\mathcal{T})}{ \rho_{+}(h(\mathcal{T}))} \Big( 1 - \partial_{x} \mathcal{C}_{E} (\mathcal{T}, \mathfrak{b}_{0}^{\epsilon=1}(\mathcal{T});K) \Big), \; \; \; \; \; \mathcal{T} \in (0,T],
\end{equation} 
\noindent a non-linear equation in $\mathfrak{b}_{0}^{\epsilon=1}(\cdot)$. Therefore, solving Equation (\ref{REFL}) numerically for $\mathcal{T} \in (0,T]$ gives $\mathfrak{b}_{0}^{\epsilon=1}(\mathcal{T})$ and subsequently allows us to recover $c_{0,0}^{+}(h(\mathcal{T}))$ via Relation (\ref{REL}) to finally obtain the $0$-th order premium $f_{0}(\cdot)$.

\subsubsection{Solution of the Higher Order Problems}
\label{SolNthPro}
\noindent We now turn to the higher order problems, i.e.~we seek, for $n\in \mathbb{N}$, a solution to (\ref{P15})-(\ref{P45}). Generalizing the form of the solution obtained in the $0$-th order problem, we make the following ansatz:
\begin{equation}
f_{n}(h(\mathcal{T}),x;K) = \bigg(c_{n,0}^{+}(h(\mathcal{T})) + \sum \limits_{j=1}^{2n} c_{n,j}^{+}(h(\mathcal{T})) \log(x)^{j} \bigg) x^{\rho_{+}(h(\mathcal{T}))}, \hspace{1.5em}  x \in (0,\mathfrak{b}_{n}^{\epsilon=1}(\mathcal{T})),  \, \mathcal{T} \in (0,T],
\label{tati}
\end{equation}
\noindent where the ``constants'' $c_{n,0}^{+}(\cdot)$ and, for $j \in \{ 1,2, \ldots , 2n \}$, $c_{n,j}^{+}(\cdot)$ are still to determine. Whether or not this ansatz provides good results is an issue that we will consider in the numerical simulations of Section~\ref{numSEC}. However, it gives a convenient way to solve (\ref{P15})-(\ref{P45}), since it allows us to obtain a system of linear equations in the coefficients $c_{n,j}^{+}(\cdot)$ that can be solved using standard numerical methods. For the derivation of this system, we substitute (\ref{tati}) into PDE (\ref{P15}), use Property (\ref{ROOT}), and match the powers of the logarithm. This gives, for $n \in \mathbb{N}$, the following system of $2n$ linear equations in the $2n$ unknowns $\big( c_{n,j}^{+}(h(\mathcal{T})) \big)_{j \in \{ 1,\ldots,2n\}}$:
\begin{align}
 2n \bigg[ \frac{\sigma^{2}}{2} \big(2 \rho_{+}(h(\mathcal{T}))-1 \big) + r-\delta + \lambda \big( \varphi e^{\rho_{+}(h(\mathcal{T})) \varphi} - (e^{\varphi}-1) \big) \bigg]  c_{n,2n}^{+}(h(\mathcal{T}))   \hspace{10em} \nonumber \\
 \hspace{5.8em} =   r (1-h(\mathcal{T})) \partial_{h}\rho_{+}(h(\mathcal{T})) c_{n-1,2(n-1)}^{+}(h(\mathcal{T})) , \hspace{7em} \label{middd1}
\end{align}
\begin{align}
 j \bigg[ \frac{\sigma^{2}}{2} \big(2 \rho_{+}(h(\mathcal{T}))-1 \big) + r-\delta + \lambda \big( \varphi e^{\rho_{+}(h(\mathcal{T})) \varphi} - (e^{\varphi}-1) \big) \bigg]  c_{n,j}^{+}(h(\mathcal{T}))  \hspace{12em} \nonumber  \\
\hspace{2em} +  j (j+1) \frac{\sigma^{2}}{2} c_{n,j+1}^{+}(h(\mathcal{T})) + \lambda \sum \limits_{k=j+1}^{2n} \binom{k}{j-1} \varphi^{k-(j-1)} e^{\rho_{+}(h(\mathcal{T})) \varphi}  c_{n,k}^{+}(h(\mathcal{T}))\hspace{6em} \label{midie}\\ 
\hspace{10em} = r (1-h(\mathcal{T})) \big( \partial_{h} c_{n-1,j-1}^{+}(h(\mathcal{T})) + \partial_{h}\rho_{+}(h(\mathcal{T})) c_{n-1,j-2}^{+}(h(\mathcal{T}))\big), \hspace{-1.5em} \nonumber
\end{align}
\begin{align}
\bigg[ \frac{\sigma^{2}}{2} \big(2 \rho_{+}(h(\mathcal{T}))-1 \big) + r-\delta + \lambda \big( \varphi e^{\rho_{+}(h(\mathcal{T})) \varphi} - (e^{\varphi}-1) \big) \bigg] c_{n,1}^{+}(h(\mathcal{T}))  \hspace{12em} \nonumber \\ 
\hspace{2em} + \sigma^2 c_{n,2}^{+}(h(\mathcal{T})) + \lambda \sum \limits_{k=2}^{2n}  \varphi^{k} e^{\rho_{+}(h(\mathcal{T})) \varphi} c_{n,k}^{+}(h(\mathcal{T}))  =  r (1-h(\mathcal{T})) \partial_{h} c_{n-1,0}^{+}(h(\mathcal{T})), \hspace{4em} \label{middd2} 
\end{align}
\noindent where Equation (\ref{midie}) only holds for $j \in \{2,3, \ldots, 2n-1 \}$. \vspace{1em} \\
\noindent To conclude, we proceed as in the derivation of the $0$-th order approximation and derive, for any $n \in \mathbb{N}$, an expression for both $c_{n,0}^{+}(\cdot)$ and $\mathfrak{b}_{n}^{\epsilon=1}(\cdot)$ by substituting (\ref{tati}) into Equations (\ref{P25}) and (\ref{P35}). This first allows us to express $c_{n,0}^{+}(h(\mathcal{T}))$, for $\mathcal{T} \in (0,T]$, in terms of the free boundary $\mathfrak{b}_{n}^{\epsilon=1}(\mathcal{T})$ as
\begin{align}
c_{n,0}^{+}(h(\mathcal{T})) = \frac{\Big( \mathfrak{b}_{n}^{\epsilon=1}(\mathcal{T}) \Big)^{1- \rho_{+}(h(\mathcal{T}))}}{h(\mathcal{T}) \rho_{+}(h(\mathcal{T}))} \Big( 1 - \partial_{x} \mathcal{C}_{E} (\mathcal{T}, \mathfrak{b}_{n}^{\epsilon=1}(\mathcal{T});K) - h(\mathcal{T}) \partial_{x} F_{n-1}^{\epsilon=1}(h(\mathcal{T}),\mathfrak{b}_{n}^{\epsilon=1}(\mathcal{T});K)\Big) \nonumber \\
- \frac{1}{\rho_{+}(h(\mathcal{T}))} \sum \limits_{j=1}^{2n} c_{n,j}^{+}(h(\mathcal{T}))j \log\big(\mathfrak{b}_{n}^{\epsilon=1}(\mathcal{T})\big)^{j-1} - \sum \limits_{j=1}^{2n} c_{n,j}^{+}(h(\mathcal{T})) \log\big(\mathfrak{b}_{n}^{\epsilon=1}(\mathcal{T}) \big)^{j},
\label{cn0eq}
\end{align}
\noindent and to finally obtain a characterization of the free boundary $\mathfrak{b}_{n}^{\epsilon=1}(\cdot)$ by means of the following non-linear equation:
\begin{align}
\mathfrak{b}_{n}^{\epsilon=1}(\mathcal{T}) = K + \mathcal{C}_{E}( \mathcal{T}, \mathfrak{b}_{n}^{\epsilon=1}(\mathcal{T});K) + h(\mathcal{T}) F_{n-1}^{\epsilon=1}(h(\mathcal{T}), \mathfrak{b}_{n}^{\epsilon=1}(\mathcal{T});K) \hspace{9em}\nonumber \\
+ \frac{\mathfrak{b}_{n}^{\epsilon=1}(\mathcal{T})}{\rho_{+}(h(\mathcal{T}))} \Big( 1 - \partial_{x} \mathcal{C}_{E}(\mathcal{T}, \mathfrak{b}_{n}^{\epsilon=1}(\mathcal{T});K) - h(\mathcal{T}) \partial_{x} F_{n-1}^{\epsilon=1}(h(\mathcal{T}),\mathfrak{b}_{n}^{\epsilon=1}(\mathcal{T});K)\Big) \hspace{-2em} \nonumber\\
- \frac{\left( \mathfrak{b}_{n}^{\epsilon=1}(\mathcal{T}) \right)^{\rho_{+}(h(\mathcal{T}))}h(\mathcal{T})}{\rho_{+}(h(\mathcal{T}))} \sum \limits_{j=1}^{2n} c_{n,j}^{+}(h(\mathcal{T})) j \log\big(\mathfrak{b}_{n}^{\epsilon=1}(\mathcal{T}) \big)^{j-1}. \hspace{4.7em}
\label{bboundeq}
\end{align}
\noindent Once again, this equation can be solved for any $\mathcal{T} \in (0,T]$ using standard numerical techniques to derive $\mathfrak{b}_{n}^{\epsilon=1}(\mathcal{T})$, $c_{n,0}^{+}(h(\mathcal{T}))$ and ultimately the $n$-th order premium $f_{n}(\cdot)$. \vspace{1.5em} \\
\noindent \underline{\bf Remark 1.}
\begin{itemize} \setlength \itemsep{-0.1em}
\item[i)] As seen from Equations (\ref{middd1})-(\ref{middd2}), our higher order approximations ($n \in \mathbb{N}$) depend on the derivatives $\partial_{h} \rho_{+}(\cdot)$ and $\partial_{h} c_{k,j}^{+}(\cdot)$ for $j=0, \ldots, 2(n-1)$ and $k =0,\ldots, n-1$. Although these derivatives can be implemented via (central) finite differencing, one may want to increase the stability of certain results. This can be achieved by deriving corresponding (non-linear) equations from (\ref{ROOT})-(\ref{REFL}) and (\ref{middd1})-(\ref{bboundeq}). For instance, differentiating Equation (\ref{ROOT}) gives that $\partial_{h} \rho_{+}(h(\mathcal{T}))$ solves, for any $\mathcal{T} \in (0,T]$, the following equation:
$$\big(r-\delta-\lambda(e^{\varphi}-1) - \frac{1}{2} \sigma^{2}\big) \partial_{h} \rho_{+}(h(\mathcal{T})) + \sigma^2 \rho_{+}(h(\mathcal{T})) \partial_{h} \rho_{+}(h(\mathcal{T})) + \lambda \varphi \partial_{h} \rho_{+}(h(\mathcal{T})) e^{\varphi \rho_{+}(h(\mathcal{T}))} = - \frac{r}{h(\mathcal{T})^2} .$$
\noindent Similarly, one obtains from Equations (\ref{REL}) and (\ref{REFL}) that, for any $\mathcal{T} \in (0,T]$, 
\begin{align*}
\partial_{h}c_{0,0}^{+}(h(\mathcal{T})) & = \frac{\Big(\mathfrak{b}_{0}^{\epsilon=1}(\mathcal{T}) \Big)^{1-\rho_{+}(h(\mathcal{T}))}}{re^{-r\mathcal{T}}} \Bigg[ \frac{ \frac{\partial_{\mathcal{T}} \mathfrak{b}_{0}^{\epsilon=1}(\mathcal{T})}{\mathfrak{b}_{0}^{\epsilon=1}(\mathcal{T})}\big(1-\rho_{+}(h(\mathcal{T}))\big)-\partial_{h}\rho_{+}(h(\mathcal{T})) re^{-r\mathcal{T}} \log(\mathfrak{b}_{0}^{\epsilon=1}(\mathcal{T}))}{h(\mathcal{T}) \rho_{+}(h(\mathcal{T}))} \\
& \hspace{6em} - \frac{re^{-r\mathcal{T}}\big( \rho_{+}(h(\mathcal{T})) + h(\mathcal{T}) \partial_{h} \rho_{+}(h(\mathcal{T})) \big)}{\big(h(\mathcal{T}) \rho_{+}(h(\mathcal{T})) \big)^{2}} \Bigg] \Big( 1 - \partial_{x} \mathcal{C}_{E} (\mathcal{T}, \mathfrak{b}_{0}^{\epsilon=1}(\mathcal{T});K) \Big) \\
&  - \frac{\Big(\mathfrak{b}_{0}^{\epsilon=1}(\mathcal{T}) \Big)^{1-\rho_{+}(h(\mathcal{T}))}}{re^{-r\mathcal{T}} h(\mathcal{T}) \rho_{+}(h(\mathcal{T}))} \Big(\partial_{\mathcal{T}} \partial_{x} \mathcal{C}_{E} (\mathcal{T}, \mathfrak{b}_{0}^{\epsilon=1}(\mathcal{T});K) + \partial_{x}^{2} \mathcal{C}_{E} (\mathcal{T}, \mathfrak{b}_{0}^{\epsilon=1}(\mathcal{T});K) \partial_{\mathcal{T}} \mathfrak{b}_{0}^{\epsilon=1}(\mathcal{T}) \Big),
\end{align*}
\noindent where $\partial_{\mathcal{T}} \mathfrak{b}_{0}^{\epsilon=1}(\mathcal{T})$ satisfies the following equation:
\begin{align*}
\partial_{\mathcal{T}} \mathfrak{b}_{0}^{\epsilon=1}(\mathcal{T}) & = \partial_{\mathcal{T}} \mathcal{C}_{E}(\mathcal{T},\mathfrak{b}_{0}^{\epsilon=1}(\mathcal{T});K) + \partial_{x} \mathcal{C}_{E}(\mathcal{T},\mathfrak{b}_{0}^{\epsilon=1}(\mathcal{T});K)\partial_{\mathcal{T}} \mathfrak{b}_{0}^{\epsilon=1}(\mathcal{T}) \\
& \hspace{2em}  + \frac{\partial_{\mathcal{T}} \mathfrak{b}_{0}^{\epsilon=1}(\mathcal{T}) \rho_{+}(h(\mathcal{T})) - \mathfrak{b}_{0}^{\epsilon=1}(\mathcal{T}) \partial_{h} \rho_{+}(h(\mathcal{T}))re^{-r\mathcal{T}}}{\rho_{+}(h(\mathcal{T}))^{2}} \Big( 1 - \partial_{x} \mathcal{C}_{E} (\mathcal{T}, \mathfrak{b}_{0}^{\epsilon=1}(\mathcal{T});K) \Big) \\
& \hspace{4em} - \frac{\mathfrak{b}_{0}^{\epsilon=1}(\mathcal{T})}{\rho_{+}(h(\mathcal{T}))} \Big( \partial_{\mathcal{T}} \partial_{x} \mathcal{C}_{E} (\mathcal{T}, \mathfrak{b}_{0}^{\epsilon=1}(\mathcal{T});K) + \partial_{x}^2 \mathcal{C}_{E} (\mathcal{T}, \mathfrak{b}_{0}^{\epsilon=1}(\mathcal{T});K) \partial_{\mathcal{T}} \mathfrak{b}_{0}^{\epsilon=1}(\mathcal{T})\Big).
\end{align*}
\noindent These results can now be used while implementing higher order algorithms ($n \in \mathbb{N}$). In particular, this allows to improve the stability of subsequent derivatives.\footnote{Without the use of such equations, subsequent derivatives would depend on the finite difference steps chosen in the computation of $\partial_{h} c_{0,0}^{+}(\cdot)$, which clearly lower their stability.}
\item[ii)] Implementing our approximations as well as the stability results described in i) requires some (analytical) tractability of the European call $\mathcal{C}_{E}(\cdot)$ (and of its derivatives) under the respective model. To keep this article self-contained, we therefore recall few results for $\mathcal{C}_{E}(\cdot)$ under Model (\ref{mixed}) in Appendix~A.
\end{itemize}
\hspace{44em} \scalebox{0.75}{$\blacklozenge$}
\subsection{Extension to Merton's Jump-Diffusion Model}
 \label{SECMERT}
\noindent We next combine the ansatz taken in \cite{ba91} with the ideas discussed previously to extend our $N$-th order algorithms to Merton's jump-diffusion model (cf.~\cite{me76}). We assume in the rest of this section that $J_{1}$ is normally distributed with mean $\mu_{\mathcal{M}}$ and variance $\sigma_{\mathcal{M}}^{2}$ or, equivalently, that $\Pi_{X}$ is given by
\begin{equation}
\lambda f_{J_{1}}(y) dy = \Pi_{X}(dy)  =  \frac{\lambda}{\sqrt{2 \pi \sigma_{\mathcal{M}}^{2}}} \exp \left( -\frac{(y-\mu_{\mathcal{M}})^{2}}{2\sigma_{\mathcal{M}}^{2}} \right) dy,
\label{MertMod}
\end{equation}
\noindent with $\mu_{\mathcal{M}} \in \mathbb{R}$ and $\sigma_{\mathcal{M}} > 0$ and obtain that $ \zeta = e^{\mu_{\mathcal{M}} + \frac{1}{2} \sigma_{\mathcal{M}}^{2}} - 1$ and that the Laplace exponent equals, for any $\theta \in \mathbb{R}$,
\begin{equation}
\Phi_{X}(\theta) = \Big(r-\delta - \lambda \zeta - \frac{1}{2}\sigma^{2} \Big) \theta + \frac{1}{2} \sigma^2 \theta^2  + \lambda \big( e^{\theta \mu_{\mathcal{M}} + \frac{1}{2}\theta^{2} \sigma_{\mathcal{M}}^{2}} -1\big).
\label{MertLaplace}
\end{equation}
\noindent Additionally, we point out that, by the very same arguments as the ones provided in Section \ref{SEC2}.\ref{Sol0thPro}, the equation $ \Phi_{X}(\theta) = y$ has for any $y >0$ two solutions, a positive and a negative root. We follow the notation used in the previous sections and denote by $\Phi_{X}^{-1,+}(y)$ its positive root.
\subsubsection{Solution of the $0$-th Order Problem}
\noindent As noted earlier, finding an exact solution to the $0$-th order problem, i.e.~to Equations (\ref{P14})-(\ref{P44}), is not anymore an easy task, as there is no known solution to Equation (\ref{P14}). Whenever $\mu_{\mathcal{M}}$ and $\sigma_{\mathcal{M}}$ are ``sufficiently small'', it seems however reasonable to follow our previous considerations and to use the following approximate solution 
\begin{equation}
f_{0}(h(\mathcal{T}),x;K) = c_{0,0}^{+}(h(\mathcal{T})) x^{\rho_{+}(h(\mathcal{T}))}, \hspace{1.5em}  x \in (0,\mathfrak{b}_{0}^{\epsilon=1}(\mathcal{T})),  \; \mathcal{T} \in (0,T],
\end{equation}
\noindent with
\begin{equation}
\rho_{+}(h(\mathcal{T})) := \Phi_{X}^{-1,+}\Big(\frac{r}{h(\mathcal{T})}\Big), \hspace{1.5em} \mathcal{T} \in (0,T].
\label{NEWRoot}
\end{equation}
\noindent This subsequently allows us to compute $c_{0,0}^{+}(\cdot)$ and $\mathfrak{b}_{0}^{\epsilon=1}(\cdot)$ via the same approach as the one used in Section~\ref{SEC2}.\ref{Sol0thPro} and to arrive at Equations (\ref{REL}), (\ref{REFL}), recovering so the $0$-th order premium $f_{0}(\cdot)$.

\subsubsection{Solution of the Higher Order Problems}
\noindent Solving the higher order problems can be done via the same method as the one introduced in Section \ref{SEC2}.\ref{SolNthPro} Indeed, assuming that the $n$-th order premium $f_{n}(\cdot)$ has the functional form described by (\ref{tati}), allows us to derive the following system of $2n$ equations in the $2n$ unknowns $\big( c_{n,j}^{+}(h(\mathcal{T})) \big)_{j \in \{ 1,\ldots,2n\}}$:
\begin{align}
 2n \bigg[ \frac{\sigma^{2}}{2} \big(2 \rho_{+}(h(\mathcal{T}))-1 \big) + r-\delta + \lambda (\mathfrak{I}_{1}(h(\mathcal{T})) - \zeta ) \bigg]  c_{n,2n}^{+}(h(\mathcal{T}))   \hspace{15em} \nonumber \\
 \hspace{5em} =   r (1-h(\mathcal{T})) \partial_{h}\rho_{+}(h(\mathcal{T})) c_{n-1,2(n-1)}^{+}(h(\mathcal{T})) , \hspace{6.4em} \label{NEWmiddd1}
\end{align}
\begin{align}
 j \bigg[ \frac{\sigma^{2}}{2} \big(2 \rho_{+}(h(\mathcal{T}))-1 \big) + r-\delta + \lambda (\mathfrak{I}_{1}(h(\mathcal{T})) - \zeta) \bigg]  c_{n,j}^{+}(h(\mathcal{T}))  \hspace{17em} \nonumber  \\
\hspace{5em} +  j (j+1) \frac{\sigma^{2}}{2} c_{n,j+1}^{+}(h(\mathcal{T})) + \lambda \sum \limits_{k=j+1}^{2n} \binom{k}{j-1} \mathfrak{I}_{k-(j-1)}(h(\mathcal{T}))  c_{n,k}^{+}(h(\mathcal{T}))\hspace{7em} \label{NEWmidie}\\ 
\hspace{14em} = r (1-h(\mathcal{T})) \big( \partial_{h} c_{n-1,j-1}^{+}(h(\mathcal{T})) + \partial_{h}\rho_{+}(h(\mathcal{T})) c_{n-1,j-2}^{+}(h(\mathcal{T}))\big), \hspace{-1.5em} \nonumber
\end{align}
\begin{align}
\bigg[ \frac{\sigma^{2}}{2} \big(2 \rho_{+}(h(\mathcal{T}))-1 \big) + r-\delta + \lambda \big(\mathfrak{I}_{1}(h(\mathcal{T})) - \zeta) \bigg] c_{n,1}^{+}(h(\mathcal{T}))  \hspace{16em} \nonumber \\ 
\hspace{4em} + \sigma^2 c_{n,2}^{+}(h(\mathcal{T})) + \lambda \sum \limits_{k=2}^{2n} \mathfrak{I}_{k}(h(\mathcal{T})) c_{n,k}^{+}(h(\mathcal{T}))  =  r (1-h(\mathcal{T})) \partial_{h} c_{n-1,0}^{+}(h(\mathcal{T})), \hspace{4.7em} \label{NEWmiddd2} 
\end{align}
\noindent where Equation (\ref{NEWmidie}) only holds for $j \in \{2,3, \ldots, 2n-1 \}$ and $\mathfrak{I}_{\ell}(\cdot)$ is given, for any $\ell \in \mathbb{N}$, by
\begin{equation}
\mathfrak{I}_{\ell}(h(\mathcal{T})) := \int \limits_{\mathbb{R}} y^{\ell} e^{\rho_{+}(h(\mathcal{T}))y} f_{J_{1}}(y) dy.
\end{equation}
\noindent Using standard calculus, the latter integral can be re-expressed as
\begin{equation}
\mathfrak{I}_{\ell}(h(\mathcal{T})) = \exp \bigg\{ \rho_{+}(h(\mathcal{T})) \Big(\mu_{\mathcal{M}} + \frac{\sigma_{\mathcal{M}}^{2}}{2} \rho_{+}(h(\mathcal{T})) \Big) \bigg\} \cdot \sum \limits_{k=0}^{\ell} \binom{\ell}{k} \mu_{\mathcal{M}}^{\ell - k} \; \mathfrak{M}\big(k, \sigma_{\mathcal{M}}^{2} \rho_{+}(h(\mathcal{T})), \sigma_{\mathcal{M}}^{2} \big),
\label{SUUUPer1}
\end{equation}
\noindent where
\begin{equation}
\mathfrak{M}(k,m,s^{2}) := \int \limits_{\mathbb{R}} z^{k} \frac{1}{\sqrt{2 \pi s^{2}}} e^{-\frac{(z-m)^{2}}{2 s^{2}} } dz
\label{SUUUPer2}
\end{equation}
denotes the $k$-th order non-central moment of the normal distribution having mean $m$ and variance $s^{2}$. Therefore, combining (\ref{NEWmiddd1})-(\ref{NEWmiddd2}) with (\ref{SUUUPer1}) and (\ref{SUUUPer2}) allows us to recover $\big( c_{n,j}^{+}(h(\mathcal{T})) \big)_{j \in \{ 1,\ldots,2n\}}$ for any $\mathcal{T} \in (0,T]$ via standard numerical techniques. To derive an expression for $c_{n,0}^{+}(\cdot)$ and $ \mathfrak{b}_{n}^{\epsilon=1}(\cdot)$ we follow the steps outlined in Section \ref{SEC2}.\ref{SolNthPro} This finally leads to Equations (\ref{cn0eq}) and (\ref{bboundeq}), from which the $n$-th order premium $f_{n}(\cdot)$ is ultimately recovered. \vspace{1em} \\
\noindent \underline{\bf Remark 2.}
\begin{itemize} \setlength \itemsep{-0.1em}
\item[i)] As in the model of constant jumps, one can derive (non-linear) equations that help stabilizing higher order approximations. This can be done using Equations (\ref{REL})-(\ref{REFL}), (\ref{cn0eq})-(\ref{bboundeq}) and (\ref{NEWRoot}), (\ref{NEWmiddd1})-(\ref{NEWmiddd2}). 
\item[ii)] Implementing our approximations as well as the stability results described in i) requires some (analytical) tractability of the European call $\mathcal{C}_{E}(\cdot)$ (and of its derivatives) under the respective model. In Appendix B, we therefore recall few results for $\mathcal{C}_{E}(\cdot)$ under Model (\ref{MertMod}).
\item[iii)] Although this article does not investigate jump-diffusion models behind the model of Merton, we believe that the general ideas underlying our method can be combined with the results obtained in \cite{kw04} and \cite{cs14} to derive $N$-th order approximations to the pricing of American-type options within the whole class of hyper-exponential jump-diffusions. This could be investigated as part of future research.
\end{itemize}
\hspace{44em} \scalebox{0.75}{$\blacklozenge$}


\section{Approximation of American Barrier Options}
\label{BARRR}
We next adapt the previous method to deal with American barrier options. As an extension of the Barone-Adesi \& Whaley algorithm, our ansatz relies once again on the (analytical) tractability of the corresponding European-type options (and Greeks). However, since analytical results for European barrier-type options are mainly known in the setting of Black \& Scholes, we focus in the sequel on this model, i.e.~we assume from now on that $(S_{t})_{t \geq 0}$ evolves according to (\ref{mixed}) with $\varphi = 0$. Investigating the applicability of our method under other asset dynamics (e.g.~under certain jump-diffusion dynamics) could be part of future work. Additionally, our derivations will focus on the American down-and-out call (DOC). Nevertheless, we note that our method can be slightly adapted to deal with any other type of (single) barrier options.\footnote{More details on barrier options, their relations, and on how to adapt our method to deal with other types of barriers can be found in \cite{jy06}, \cite{gh00} and \cite{ch07}.}~To illustrate this point, the numerical discussion in Section \ref{numSEC} also provides simulation results for the American up-and-out put (UOP).
  
\subsection{Pricing Problem and Perturbation Expansion}
\label{pwr1}
\subsubsection{Pricing with Rebates}
\label{pwr2}
Let us start by reviewing well-known facts on American down-and-out call options. To keep our derivations applicable in a wide range of problems, we consider barrier options with strike-and-barrier-dependent rebates, i.e.~we consider the following American-type down-and-out call option having maturity $\mathcal{T} \geq 0$, initial value $S_{0}= x \geq 0$, strike price $K \geq 0$, (lower) barrier level $L \geq 0$ and rebate $\mathcal{R}(K,L)$:
\begin{align}
\mathcal{DOC}_{A}(\mathcal{T},x;K,L,\mathcal{R}) & := \sup \limits_{ \tau \in \mathfrak{T}_{[0,\mathcal{T}]} } \mathbb{E}^{\mathbb{Q}}_{x} \left[ B_{\tau}(r)^{-1} \left(S_{\tau} -K \right)^{+} \mathds{1}_{\{ \tau_{L} > \tau \}} + B_{\tau_{L}}(r)^{-1} \,\mathcal{R}(K,L) \mathds{1}_{ \{ \tau_{L} \leq \tau \}} \right] \nonumber \\
& = \sup \limits_{ \tau \in \mathfrak{T}_{[0,\mathcal{T}]} } \Big( \mathbb{E}^{\mathbb{Q}}_{x} \left[ B_{\tau}(r)^{-1} \left(S_{\tau} -K \right)^{+} \mathds{1}_{\{ \tau_{L} > \tau \}}\right] + \mathcal{R}(K,L) \mathbb{E}^{\mathbb{Q}}_{x} \left[ B_{\tau_{L}}(r)^{-1} \,\mathds{1}_{ \{\tau_{L} \leq \tau \} } \right] \Big) .
\label{BarrAMERO}
\end{align}
\noindent Here $ \tau_{L} : = \inf \{ t >0 : \; S_{t} \leq L \}$ denotes the first passage time of the process $(S_{t})_{t \geq 0}$ below the (lower) barrier level $L$, while $\mathfrak{T}_{[0,\mathcal{T}]}$ refers, as earlier, to the set of stopping times that take values in the interval~$[0,\mathcal{T}]$. Additionally, we define the European counterpart to (\ref{BarrAMERO}) via 
\begin{equation}
\mathcal{DOC}_{E}(\mathcal{T},x;K,L,\mathcal{R}) := \mathbb{E}^{\mathbb{Q}}_{x} \left[ B_{\mathcal{T}}(r)^{-1} \left(S_{\mathcal{T}} -K \right)^{+} \mathds{1}_{\{ \tau_{L} > \mathcal{T} \}}\right] + \mathcal{R}(K,L) \mathbb{E}^{\mathbb{Q}}_{x} \left[ B_{\tau_{L}}(r)^{-1} \,\mathds{1}_{ \{\tau_{L} \leq \mathcal{T} \} } \right],
\label{BarrEURO}
\end{equation}
and note that, in the above definitions (\ref{BarrAMERO}) and (\ref{BarrEURO}), the rebates are implicitly understood to be paid immediately. \vspace{1em} \\
\noindent As for standard American options, decomposition techniques are popular methods to price American barrier options. Following this ansatz as well as the line of arguments provided in Section \ref{PricProbl}, we therefore define the down-and-out early exercise premium, $\mathcal{E}_{\mathcal{DOC}}(\cdot)$, via
\begin{equation}
\mathcal{E}_{\mathcal{DOC}}(\mathcal{T},x;K,L,\mathcal{R}) := \mathcal{DOC}_{A}(\mathcal{T},x;K,L,\mathcal{R}) - \mathcal{DOC}_{E}(\mathcal{T},x;K,L,\mathcal{R}),
\label{EEprem}
\end{equation}
\noindent and focus on the respective pricing problem for (\ref{EEprem}). Here, we first note that the American-type option (\ref{BarrAMERO}) should not be exercised before maturity whenever $\delta \leq 0$ and consequently reduces in this case to its European counterpart (\ref{BarrEURO}).\footnote{This is in line with the analysis provided in \cite{gh00}.}~Hence, we focus in the sequel on the pricing problem in the non-trivial case, $\delta > 0$. In this case, one obtains, by slightly adapting the arguments presented in \cite{ma18},\footnote{See also \cite{gh00}, \cite{ga07}, and \cite{al14} for corresponding results.}~that the pair $\big( \mathcal{E}_{\mathcal{DOC}}(\cdot), \mathfrak{b}_{\mathcal{DOC}}(\cdot) \big)$, where $\mathfrak{b}_{\mathcal{DOC}}(\cdot)$ denotes the down-and-out early exercise boundary, is a solution to the following free-boundary problem:
\begin{equation}
-\partial_{\mathcal{T}} \mathcal{E}_{\mathcal{DOC}}(\mathcal{T},x;K,L,\mathcal{R}) + \mathcal{A}_{S} \mathcal{E}_{\mathcal{DOC}} (\mathcal{T},x;K,L,\mathcal{R}) - r \mathcal{E}_{\mathcal{DOC}}(\mathcal{T},x;K,L,\mathcal{R})  =  0 , \; \; \;  x \in (0,\mathfrak{b}_{\mathcal{DOC}}(\mathcal{T})), \; \mathcal{T} \in (0,T], \label{PBa1}
\end{equation}
\noindent subject to the boundary conditions
\begin{align}
\mathcal{E}_{\mathcal{DOC}}(\mathcal{T}, \mathfrak{b}_{\mathcal{DOC}}(\mathcal{T});K,L,\mathcal{R}) & = \mathfrak{b}_{\mathcal{DOC}}(\mathcal{T}) -K - \mathcal{DOC}_{E}(\mathcal{T}, \mathfrak{b}_{\mathcal{DOC}}(\mathcal{T});K,L,\mathcal{R}), \; \; \; \; \; \mathcal{T} \in (0,T], \label{PBa2} \\
\partial_{x} \mathcal{E}_{\mathcal{DOC}}( \mathcal{T}, \mathfrak{b}_{\mathcal{DOC}}(\mathcal{T});K,L,\mathcal{R}) & =  1 - \partial_{x} \mathcal{DOC}_{E} (\mathcal{T}, \mathfrak{b}_{\mathcal{DOC}}(\mathcal{T});K,L,\mathcal{R}), \; \; \; \; \; \mathcal{T} \in (0,T], \label{PBa3} \\
\mathcal{E}_{\mathcal{DOC}}(\mathcal{T},x;K,L,\mathcal{R}) & =  0, \; \; \; \; \; x \in [0,L], \; \mathcal{T} \in (0,T], \label{PBa4}
\end{align}
\noindent and initial condition
\begin{equation}
\mathcal{E}_{\mathcal{DOC}}(0, x;K,L,\mathcal{R}) =  0, \; \; \; \; \; x \in (0,\mathfrak{b}_{\mathcal{DOC}}(\mathcal{T})).
\label{PBa5}
\end{equation}
\subsubsection{Perturbation Ansatz}
\noindent We next repeat the ansatz adopted by Barone-Adesi \& Whaley (cf.~\cite{ba87}) and assume that the early exercise premium $\mathcal{E}_{\mathcal{DOC}}(\cdot)$ takes the form
\begin{equation}
\mathcal{E}_{\mathcal{DOC}}(\mathcal{T},x;K,L,\mathcal{R}) = h(\mathcal{T})F_{\mathcal{DOC}}(h(\mathcal{T}),x;K,L,\mathcal{R}).
\end{equation}
\noindent This allows us (to transform (\ref{PBa1})-(\ref{PBa4}) into an analogue of (\ref{P12})-(\ref{P42}) and) to arrive at the following alternative perturbation problem:
\begin{equation}
- \frac{r}{h(\mathcal{T})} F_{\mathcal{DOC}}^{\epsilon}(h(\mathcal{T}),x;K,L,\mathcal{R}) + \mathcal{A}_{S} F_{\mathcal{DOC}}^{\epsilon}(h(\mathcal{T}),x;K,L,\mathcal{R}) - \epsilon r \big(1-h(\mathcal{T})\big) \partial_{h}F_{\mathcal{DOC}}^{\epsilon}(h(\mathcal{T}),x;K,L,\mathcal{R}) =0, \label{PBa13}
\end{equation}
\noindent on $(\mathcal{T},x) \in (0,T] \times (0,\mathfrak{b}_{\mathcal{DOC}}^{\epsilon}(\mathcal{T}))$ and with boundary conditions
\begin{align}
F_{\mathcal{DOC}}^{\epsilon}(h(\mathcal{T}), \mathfrak{b}_{\mathcal{DOC}}^{\epsilon}(\mathcal{T});K,L,\mathcal{R}) & = \frac{1}{h(\mathcal{T})} \Big( \mathfrak{b}_{\mathcal{DOC}}^{\epsilon}(\mathcal{T}) - K - \mathcal{DOC}_{E}(\mathcal{T}, \mathfrak{b}_{\mathcal{DOC}}^{\epsilon}(\mathcal{T});K,L,\mathcal{R}) \Big), \; \; \; \; \; \mathcal{T} \in (0,T], \label{PBa23} \\
\partial_{x} F_{\mathcal{DOC}}^{\epsilon}( h(\mathcal{T}), \mathfrak{b}_{\mathcal{DOC}}^{\epsilon}(\mathcal{T});K,L,\mathcal{R}) & = \frac{1}{h(\mathcal{T})} \Big( 1 - \partial_{x} \mathcal{DOC}_{E} (\mathcal{T}, \mathfrak{b}_{\mathcal{DOC}}^{\epsilon}(\mathcal{T});K,L,\mathcal{R}) \Big), \;  \; \; \; \; \mathcal{T} \in (0,T], \label{PBa33} \\
F_{\mathcal{DOC}}^{\epsilon}(h(\mathcal{T}),x;K,L,\mathcal{R}) & =  0,  \; \; \; \; \; x \in [0,L], \; \mathcal{T} \in (0,T]. \label{PBa43}
\end{align}
\noindent As earlier, we assume the existence of functions $\big(f^{\mathcal{DOC}}_{n}(\cdot)\big)_{n  \geq 0}$ and $\big(b^{\mathcal{DOC}}_{n}(\cdot)\big)_{n  \geq 0}$ such that the solution pair $\big(F_{\mathcal{DOC}}^{\epsilon}(\cdot),\mathfrak{b}_{\mathcal{DOC}}^{\epsilon}(\cdot)\big)$ to (\ref{PBa13})-(\ref{PBa43}) has a representation as ``well-behaved'' series expansion of the form
\begin{align}
 F^{\epsilon}_{\mathcal{DOC}}(h(\mathcal{T}),x;K,L,\mathcal{R}) & = \sum \limits_{n=0}^{\infty} \epsilon^{n}  f^{\mathcal{DOC}}_{n}(h(\mathcal{T}),x;K,L,\mathcal{R}), \; \; \; \; \; \; \;  x \in (0,\mathfrak{b}_{\mathcal{DOC}}^{\epsilon}(\mathcal{T})), \mathcal{T} \in [0,T],  \label{Sum1Ba}\\
  \mathfrak{b}_{\mathcal{DOC}}^{\epsilon}(\mathcal{T}) & =  \sum \limits_{n=0}^{\infty} \epsilon^{n}  b^{\mathcal{DOC}}_{n}(\mathcal{T}), \; \; \; \; \; \; \; \mathcal{T} \in [0,T], \label{Sum2Ba}
\end{align} 
\noindent and define corresponding partial sums of $N$-th order via 
\begin{align}
 F_{\mathcal{DOC},N}^{\epsilon}(h(\mathcal{T}),x;K,L,\mathcal{R}) & =  \sum \limits_{n=0}^{N} \epsilon^{n}  f^{\mathcal{DOC}}_{n}(h(\mathcal{T}),x;K,L,\mathcal{R}), \; \; \; \; \; \; \;  x \in (0,\mathfrak{b}_{\mathcal{DOC},N}^{\epsilon}(\mathcal{T})), \mathcal{T} \in [0,T], \label{susuBa} \\
  \mathfrak{b}_{\mathcal{DOC},N}^{\epsilon}(\mathcal{T}) & =  \sum \limits_{n=0}^{N} \epsilon^{n}  b^{\mathcal{DOC}}_{n}(\mathcal{T}), \; \; \; \; \; \; \; \mathcal{T} \in [0,T].
\end{align}
\noindent Hence, arguing again as in Section~\ref{PricProbl}.\ref{GenBAW} leads us to $n$-th order analogues to Problems (\ref{P14})-(\ref{P44}) and (\ref{P15})-(\ref{P45}): For $n=0$, the $0$-th order problem reads
\begin{equation}
- \frac{r}{h(\mathcal{T})} f^{\mathcal{DOC}}_{0}(h(\mathcal{T}),x;K,L,\mathcal{R}) + \mathcal{A}_{S} f^{\mathcal{DOC}}_{0}(h(\mathcal{T}),x;K,L,\mathcal{R}) =0, \; \; \; \; \;  x \in (0,\mathfrak{b}_{\mathcal{DOC},0}^{\epsilon}(\mathcal{T})), \mathcal{T} \in (0,T], \label{PBa14}
\end{equation}
\noindent with boundary conditions
\begin{align}
f^{\mathcal{DOC}}_{0}(h(\mathcal{T}), \mathfrak{b}_{\mathcal{DOC},0}^{\epsilon}(\mathcal{T});K,L,\mathcal{R}) & = \frac{1}{h(\mathcal{T})} \Big( \mathfrak{b}_{\mathcal{DOC},0}^{\epsilon}(\mathcal{T}) -K - \mathcal{DOC}_{E}(\mathcal{T}, \mathfrak{b}_{\mathcal{DOC},0}^{\epsilon}(\mathcal{T});K,L,\mathcal{R}) \Big), \; \; \; \; \; \mathcal{T} \in (0,T], \label{PBa24} \\
\partial_{x} f^{\mathcal{DOC}}_{0}( h(\mathcal{T}), \mathfrak{b}_{\mathcal{DOC},0}^{\epsilon}(\mathcal{T});K,L,\mathcal{R}) & = \frac{1}{h(\mathcal{T})} \Big( 1 - \partial_{x} \mathcal{DOC}_{E} (\mathcal{T}, \mathfrak{b}_{\mathcal{DOC},0}^{\epsilon}(\mathcal{T});K,L,\mathcal{R}) \Big), \; \; \; \; \; \mathcal{T} \in (0,T], \label{PBa34} \\
f_{0}(h(\mathcal{T}),x;K,L,\mathcal{R}) & = 0, \; \; \; \; \; x \in [0,L], \; \mathcal{T} \in (0,T]. \label{PBa44}
\end{align}
\noindent Additionally, the following higher order problems ($n\in \mathbb{N}$) are obtained:
 \begin{equation}
- \frac{r}{h(\mathcal{T})} f_{n}^{\mathcal{DOC}}(h(\mathcal{T}),x;K,L,\mathcal{R}) + \mathcal{A}_{S} f_{n}^{\mathcal{DOC}}(h(\mathcal{T}),x;K,L,\mathcal{R}) - r\big(1-h(\mathcal{T})\big) \partial_{h}f_{n-1}^{\mathcal{DOC}}(h(\mathcal{T}),x;K,L,\mathcal{R}) =0, \label{PBa15}
\end{equation}
\noindent on $(\mathcal{T},x) \in (0,T] \times (0,\mathfrak{b}_{\mathcal{DOC},n}^{\epsilon}(\mathcal{T}))$ and with boundary conditions, for $\mathcal{T} \in (0,T]$:
\begin{align}
f_{n}^{\mathcal{DOC}}(h(\mathcal{T}), \mathfrak{b}_{\mathcal{DOC},n}^{\epsilon}(\mathcal{T});K,L,\mathcal{R}) & = \frac{\epsilon^{-n}}{ h(\mathcal{T})} \Big( \mathfrak{b}_{\mathcal{DOC},n}^{\epsilon}(\mathcal{T}) -K - \mathcal{DOC}_{E}(\mathcal{T}, \mathfrak{b}_{\mathcal{DOC},n}^{\epsilon}(\mathcal{T});K,L,\mathcal{R}) \nonumber \\
& \hspace{6.5em} - h(\mathcal{T})F_{\mathcal{DOC},n-1}^{\epsilon}(h(\mathcal{T}),\mathfrak{b}_{\mathcal{DOC},n}^{\epsilon}(\mathcal{T});K,L,\mathcal{R})\Big), \label{PBa25} \\
\partial_{x} f_{n}^{\mathcal{DOC}}( h(\mathcal{T}), \mathfrak{b}_{\mathcal{DOC},n}^{\epsilon}(\mathcal{T});K,L,\mathcal{R}) & =  \frac{\epsilon^{-n}}{h(\mathcal{T})} \Big( 1 - \partial_{x} \mathcal{DOC}_{E} (\mathcal{T}, \mathfrak{b}_{\mathcal{DOC},n}^{\epsilon}(\mathcal{T});K,L,\mathcal{R}) \nonumber \\
& \hspace{6.5em} - h(\mathcal{T}) \partial_{x} F_{\mathcal{DOC},n-1}^{\epsilon}(h(\mathcal{T}),\mathfrak{b}_{\mathcal{DOC},n}^{\epsilon}(\mathcal{T});K,L,\mathcal{R}) \Big), \label{PBa35} \\
f_{n}^{\mathcal{DOC}}(h(\mathcal{T}),x;K,L,\mathcal{R}) & = 0, \; \; \; \; \; x \in [0,L].\label{PBa45}
\end{align}
Solving these problems for $\epsilon = 1$ clearly allows us to recover $N$-th order approximations of the down-and-out early exercise premium via
\begin{equation}
\mathcal{E}_{\mathcal{DOC},N}^{\epsilon=1}(\mathcal{T},x;K,L,\mathcal{R})  :=  h(\mathcal{T})F_{\mathcal{DOC},N}^{\epsilon=1}(h(\mathcal{T}),x;K,L,\mathcal{R})  \hspace{2em} x \in [0,\mathfrak{b}_{\mathcal{DOC},N}^{\epsilon=1}(\mathcal{T})), \label{aboveBa}
\end{equation}
\begin{equation}
\mathcal{E}_{\mathcal{DOC},N}^{\epsilon=1}(\mathcal{T},x;K,L,\mathcal{R}) := x-K - \mathcal{DOC}_{E}(\mathcal{T},x;K,L,\mathcal{R}), \hspace{2em}  x \in [\mathfrak{b}_{\mathcal{DOC},N}^{\epsilon=1}(\mathcal{T}),\infty),
\label{above2Ba}
\end{equation}
\noindent where $\mathcal{T} \in [0,T]$. Therefore, we focus in the sequel on the corresponding problems for $\epsilon =1$.
\subsection{Derivation of the Solutions}
\subsubsection{Solution of the $0$-th Order Problem}
\noindent To derive a solution to the $0$-th order problem, we decompose the (state-)domain of Equation (\ref{PBa14}) for any $\mathcal{T} \in (0,T]$ into two intervals, $I_{0} := [0,L]$ and $I_{1} := (L, \mathfrak{b}_{\mathcal{DOC},0}^{\epsilon=1}(\mathcal{T}))$, derive solutions $V_{0}^{\mathcal{DOC}}(\cdot)$, $V_{1}^{\mathcal{DOC}}(\cdot)$ on these domains and combine them to recover $f_{0}^{\mathcal{DOC}}(\cdot)$ via
\begin{equation}
f_{0}^{\mathcal{DOC}}(h(\mathcal{T}),x;K,L,\mathcal{R}) = \left \{ \begin{array}{cc}
V_{0}^{\mathcal{DOC}}(h(\mathcal{T}),x;K,L,\mathcal{R}), & x \in I_{0}, \\
V_{1}^{\mathcal{DOC}}(h(\mathcal{T}),x;K,L,\mathcal{R}), & x \in I_{1}.
\end{array} \right. 
\label{FnullDef}
\end{equation}
\noindent First, it is clear that $V_{0}^{\mathcal{DOC}}(h(\mathcal{T}),x;K,L,\mathcal{R}) \equiv 0$ must hold for $x \in I_{0}$. Therefore, we only need to derive an expression for $V_{1}^{\mathcal{DOC}}(\cdot)$. Here, following the arguments provided in Section \ref{SEC2}.\ref{Sol0thPro}, the general solution of the homogeneous equation (\ref{PBa14}) on $I_{1}$ is obtained as
\begin{equation}
V_{1}^{\mathcal{DOC}}(h(\mathcal{T}),x;K,L,\mathcal{R}) = c_{0,0}^{\mathcal{DOC},+}(h(\mathcal{T})) x^{\rho_{+}(h(\mathcal{T}))} + c_{0,0}^{\mathcal{DOC},-}(h(\mathcal{T})) x^{\rho_{-}(h(\mathcal{T}))}, \hspace{1.5em}  x \in I_{1}, \; \mathcal{T} \in (0,T],
\end{equation} 
\noindent where $\rho_{+}(\cdot)$ and $\rho_{-}(\cdot)$ are defined as in (\ref{ROOT}) but with $\varphi =0$ and $c_{0,0}^{\mathcal{DOC},+}(\cdot)$, $c_{0,0}^{\mathcal{DOC},-}(\cdot)$ are ``constants'' to be determined. To conclude, we therefore need to determine $c_{0,0}^{\mathcal{DOC},+}(\cdot)$, $c_{0,0}^{\mathcal{DOC},-}(\cdot)$ as well as $\mathfrak{b}_{\mathcal{DOC},0}^{\epsilon=1}(\cdot)$. This is done by combining Conditions (\ref{PBa24})-(\ref{PBa44}). Indeed, Condition (\ref{PBa44}) first implies that
\begin{equation}
c_{0,0}^{\mathcal{DOC},-}(h(\mathcal{T})) = - L^{\rho_{+}(h(\mathcal{T}))-\rho_{-}(h(\mathcal{T}))} \cdot c_{0,0}^{\mathcal{DOC},+}(h(\mathcal{T})), \hspace{1.5em} \mathcal{T} \in (0,T].
\label{LATTER}
\end{equation}
\noindent Then, combining (\ref{LATTER}) with Condition (\ref{PBa34}) allows us to derive, for $\mathcal{T} \in (0,T]$, that
\begin{equation}
c_{0,0}^{\mathcal{DOC},+}(h(\mathcal{T})) = \frac{ 1 - \partial_{x} \mathcal{DOC}_{E} (\mathcal{T}, \mathfrak{b}_{\mathcal{DOC},0}^{\epsilon=1}(\mathcal{T});K,L,\mathcal{R}) }{h(\mathcal{T}) \Big(\rho_{+}(h(\mathcal{T})) \big(\mathfrak{b}_{\mathcal{DOC},0}^{\epsilon=1} (\mathcal{T}) \big)^{\rho_{+}(h(\mathcal{T}))-1} - \rho_{-}(h(\mathcal{T})) L^{\rho_{+}(h(\mathcal{T})) - \rho_{-}(h(\mathcal{T}))} \big(\mathfrak{b}_{\mathcal{DOC},0}^{\epsilon=1} (\mathcal{T}) \big)^{\rho_{-}(h(\mathcal{T}))-1} \Big)}
\label{CSTAR}
\end{equation}
\noindent and inserting the latter expression into (\ref{PBa24}) finally gives that $\mathfrak{b}_{\mathcal{DOC},0}^{\epsilon=1}(\mathcal{T})$ solves, for any $\mathcal{T} \in (0,T]$, the following non-linear equation
\begin{align}
\mathfrak{b}_{\mathcal{DOC},0}^{\epsilon=1}(\mathcal{T}) = K + \mathcal{DOC}_{E} (\mathcal{T}, \mathfrak{b}_{\mathcal{DOC},0}^{\epsilon=1}(\mathcal{T});K,L,\mathcal{R})  \hspace{26em}\nonumber \\
 \hspace{0.75em} + \frac{\Big( 1 - \partial_{x} \mathcal{DOC}_{E} (\mathcal{T}, \mathfrak{b}_{\mathcal{DOC},0}^{\epsilon=1}(\mathcal{T});K,L,\mathcal{R}) \Big) \Big( \big(\mathfrak{b}_{\mathcal{DOC},0}^{\epsilon=1}(\mathcal{T})\big)^{\rho_{+}(h(\mathcal{T}))} -  L^{\rho_{+}(h(\mathcal{T})) - \rho_{-}(h(\mathcal{T}))} \big(\mathfrak{b}_{\mathcal{DOC},0}^{\epsilon=1}(\mathcal{T}) \big)^{\rho_{-}(h(\mathcal{T}))} \Big)}{\rho_{+}(h(\mathcal{T})) \big(\mathfrak{b}_{\mathcal{DOC},0}^{\epsilon=1} (\mathcal{T}) \big)^{\rho_{+}(h(\mathcal{T}))-1} - \rho_{-}(h(\mathcal{T})) L^{\rho_{+}(h(\mathcal{T})) - \rho_{-}(h(\mathcal{T}))} \big(\mathfrak{b}_{\mathcal{DOC},0}^{\epsilon=1} (\mathcal{T}) \big)^{\rho_{-}(h(\mathcal{T}))-1}} .
\label{LAAST}
\end{align}
\noindent Therefore, solving for any $\mathcal{T} \in (0,T]$ Equation (\ref{LAAST}) for $\mathfrak{b}_{\mathcal{DOC},0}^{\epsilon=1}(\mathcal{T}) $ numerically allows us to recover $c_{0,0}^{\mathcal{DOC},+}(h(\mathcal{T}))$, $c_{0,0}^{\mathcal{DOC},-}(h(\mathcal{T}))$, and subsequently $f_{0}^{\mathcal{DOC}}(\cdot)$ via (\ref{FnullDef}).

\subsubsection{Solution of the Higher Order Problems}
\noindent We finally seek, for $n \in \mathbb{N}$, a solution to Problem (\ref{PBa15})-(\ref{PBa45}). As in the previous section, we define $I_{0} := [0,L]$ and $I_{1} := (L,\mathfrak{b}_{\mathcal{DOC},n}^{\epsilon=1}(\mathcal{T}))$, decompose $f_{n}^{\mathcal{DOC}}(\cdot)$, for any $\mathcal{T} \in (0,T]$, via
\begin{equation}
f_{n}^{\mathcal{DOC}}(h(\mathcal{T}),x;K,L,\mathcal{R}) = \left \{ \begin{array}{cc}
V_{n,0}^{\mathcal{DOC}}(h(\mathcal{T}),x;K,L,\mathcal{R}), & x \in I_{0}, \\
V_{n,1}^{\mathcal{DOC}}(h(\mathcal{T}),x;K,L,\mathcal{R}), & x \in I_{1},
\end{array} \right. 
\label{FHigherDef}
\end{equation}
\noindent and derive expressions for the relevant functions. In view of Equation (\ref{PBa45}), it directly follows that $V_{n,0}^{\mathcal{DOC}}(h(\mathcal{T}),x;K) \equiv 0$ must hold on $I_{0}$.
\noindent Furthermore, following the ansatz taken in Section \ref{SEC2}.\ref{SolNthPro}, we now assume that $V_{n,1}^{\mathcal{DOC}}(\cdot)$ takes for $x \in (0,\mathfrak{b}_{\mathcal{DOC},n}^{\epsilon=1}(\mathcal{T}))$, $\mathcal{T} \in (0,T]$, the following form
\begin{align}
V_{n,1}^{\mathcal{DOC}}(h(\mathcal{T}),x;K,L,\mathcal{R}) & = \bigg(c_{n,0}^{\mathcal{DOC},+}(h(\mathcal{T})) + \sum \limits_{j=1}^{2n} c_{n,j}^{\mathcal{DOC},+}(h(\mathcal{T})) \log(x)^{j} \bigg) x^{\rho_{+}(h(\mathcal{T}))} \hspace{10em} \nonumber\\
& \hspace{5em} + \bigg(c_{n,0}^{\mathcal{DOC},-}(h(\mathcal{T})) + \sum \limits_{j=1}^{2n} c_{n,j}^{\mathcal{DOC},-}(h(\mathcal{T})) \log(x)^{j} \bigg) x^{\rho_{-}(h(\mathcal{T}))}, 
\label{ExprEE1}
\end{align}
\noindent and derive a system of equations in the coefficients $\Big(c_{n,j}^{\mathcal{DOC},\pm}(\cdot)\Big)_{j \in \{1, \ldots, 2n \}}$. Indeed, proceeding as in Section~\ref{SEC2}.\ref{SolNthPro} gives that the coefficients $\Big(c_{n,j}^{\mathcal{DOC},\pm}(h(\mathcal{T}))\Big)_{j \in \{1, \ldots, 2n \}}$ solve for any $\mathcal{T} \in (0,T]$ System (\ref{middd1})-(\ref{middd2}), where $\rho_{+}(h(\mathcal{T})$ is then replaced by $\rho_{\pm}(h(\mathcal{T}))$ and $\varphi = 0$. To conclude, we therefore need to determine $c_{n,0}^{\mathcal{DOC},+}(\cdot)$, $c_{n,0}^{\mathcal{DOC},-}(\cdot)$ and $\mathfrak{b}_{\mathcal{DOC},n}^{\epsilon=1}(\cdot)$ and this is done via the same methods as the ones used in the previous section: First, we obtain from Condition (\ref{PBa45}) that
\begin{equation}
c_{n,0}^{\mathcal{DOC},-}(h(\mathcal{T})) = -  \Big( L^{\rho_{+}(h(\mathcal{T}))-\rho_{-}(h(\mathcal{T}))} \cdot c_{0,0}^{\mathcal{DOC},+}(h(\mathcal{T})) + \mathfrak{R}_{n}^{\star}(h(\mathcal{T}),L;K,L,\mathcal{R}) \Big), \hspace{1.5em} \mathcal{T} \in (0,T],
\label{TheRep}
\end{equation}
\noindent where, for $x \in [0,\mathfrak{b}_{\mathcal{DOC},n}^{\epsilon=1} (\mathcal{T})]$ and $\mathcal{T} \in (0,T]$, the ``rest term'', $\mathfrak{R}_{n}^{\star}(\cdot)$, equals
\begin{equation}
\mathfrak{R}_{n}^{\star}(h(\mathcal{T}),x;K,L,\mathcal{R}) := L^{-\rho_{-}(h(\mathcal{T}))} \cdot V_{n,1}^{\mathcal{DOC},\star}(h(\mathcal{T}),x;K,L,\mathcal{R}),
\end{equation}
and $V_{n,1}^{\mathcal{DOC},\star}(\cdot)$ is defined via
\begin{align}
V_{n,1}^{\mathcal{DOC},\star}(h(\mathcal{T}),x;K,L,\mathcal{R}) := V_{n,1}^{\mathcal{DOC}}(h(\mathcal{T}),x;K,L,\mathcal{R}) - \left( c_{n,0}^{\mathcal{DOC},+}(h(\mathcal{T})) x^{\rho_{+}(h(\mathcal{T}))} + c_{n,0}^{\mathcal{DOC},-}(h(\mathcal{T})) x^{\rho_{-}(h(\mathcal{T}))}\right).
\end{align}
\noindent Then, rewriting (\ref{PBa35}) using Representation (\ref{TheRep}) leads to
\begin{align}
c_{n,0}^{\mathcal{DOC},+}(h(\mathcal{T})) & = \frac{1 - \partial_{x} \mathcal{DOC}_{E} (\mathcal{T}, \mathfrak{b}_{\mathcal{DOC},n}^{\epsilon=1}(\mathcal{T});K,L,\mathcal{R}) - h(\mathcal{T}) \partial_{x} F_{\mathcal{DOC},n-1}^{\epsilon=1}(h(\mathcal{T}),\mathfrak{b}_{\mathcal{DOC},n}^{\epsilon=1}(\mathcal{T});K,L,\mathcal{R})}{h(\mathcal{T}) \Big(\rho_{+}(h(\mathcal{T})) \big(\mathfrak{b}_{\mathcal{DOC},n}^{\epsilon=1} (\mathcal{T}) \big)^{\rho_{+}(h(\mathcal{T}))-1} - \rho_{-}(h(\mathcal{T})) L^{\rho_{+}(h(\mathcal{T}))-\rho_{-}(h(\mathcal{T}))} \big(\mathfrak{b}_{\mathcal{DOC},n}^{\epsilon=1} (\mathcal{T}) \big)^{\rho_{-}(h(\mathcal{T}))-1} \Big)} \nonumber \\
& - \frac{ \partial_{x} V_{n,1}^{\mathcal{DOC},\star}(h(\mathcal{T}),\mathfrak{b}_{\mathcal{DOC},n}^{\epsilon=1}(\mathcal{T});K,L,\mathcal{R}) - \rho_{-}(h(\mathcal{T})) \big( \mathfrak{b}_{\mathcal{DOC},n}^{\epsilon=1}(\mathcal{T})\big)^{\rho_{-}(h(\mathcal{T}))-1} \mathfrak{R}^{\star}_{n}(h(\mathcal{T}),L;K,L,\mathcal{R})}{ \rho_{+}(h(\mathcal{T})) \big(\mathfrak{b}_{\mathcal{DOC},n}^{\epsilon=1} (\mathcal{T}) \big)^{\rho_{+}(h(\mathcal{T}))-1} - \rho_{-}(h(\mathcal{T})) L^{\rho_{+}(h(\mathcal{T}))-\rho_{-}(h(\mathcal{T}))} \big(\mathfrak{b}_{\mathcal{DOC},n}^{\epsilon=1} (\mathcal{T}) \big)^{\rho_{-}(h(\mathcal{T}))-1}}
\label{EAQ2}
\end{align}
\noindent and inserting the latter expression into (\ref{PBa25}) finally gives that $\mathfrak{b}_{\mathcal{DOC},n}^{\epsilon=1}(h(\mathcal{T}))$ solves, for any $\mathcal{T} \in (0,T]$, the following non-linear equation
\begin{align}
\mathfrak{b}_{\mathcal{DOC},n}^{\epsilon=1}(\mathcal{T}) & = K + \mathcal{DOC}_{E}(\mathcal{T},\mathfrak{b}_{\mathcal{DOC},n}^{\epsilon=1}(\mathcal{T});K,L,\mathcal{R}) + h(\mathcal{T}) F_{\mathcal{DOC},n-1}^{\epsilon=1}(h(\mathcal{T}),\mathfrak{b}_{\mathcal{DOC},n}^{\epsilon=1}(\mathcal{T});K,L,\mathcal{R}) \nonumber \\
& \hspace{2.2em} + h(\mathcal{T}) V_{n,1}^{\mathcal{DOC},\star}(h(\mathcal{T}),\mathfrak{b}_{\mathcal{DOC},n}^{\epsilon=1}(\mathcal{T});K,L,\mathcal{R}) - h(\mathcal{T}) \big( \mathfrak{b}_{\mathcal{DOC},n}^{\epsilon=1}(\mathcal{T})\big)^{\rho_{-}(h(\mathcal{T}))} \mathfrak{R}^{\star}_{n}(h(\mathcal{T}),L;K,L,\mathcal{R}) \nonumber\\
& \hspace{2.2em} + \mathcal{Q}(h(\mathcal{T}),\mathfrak{b}_{\mathcal{DOC},n}^{\epsilon=1}(\mathcal{T})) \bigg[ 1 - \partial_{x} \mathcal{DOC}_{E} (\mathcal{T}, \mathfrak{b}_{\mathcal{DOC},n}^{\epsilon=1}(\mathcal{T});K) \nonumber \\
& \hspace{4em} - h(\mathcal{T}) \Big( \partial_{x} F_{\mathcal{DOC},n-1}^{\epsilon=1}(h(\mathcal{T}),\mathfrak{b}_{\mathcal{DOC},n}^{\epsilon=1}(\mathcal{T});K,L,\mathcal{R})+ \partial_{x}V_{n,1}^{\mathcal{DOC},\star}(h(\mathcal{T}),\mathfrak{b}_{\mathcal{DOC},n}^{\epsilon=1}(\mathcal{T});K,L,\mathcal{R}) \nonumber \\
& \hspace{15em} - \rho_{-}(h(\mathcal{T})) \big( \mathfrak{b}_{\mathcal{DOC},n}^{\epsilon=1}(\mathcal{T})\big)^{\rho_{-}(h(\mathcal{T}))-1} \mathfrak{R}_{n}^{\star}(h(\mathcal{T}),L;K,L,\mathcal{R}) \Big) \bigg]
\label{EAQ1}
\end{align}
\noindent with
\begin{equation}
\mathcal{Q}(h(\mathcal{T}),x) := \frac{ x^{\rho_{+}(h(\mathcal{T}))} - L^{\rho_{+}(h(\mathcal{T}))-\rho_{-}(h(\mathcal{T}))} x^{\rho_{-}(h(\mathcal{T}))}}{ \rho_{+}(h(\mathcal{T})) x^{\rho_{+}(h(\mathcal{T}))-1} - \rho_{-}(h(\mathcal{T})) L^{\rho_{+}(h(\mathcal{T}))-\rho_{-}(h(\mathcal{T}))} x^{\rho_{-}(h(\mathcal{T}))-1}}, \hspace{1.5em} x \in [0,\mathfrak{b}_{\mathcal{DOC},n}^{\epsilon=1} (\mathcal{T})]. 
\end{equation}
\noindent Therefore, using Equations (\ref{EAQ1}), (\ref{EAQ2}) and (\ref{TheRep}), we can deduce all the remaining unknowns and recover $f_{n}^{\mathcal{DOC}}(\cdot)$ via (\ref{FHigherDef}).

\section{Numerical Results}
\label{numSEC}
In this section, our approximations of up to order three are tested via numerical experiments. We combine a variety of parameters that were used in similar simulation studies provided in \cite{ba87}, \cite{ba91}, \cite{gh00}, \cite{ai03}, \cite{jc02}, \cite{ch07} and \cite{fa15}. Although the resulting parameter constellations do not reflect the current market situation, testing option pricing problems with these parameters allows for a direct comparison of the results across articles and has therefore become a standard over the years. For this reason we also stick with these parameters here. We discuss the accuracy and efficiency of our approximations via classical methods. In particular, we use the root mean squared error (RMSE) as measure of accuracy, while the total CPU time (in seconds) required to execute the algorithms is considered as measure of efficiency. All our numerical experiments are obtained using Matlab R2017b on an Intel CORE i7 processor.
\begin{center}
\captionof{table}{Theoretical call values for $K=100$, $r-\delta=-0.04$, $\lambda=2.5$, $\mu_{\mathcal{M}}=0.05$, $\sigma_{\mathcal{M}} = 0.03$.}
\label{table 1}
\scalebox{0.764}{
\begin{tabular}{lrrrrrrrrrrrrr}  
\toprule
\multicolumn{14}{c}{\bf Call Option Prices} \\
\bottomrule
  &         &      \multicolumn{6}{c}{\it Model of Constant Jumps}   & \multicolumn{6}{c}{\it Merton's Jump-Diffusion Model} \\
\cmidrule(r){3-8} \cmidrule(l){9-14}
 \multicolumn{2}{c}{\it Parameters}   & \it European & \multicolumn{5}{c}{\it American} & \it European & \multicolumn{5}{c}{\it American} \\
\cmidrule(r){1-2} \cmidrule(r){3-3} \cmidrule(r){4-8} \cmidrule(l){9-9} \cmidrule(l){10-14} 
  &          &                 &     &    \multicolumn{4}{c}{\it $N$-th Order Approx.}      &              & & \multicolumn{4}{c}{\it $N$-th Order Approx.} \\
\cmidrule{5-8} \cmidrule{11-14}
  &         &  \it Europ.  &   \it Bench-       &        &         &         &     & \it Europ.       &  \it Bench-           &         &    &       \\
	& $S_{0}$ &   \it Price &  \it mark  & $N=0$  &  $N=1$  &  $N=2$  & $N=3$  &  \it Price   & \it mark & $N=0$  &  $N=1$  &  $N=2$ &   $N=3$  \\
	\midrule
	\midrule
(1) & $80$ & $0.061$ & $0.062$ & $0.065$ & $0.057$ & $0.064$ & $0.062$ & $0.084$ & $0.086$ & $0.090$ & $0.081$ & $0.089$ & $0.086$  \\
 $r = 0.08$ & $90$ & $0.749$ & $0.764$ & $0.773$ & $0.757$ & $0.766$ & $0.766$ & $0.831$ & $0.849$ & $0.860$ & $0.843$ & $0.852$ & $0.851$ \\
$\sigma=0.2$ & $100$ & $3.719$ & $3.833$ & $3.831$ & $3.822$ & $3.834$ & $3.835$ & $3.821$ & $3.939$ & $3.941$ & $3.932$ & $3.943$ & $3.944$ \\
 $\mathcal{T}=0.25$ & $110$ & $10.043$ & $10.525$ & $10.483$ & $10.516$ & $10.527$ & $10.527$ & $10.098$ & $10.572$ & $10.541$ & $10.571$ & $10.583$ & $10.583$ \\
 & $120$ & $18.681$ & $20.000$ & $20.000$ & $20.000$ & $20.000$ & $20.000$ & $18.697$ & $20.000$ & $20.000$ & $20.000$ & $20.000$ & $20.000$ \\
\midrule 
\midrule
(2) & $80$ & $0.643$ & $0.671$ & $0.704$ & $0.650$ & $0.671$ & $0.680$ & $0.730$ & $0.763$ & $0.799$ & $0.742$ & $0.764$ & $0.772$  \\
 $r = 0.08$ & $90$ & $2.262$ & $2.394$ & $2.441$ & $2.368$ & $2.388$ & $2.401$ & $2.411$ & $2.555$ & $2.604$ & $2.530$ & $2.551$ & $2.563$ \\
$\sigma=0.2$ & $100$ & $5.597$ & $6.035$ & $6.061$ & $6.001$ & $6.023$ & $6.037$ & $5.773$ & $6.225$ & $6.257$ & $6.196$ & $6.219$ & $6.232$ \\
 $\mathcal{T}=0.75$ & $110$ & $10.834$ & $11.972$ & $11.936$ & $11.935$ & $11.959$ & $11.970$ & $10.991$ & $12.126$ & $12.101$ & $12.098$ & $12.123$ & $12.133$ \\
 & $120$ & $17.676$ & $20.149$ & $20.102$ & $20.138$ & $20.148$ & $20.151$ & $17.787$ & $20.201$ & $20.161$ & $20.200$ & $20.212$ & $20.215$ \\
\midrule
\midrule
(3) & $80$ & $1.482$ & $1.623$ & $1.714$ & $1.587$ & $1.601$ & $1.637$ & $1.622$ & $1.779$ & $1.875$ & $1.743$ & $1.757$ & $1.795$  \\
 $r = 0.08$ & $90$ & $3.480$ & $3.901$ & $4.009$ & $3.859$ & $3.867$ & $3.906$ & $3.678$ & $4.126$ & $4.239$ & $4.086$ & $4.095$ & $4.135$ \\
$\sigma=0.2$ & $100$ & $6.693$ & $7.718$ & $7.798$ & $7.667$ & $7.675$ & $7.713$ & $6.924$ & $7.977$ & $8.065$ & $7.931$ & $7.941$ & $7.979$ \\
 $\mathcal{T}=1.50$ & $110$ & $11.147$ & $13.292$ & $13.297$ & $13.236$ & $13.249$ & $13.279$ & $11.379$ & $13.530$ & $13.549$ & $13.482$ & $13.497$ & $13.527$ \\
 & $120$ & $16.704$ & $20.712$ & $20.654$ & $20.675$ & $20.686$ & $20.702$ & $16.913$ & $20.857$ & $20.814$ & $20.830$ & $20.843$ & $20.861$ \\
\bottomrule
{\bf RMSE} &   &   & \bf --  & \bf 0.051  &  \bf 0.031  & \bf 0.021  & \bf 0.007 &   & \bf --  &  \bf 0.052 &  \bf 0.027  & \bf 0.017  & \bf 0.008  \\
{\bf CPU (sec.)} &   &   & \bf 2306.07  &  \bf 0.07  & \bf 0.47  & \bf 1.41  &  \bf 3.19  &   & \bf 2381.86  &  \bf 0.07  &  \bf 0.48  &  \bf 1.42  &  \bf 3.21 \\
\bottomrule
\end{tabular}
}
\end{center}
$\mbox{}$ \\
\subsection{Standard American Options}
We start by discussing our approximations for standard American options under the model of constant jumps as well as under Merton's jump-diffusion model (cf.~\cite{me76}). For each set of parameters, our approximations are tested as follows: We first compute the true European value of the option in the respective model and subsequently determine the early exercise premium via an explicit finite difference scheme.\footnote{Our finite difference scheme corresponds to a fully explicit (American) version of the explicit-implicit method presented in \cite{cv05b}. Instead of working with PIDEs in price coordinates, this method is based on the corresponding PIDEs in log-moneyness coordinate. For an American call, this means that we first transform the pricing problem via
\begin{eqnarray*}
u(\mathcal{T},x)  :=  \sup \limits_{ \tau \in \mathfrak{T}_{[0,\mathcal{T}]}} \mathbb{E}^{\mathbb{Q}} \left[e^{-r\tau} \left( e^{x+X_{\tau}} -1 \right)^{+} \right] ,\\
u(\mathcal{T},\mathbf{x})  =  K \cdot \mathcal{C}_{A}(\mathcal{T},x;K), \hspace{2em} \mathbf{x}  =  \log \left( \frac{x}{K}\right),
\end{eqnarray*}
\noindent and solve the resulting early exercise problem. Hence, in the continuation region the PIDE considered so far 
$$ -\partial_{\mathcal{T}} \mathcal{C}_{A}(\mathcal{T},x;K) + \mathcal{A}_{S} \mathcal{C}_{A}(\mathcal{T},x;K) - r \mathcal{C}_{A}(\mathcal{T},x;K)  =  0  $$
\noindent transforms to the following log-moneyness equation
$$\partial_{\mathcal{T}} u(\mathcal{T},{\bf x}) = \frac{1}{2} \sigma^{2} \partial_{\bf x}^{2} u(\mathcal{T},{\bf x}) + \Big(r-\delta - \lambda \zeta - \frac{1}{2} \sigma^{2} \Big) \partial_{\bf x} u(\mathcal{T},{\bf x}) + \lambda \int \limits_{\mathbb{R}} \big( u(\mathcal{T},{\bf x}+y) - u(\mathcal{T},{\bf x}) \big) f_{J_{1}}(y) dy - r u(\mathcal{T},{\bf x}) $$
\noindent and the corresponding (early-exercise) free-boundary problem is solved using a fully explicit finite difference scheme.}~Adding this premium to the corresponding European value allows us to build a benchmark for the American option price against which the approximations are finally tested. Compared with a direct application of our explicit scheme to the American option, this decomposition approach has some benefits. In particular, applying the explicit scheme to the early exercise premium instead substantially reduces the pricing errors and therefore leads to more accuracy in our benchmark. 

\begin{figure}[!b]
\begin{subfigure}{.5\linewidth}
\centering
\includegraphics[scale=.16]{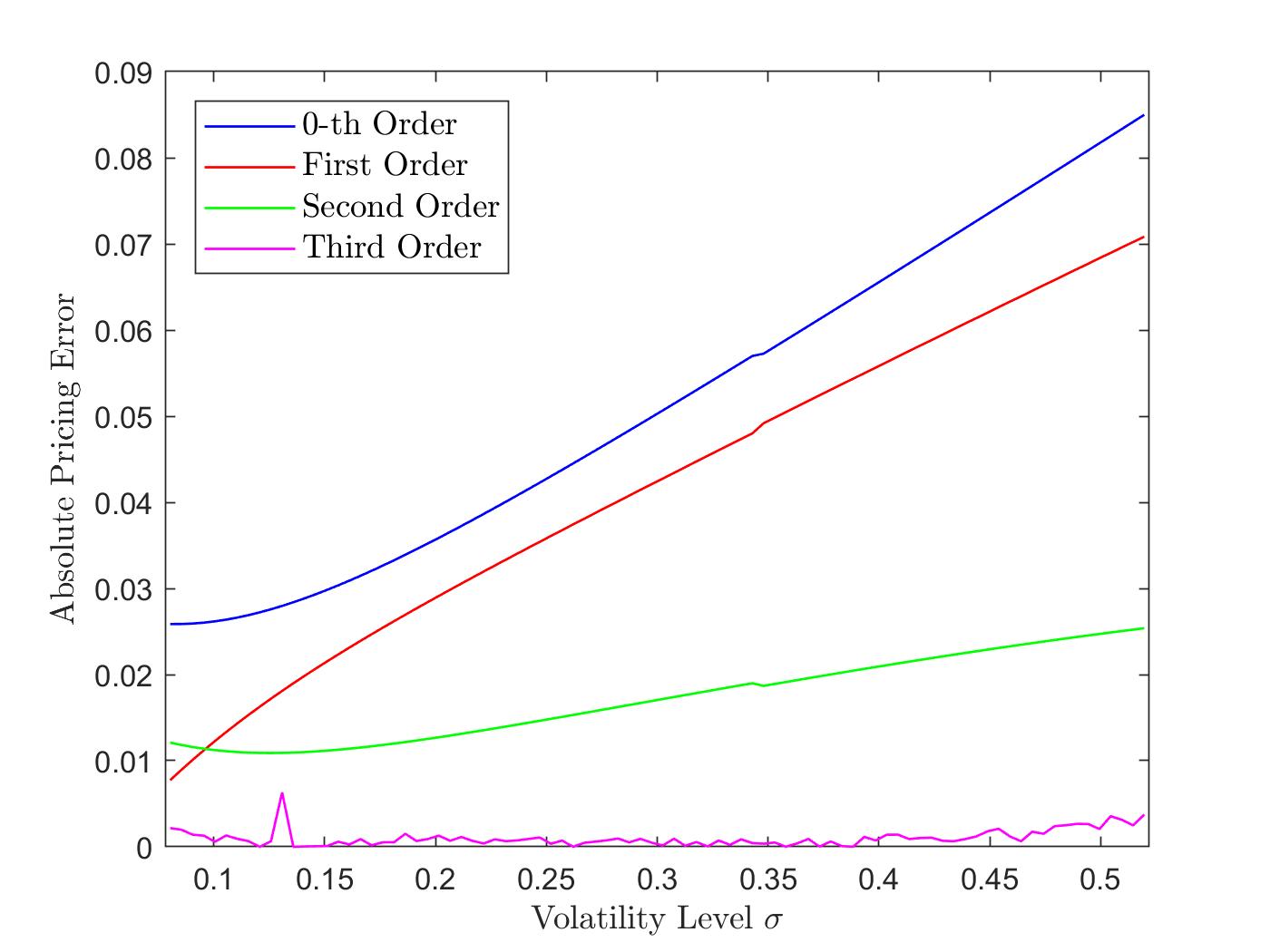}
\caption{Volatility Graph: $\sigma \in (0.075,0.525)$.}
\label{fig:sub31}
\end{subfigure}%
\begin{subfigure}{.5\linewidth}
\centering
\includegraphics[scale=.16]{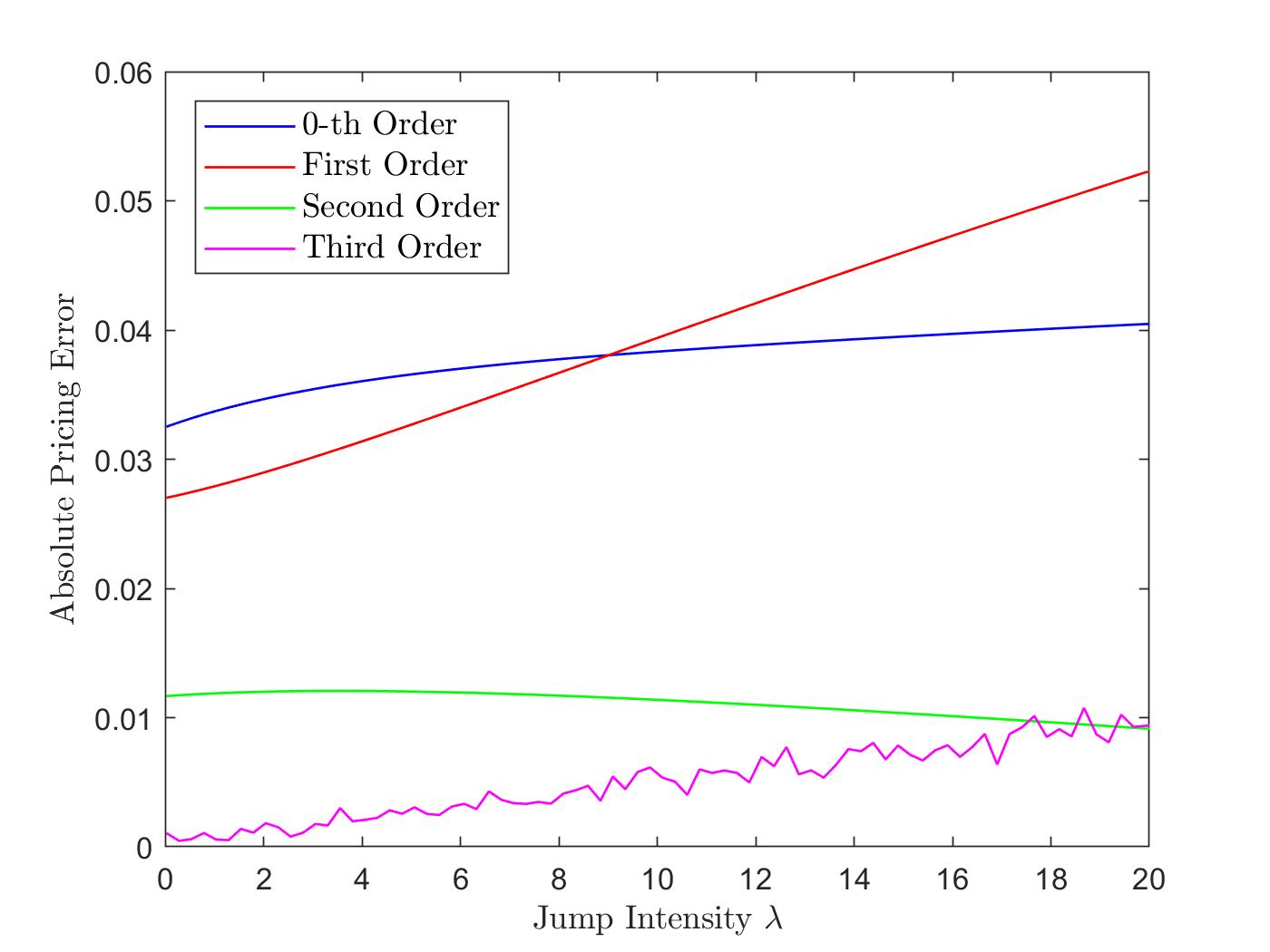}
\caption{Jump Intensity Graph: $\lambda \in (0,20)$.}
\label{fig:sub32}
\end{subfigure}\\[1ex]
\begin{subfigure}{\linewidth}
\centering
\includegraphics[scale=.16]{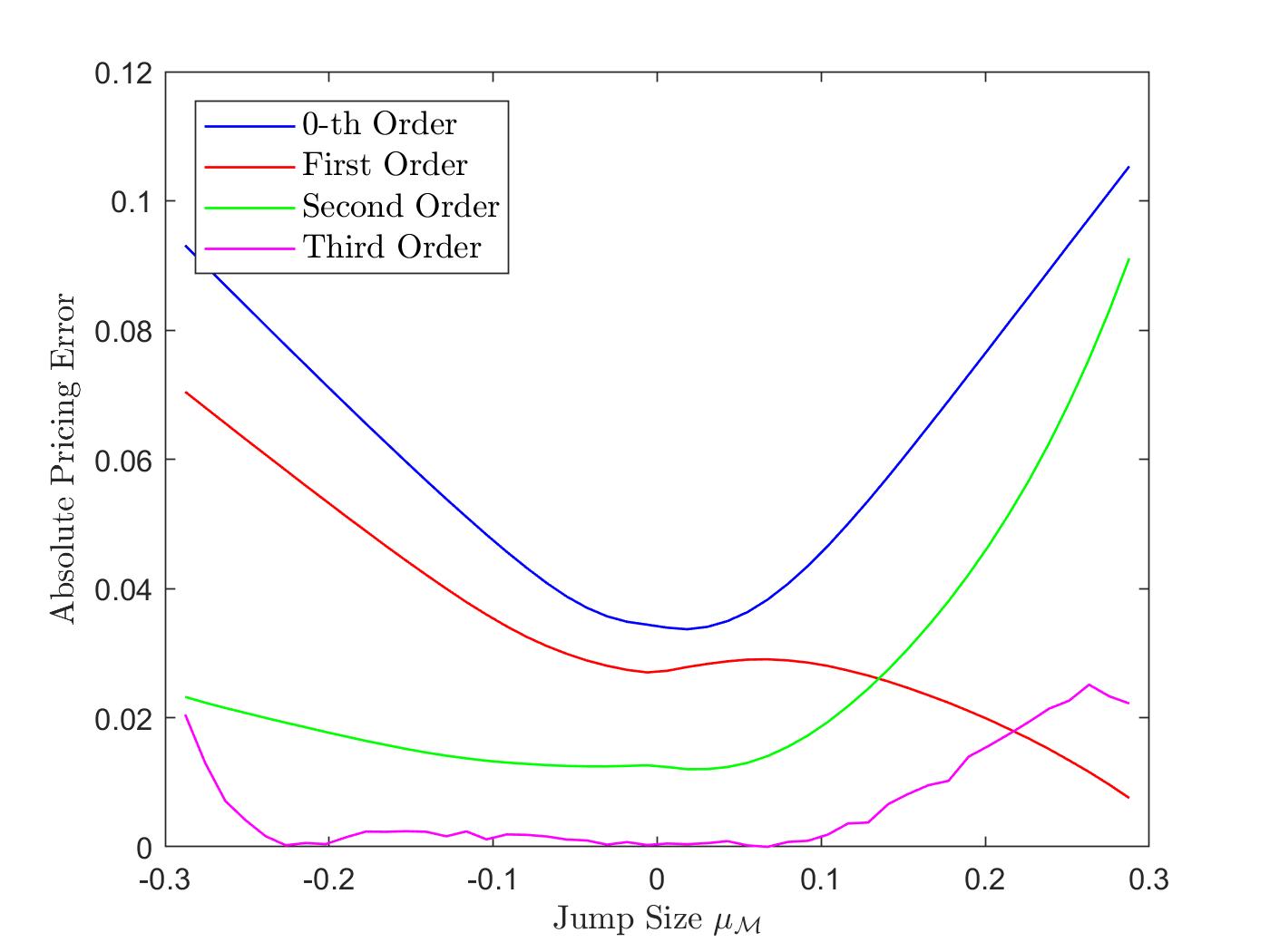}
\caption{Jump Size Graph: $\mu_{\mathcal{M}} \in (-0.3,0.3)$.}
\label{fig:sub33}
\end{subfigure}
\caption{Absolute call option pricing errors as functions of the volatility $\sigma \in (0.075,0.525)$, the jump intensity $\lambda \in (0,20)$ and the jump size $\mu_{\mathcal{M}} \in (-0.3,0.3)$, when the remaining parameters are chosen as: $\sigma = 0.2$, $r=0.08$, $r-\delta = 0.00$, $\lambda = 2.5$, $\sigma_{\mathcal{M}} = 0.03$, $S_{0} = 100$, $K = 100$, $\mathcal{T} = 0.75$.}
\label{fig:test3}
\end{figure}

\begin{center}
\captionof{table}{Theoretical call and put values for $K=100$, $r-\delta=0.00$, $\lambda=2.5$, $\mu_{\mathcal{M}}=0.05$, $\sigma_{\mathcal{M}} = 0.03$.}
\label{table 2}
\scalebox{0.764}{
\begin{tabular}{lrrrrrrrrrrrrr}  
\toprule
\multicolumn{14}{c}{\bf Call and Put Option Prices under Merton's Jump-Diffusion Model} \\
\bottomrule
  &         &      \multicolumn{6}{c}{\it Call Option Prices}   & \multicolumn{6}{c}{\it Put Option Prices} \\
\cmidrule(r){3-8} \cmidrule(l){9-14}
 \multicolumn{2}{c}{\it Parameters}   & \it European & \multicolumn{5}{c}{\it American} & \it European & \multicolumn{5}{c}{\it American} \\
\cmidrule(r){1-2} \cmidrule(r){3-3} \cmidrule(r){4-8} \cmidrule(l){9-9} \cmidrule(l){10-14} 
  &          &                 &     &    \multicolumn{4}{c}{\it $N$-th Order Approx.}      &              & & \multicolumn{4}{c}{\it $N$-th Order Approx.} \\
\cmidrule{5-8} \cmidrule{11-14}
  &         &  \it Europ.  &   \it Bench-       &        &         &         &     & \it Europ.       &  \it Bench-           &         &    &       \\
	& $S_{0}$ &   \it Price &  \it mark  & $N=0$  &  $N=1$  &  $N=2$  & $N=3$  &  \it Price   & \it mark & $N=0$  &  $N=1$  &  $N=2$ &   $N=3$  \\
	\midrule
	\midrule
(1) & $80$ & $0.105$ & $0.105$ & $0.106$ & $0.103$ & $0.107$ & $0.103$ & $19.709$ & $20.004$ & $20.000$ & $20.004$ & $20.004$ & $20.003$  \\
 $r = 0.08$ & $90$ & $0.982$ & $0.984$ & $0.988$ & $0.981$ & $0.987$ & $0.982$ & $10.784$ & $10.852$ & $10.849$ & $10.847$ & $10.855$ & $10.849$ \\
$\sigma=0.2$ & $100$ & $4.304$ & $4.323$ & $4.328$ & $4.319$ & $4.327$ & $4.322$ & $4.304$ & $4.315$ & $4.321$ & $4.311$ & $4.319$ & $4.313$ \\
 $\mathcal{T}=0.25$ & $110$ & $10.964$ & $11.049$ & $11.045$ & $11.044$ & $11.053$ & $11.049$ & $1.162$ & $1.164$ & $1.168$ & $1.160$ & $1.167$ & $1.161$ \\
 & $120$ & $19.814$ & $20.074$ & $20.063$ & $20.075$ & $20.080$ & $20.077$ & $0.210$ & $0.210$ & $0.212$ & $0.207$ & $0.214$ & $0.207$ \\
\midrule 
\midrule
(2) & $80$ & $1.012$ & $1.020$ & $1.038$ & $1.000$ & $1.034$ & $1.017$ & $19.848$ & $20.495$ & $20.481$ & $20.475$ & $20.500$ & $20.493$  \\
 $r = 0.08$ & $90$ & $3.149$ & $3.183$ & $3.213$ & $3.158$ & $3.197$ & $3.181$ & $12.567$ & $12.816$ & $12.840$ & $12.786$ & $12.825$ & $12.815$ \\
$\sigma=0.2$ & $100$ & $7.171$ & $7.283$ & $7.319$ & $7.254$ & $7.296$ & $7.282$ & $7.171$ & $7.262$ & $7.300$ & $7.232$ & $7.274$ & $7.261$ \\
 $\mathcal{T}=0.75$ & $110$ & $13.114$ & $13.401$ & $13.426$ & $13.370$ & $13.413$ & $13.401$ & $3.696$ & $3.728$ & $3.762$ & $3.699$ & $3.743$ & $3.726$ \\
 & $120$ & $20.571$ & $21.193$ & $21.190$ & $21.166$ & $21.204$ & $21.194$ & $1.736$ & $1.746$ & $1.771$ & $1.719$ & $1.763$ & $1.743$ \\
\midrule
\midrule
(3) & $80$ & $2.499$ & $2.551$ & $2.624$ & $2.490$ & $2.574$ & $2.562$ & $20.237$ & $21.488$ & $21.532$ & $21.432$ & $21.490$ & $21.491$  \\
 $r = 0.08$ & $90$ & $5.333$ & $5.479$ & $5.582$ & $5.409$ & $5.498$ & $5.491$ & $14.202$ & $14.820$ & $14.923$ & $14.748$ & $14.826$ & $14.829$ \\
$\sigma=0.2$ & $100$ & $9.542$ & $9.885$ & $10.003$ & $9.808$ & $9.898$ & $9.896$ & $9.542$ & $9.846$ & $9.969$ & $9.769$ & $9.859$ & $9.860$ \\
 $\mathcal{T}=1.50$ & $110$ & $15.037$ & $15.731$ & $15.841$ & $15.653$ & $15.740$ & $15.740$ & $6.168$ & $6.317$ & $6.434$ & $6.238$ & $6.337$ & $6.334$ \\
 & $120$ & $21.596$ & $22.856$ & $22.931$ & $22.782$ & $22.862$ & $22.862$ & $3.857$ & $3.930$ & $4.030$ & $3.851$ & $3.956$ & $3.947$ \\
\bottomrule
{\bf RMSE} &   &   & \bf --  & \bf 0.058 & \bf 0.045  & \bf 0.012  & \bf 0.006  &   &  \bf -- &  \bf 0.061 & \bf 0.045  & \bf 0.012  & \bf 0.008  \\
{\bf CPU (sec.)} &   &   &  \bf 2359.24 &  \bf 0.08 & \bf 0.49 & \bf 1.58  &  \bf 3.51 &   & \bf 2398.87 &  \bf 0.06 & \bf 0.37  & \bf 1.19  & \bf 2.66 \\
\bottomrule
\end{tabular}
}
\end{center}
$\mbox{}$ \vspace{1em} \\
\noindent To test our approximations, we combine the choices made in \cite{ba87}, \cite{ba91}, \cite{jc02}, and \cite{fa15}. For the diffusion as well as the option specific parameters, we rely on \cite{ba87}, \cite{fa15} and take $\sigma = 0.2$, $r = 0.08$, $r - \delta =:b \in \{-0.04, 0.00, 0.04 \}$, $S_{0} \in \{80,90,100,110,120 \}$ and $K=100$. For the jump parameters, we combine the choices made in \cite{ba91} and \cite{jc02}: First, we take $\lambda = 2.5$. Although this parameter is neither used in \cite{ba91} nor in \cite{jc02}, it provides a sensible choice between the conservative value of \cite{jc02}, $\lambda = 1$, and the more extreme choice in \cite{ba91}, $\lambda = 10$. In any cases, we will see that changing this parameter does not substantially alter the quality of the results obtained in this section (cf.~Figure~\ref{fig:sub32}). For the volatility of jumps, we rely on the parameters in \cite{ba91} and fix $\sigma_{\mathcal{M}} = 0.03$. Finally, we consider $\varphi = \mu_{\mathcal{M}} = 0.05$. This choice results for the model of constant jumps in jump sizes of $e^{\varphi}-1 \approx 0.051$ and for Merton's jump-diffusion model in $\zeta \approx 0.052$. Here again, we note that changing the jump sizes in a sensible range does not alter our results substantially (cf~Figure~\ref{fig:sub33}). We will further investigate the impact of the volatility level $ \sigma$, the jump intensity $\lambda$, and the jump size $\mu_{\mathcal{M}}$ on the accuracy of our methods at the end of this section. The results are summarized in Tables~\ref{table 1}-\ref{table 3}.

\begin{center}
\captionof{table}{Theoretical put values for $K=100$, $r-\delta=0.04$, $\lambda=2.5$, $\mu_{\mathcal{M}}=0.05$, $\sigma_{\mathcal{M}} = 0.03$.}
\label{table 3}
\scalebox{0.764}{
\begin{tabular}{lrrrrrrrrrrrrr}  
\toprule
\multicolumn{14}{c}{\bf Put Option Prices} \\
\bottomrule
  &         &      \multicolumn{6}{c}{\it Model of Constant Jumps}   & \multicolumn{6}{c}{\it Merton's Jump-Diffusion Model} \\
\cmidrule(r){3-8} \cmidrule(l){9-14}
 \multicolumn{2}{c}{\it Parameters}   & \it European & \multicolumn{5}{c}{\it American} & \it European & \multicolumn{5}{c}{\it American} \\
\cmidrule(r){1-2} \cmidrule(r){3-3} \cmidrule(r){4-8} \cmidrule(l){9-9} \cmidrule(l){10-14} 
  &          &                 &     &    \multicolumn{4}{c}{\it $N$-th Order Approx.}      &              & & \multicolumn{4}{c}{\it $N$-th Order Approx.} \\
\cmidrule{5-8} \cmidrule{11-14}
  &         &  \it Europ.  &   \it Bench-       &        &         &         &     & \it Europ.       &  \it Bench-           &         &    &       \\
	& $S_{0}$ &   \it Price &  \it mark  & $N=0$  &  $N=1$  &  $N=2$  & $N=3$  &  \it Price   & \it mark & $N=0$  &  $N=1$  &  $N=2$ &   $N=3$  \\
	\midrule
	\midrule
(1) & $80$ & $18.914$ & $20.000$ & $20.000$ & $20.000$ & $20.000$ & $20.000$ & $18.945$ & $20.000$ & $20.000$ & $20.000$ & $20.000$ & $20.000$  \\
 $r = 0.08$ & $90$ & $9.977$ & $10.371$ & $10.331$ & $10.365$ & $10.373$ & $10.370$ & $10.069$ & $10.430$ & $10.394$ & $10.425$ & $10.432$ & $10.429$ \\
$\sigma=0.2$ & $100$ & $3.748$ & $3.832$ & $3.831$ & $3.824$ & $3.833$ & $3.830$ & $3.843$ & $3.917$ & $3.921$ & $3.912$ & $3.920$ & $3.916$ \\
 $\mathcal{T}=0.25$ & $110$ & $0.938$ & $0.950$ & $0.960$ & $0.945$ & $0.953$ & $0.949$ & $0.981$ & $0.992$ & $1.002$ & $0.987$ & $0.994$ & $0.990$ \\
 & $120$ & $0.156$ & $0.158$ & $0.163$ & $0.152$ & $0.160$ & $0.157$ & $0.167$ & $0.168$ & $0.173$ & $0.162$ & $0.171$ & $0.167$ \\
\midrule 
\midrule
(2) & $80$ & $17.803$ & $20.000$ & $20.000$ & $20.000$ & $20.000$ & $20.000$ & $17.923$ & $20.008$ & $20.000$ & $20.000$ & $20.008$ & $20.008$  \\
 $r = 0.08$ & $90$ & $10.718$ & $11.606$ & $11.562$ & $11.578$ & $11.602$ & $11.606$ & $10.888$ & $11.736$ & $11.699$ & $11.709$ & $11.733$ & $11.736$ \\
$\sigma=0.2$ & $100$ & $5.754$ & $6.092$ & $6.112$ & $6.065$ & $6.089$ & $6.095$ & $5.922$ & $6.247$ & $6.272$ & $6.221$ & $6.245$ & $6.250$ \\
 $\mathcal{T}=0.75$ & $110$ & $2.771$ & $2.893$ & $2.935$ & $2.869$ & $2.892$ & $2.897$ & $2.898$ & $3.015$ & $3.061$ & $2.992$ & $3.016$ & $3.020$ \\
 & $120$ & $1.211$ & $1.253$ & $1.292$ & $1.230$ & $1.255$ & $1.258$ & $1.289$ & $1.329$ & $1.371$ & $1.306$ & $1.333$ & $1.335$ \\
\midrule
\midrule
(3) & $80$ & $16.883$ & $20.204$ & $20.154$ & $20.187$ & $20.198$ & $20.202$ & $17.082$ & $20.279$ & $20.227$ & $20.260$ & $20.273$ & $20.278$  \\
 $r = 0.08$ & $90$ & $11.207$ & $12.840$ & $12.831$ & $12.794$ & $12.822$ & $12.839$ & $11.440$ & $13.027$ & $13.028$ & $12.982$ & $13.011$ & $13.028$ \\
$\sigma=0.2$ & $100$ & $7.089$ & $7.888$ & $7.957$ & $7.841$ & $7.870$ & $7.892$ & $7.316$ & $8.101$ & $8.178$ & $8.054$ & $8.084$ & $8.107$ \\
 $\mathcal{T}=1.50$ & $110$ & $4.302$ & $4.691$ & $4.795$ & $4.647$ & $4.677$ & $4.701$ & $4.497$ & $4.883$ & $4.992$ & $4.838$ & $4.869$ & $4.894$ \\
 & $120$ & $2.523$ & $2.712$ & $2.817$ & $2.668$ & $2.701$ & $2.726$ & $2.674$ & $2.862$ & $2.972$ & $2.818$ & $2.853$ & $2.878$ \\
\bottomrule
{\bf RMSE} &   &   & \bf --  & \bf 0.049  & \bf 0.027  & \bf 0.008  & \bf 0.005  &   & \bf --  & \bf 0.052  &  \bf 0.027 & \bf 0.008  & \bf 0.006  \\
{\bf CPU (sec.)} &   &   &  \bf 2322.10 &  \bf 0.06  & \bf 0.39 & \bf 1.18  & \bf 2.63  &   & \bf 2401.61  &  \bf 0.06 & \bf 0.40  & \bf 1.19  &  \bf 2.65 \\
\bottomrule
\end{tabular}
}
\end{center}
$\mbox{}$ \vspace{1em} \\
\noindent Several
\begin{figure}
\begin{subfigure}{.5\linewidth}
\centering
\includegraphics[scale=0.19]{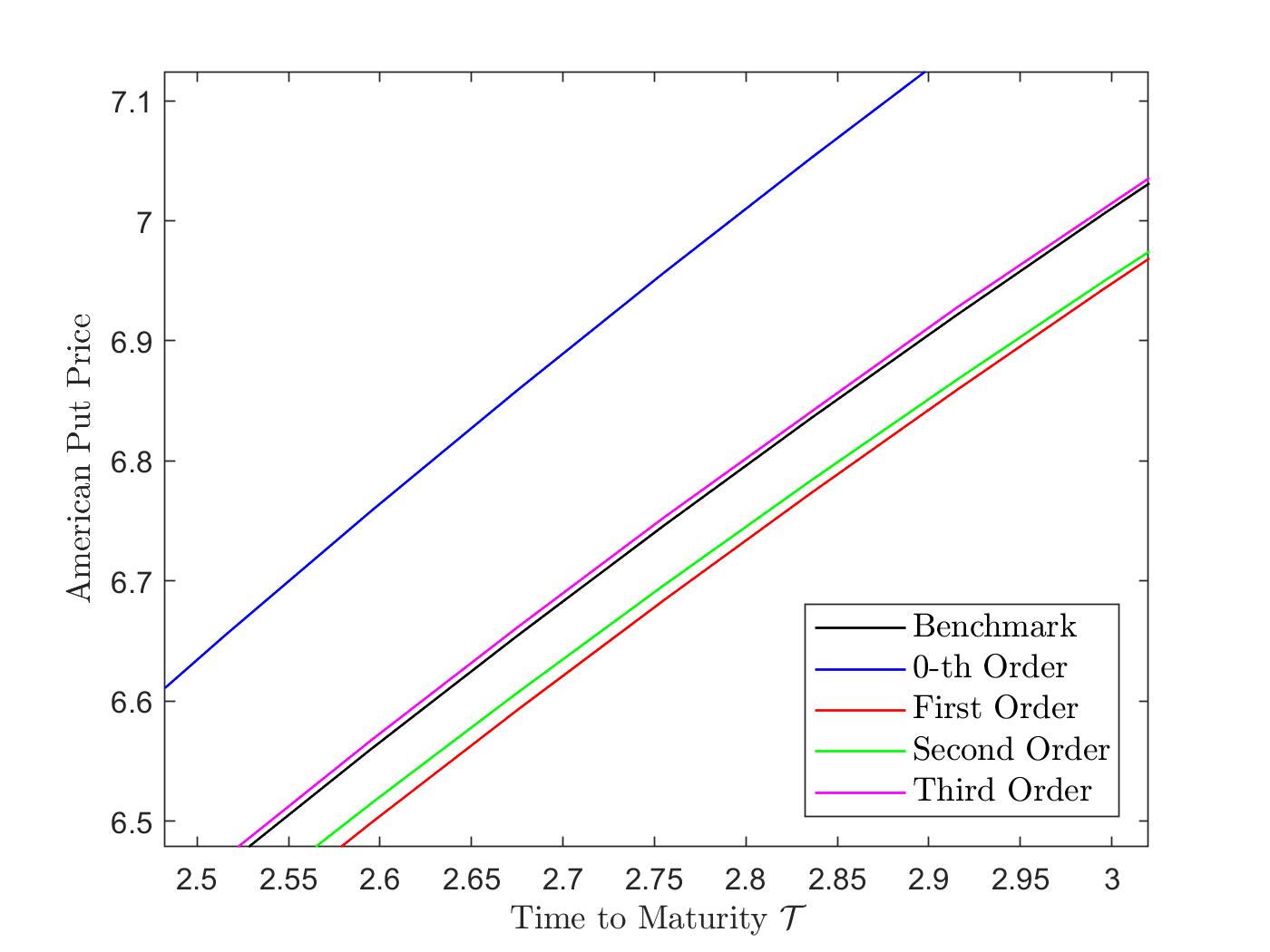}
\caption{Partial graph: $\mathcal{T} \in (2.49,3.01)$.}
\label{fig:sub21}
\end{subfigure}%
\begin{subfigure}{.5\linewidth}
\centering
\includegraphics[scale=.19]{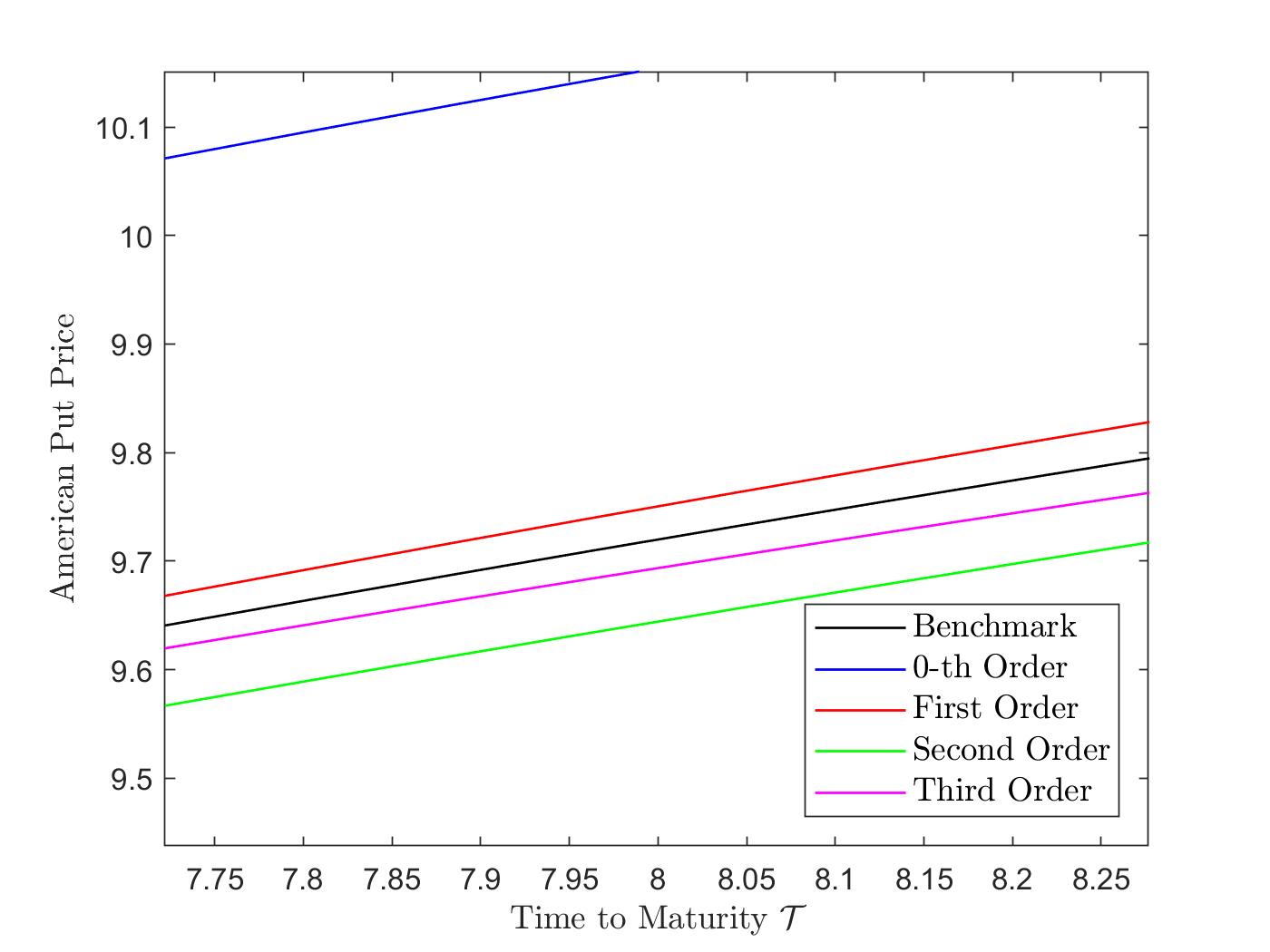}
\caption{Partial graph: $\mathcal{T} \in (7.73,8.27)$.}
\label{fig:sub22}
\end{subfigure}\\[1ex]
\begin{subfigure}{\linewidth}
\centering
\includegraphics[scale=.29]{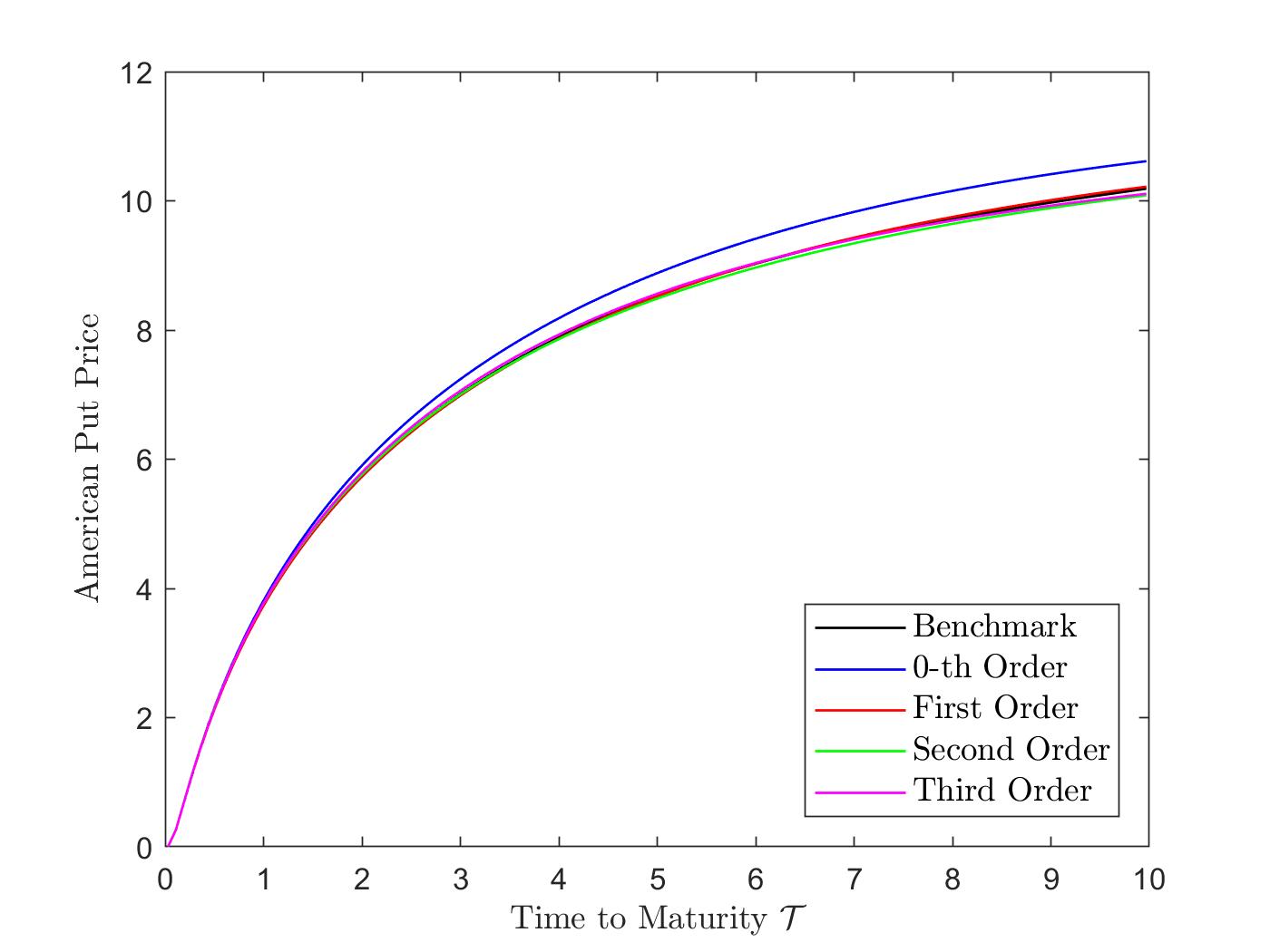}
\caption{Full graph: $\mathcal{T} \in (0,10)$.}
\label{fig:sub23}
\end{subfigure}
\caption{American put price as function of the time to maturity $\mathcal{T} \in (0,10)$ when the parameters are chosen as: $\sigma =0.2$, $r=0.08$, $r-\delta= 0.04$, $\lambda = 2.5$, $\mu_{\mathcal{M}} = 0.05$, $\sigma_{\mathcal{M}} = 0.03$, $S_0 = 110$, $K=100$. }
\label{fig:test2}
\end{figure}
\noindent facts can be observed from the numerical results reported in Tables~\ref{table 1}-\ref{table 3}. First, we observe that a high pricing accuracy can be obtained by increasing the order of our approximations. Indeed, compared to the $0$-th order approximation, i.e.~Bates' method, any higher order approximation augments the pricing accuracy significantly. In addition, increasing the order of the approximation by one roughly halves the absolute pricing errors (RMSE) made by the method. However, this happens at the expense of greater computational complexity (CPU). Secondly, Table~\ref{table 1} and Table~\ref{table 3} reveal that all our approximations exhibit a similar behavior in both models, the model of constant jumps and Merton's jump-diffusion model. This is not surprising, as the model of constant jumps can be obtained as a limiting case of Merton's jump-diffusion model, namely when $\sigma_{\mathcal{M}} \downarrow 0$. This also justifies our choice to restrict our analysis to call and put options under Merton's jump-diffusion model in Table~\ref{table 2} as well as in Figure~\ref{fig:test2} and Figure~\ref{fig:test3}. Finally, we should mention that our approximations of higher orders do not outperform the $0$-th order method when the early exercise premium becomes very small. \noindent This has shown up in numerical simulations.\footnote{In the case of a call options, this holds for $b=0.04$, whenever $T \in (0,2]$ roughly.}~In such cases, however, the European value already provides good results for the American price and relying on this value gives the best approximation. \vspace{1em} \\
\noindent We next look at the impact of an increase in time to maturity on the accuracy of our approximations. This is exemplified in Figure~\ref{fig:test2}, where we have plotted, for $\mathcal{T} \in (0,10)$ and $r-\delta = 0.04$, out-of-the money American put option prices computed via our explicit finite difference scheme (Benchmark) as well as our corresponding approximations of order up to three. The results are in line with the observations obtained for Tables~\ref{table 1}-\ref{table 3}. Indeed, as in Tables~\ref{table 1}-\ref{table 3}, increasing the order of our approximations is shown to substantially augment the accuracy of the method on $\mathcal{T} \in (0,10)$. In particular, while the $0$-th order approximation tends to move substantially away from the benchmark as time increases, higher order versions seem to be more robust and stay impressively close to the ``true'' value. \vspace{1em} \\
\noindent Lastly,
\noindent we investigate the impact of the volatility level $\sigma$, the jump intensity $\lambda$, and the jump size $\mu_{\mathcal{M}}$ on the accuracy of our method. To this end, we have plotted, for $r=0.08$, $r-\delta = 0.00$, $S_{0} = 100$, $K = 100$, and time to maturity $\mathcal{T} = 0.75$, the absolute call option pricing errors as functions of the volatility level $\sigma \in (0.075,0.525)$, the jump intensity $\lambda \in (0,20)$, and the jump size $\mu_{\mathcal{M}} \in (-0.3,0.3)$. The graphs are provided in Figure~\ref{fig:test3}. Here again, the results are in line with our previous observations. In particular, we see that increasing the order of our approximations leads to an impressive decrease of the pricing error for a very large range of parameters. With respect to the jump size, we note that this holds true for negative jumps as well as for positive jumps roughly up to the size of $\zeta \approx 0.14$. Similarly the results hold true for intensities roughly up to $\lambda = 10$. As seen in Section~\ref{SECMERT} (see also Section~\ref{SEC2}), we note however that our general solution ansatz is expected to deviate from the true solution, for call options, whenever positive jumps have a considerable impact on the asset dynamics. This is in particular the case when either ``large'' positive jumps or ``large'' jump intensities are considered. This possibly explains the loss of monotonicity in the pricing accuracy of our approximations observed in Figure~\ref{fig:sub32} and Figure~\ref{fig:sub33}. In any cases, we observe that all our higher order approximations substantially beat the $0$-th order version for a sensible range of parameters and that our approximation of order three exhibits a remarkable accuracy on the full set of parameters tested.

\subsection{American Barrier Options}
\noindent We now turn to a discussion of our approximations for American barrier options under the model of Black~\&~Scholes. For each set of parameters, our approximations are tested against Ritchken's trinomial tree method with $5000$ time steps. A similar benchmark was used in \cite{ch07}, where the authors used $10000$ time steps instead. However, we note that choosing $5000$ times steps does not alter the results for all the parameter sets considered here. Following the simulations offered in \cite{ch07}, we restrict our tests to regular down-and-out call options as well as to regular and reverse up-and-out put options. However, we note that considering other barrier types should not alter the quality of our results, as this merely requires simple adaptions. 

\begin{center}
\captionof{table}{Theoretical down-and-out call values for $K=45$, $\delta=0.025$ and barrier level $L=40$.}
\label{table 4}
\scalebox{0.75}{
\begin{tabular}{lrrrrrrrrrrrrr}  
\toprule
\multicolumn{14}{c}{\bf Down-and-Out Call Option Prices} \\
\bottomrule
  &         &      \multicolumn{6}{c}{\it Volatility Param.~$\sigma=0.2$}   & \multicolumn{6}{c}{\it Volatility Param.~$\sigma=0.4$} \\
\cmidrule(r){3-8} \cmidrule(l){9-14}
 \multicolumn{2}{c}{\it Parameters}   & \it European & \multicolumn{5}{c}{\it American} & \it European & \multicolumn{5}{c}{\it American} \\
\cmidrule(r){1-2} \cmidrule(r){3-3} \cmidrule(r){4-8} \cmidrule(l){9-9} \cmidrule(l){10-14} 
  &          &                 &     &    \multicolumn{4}{c}{\it $N$-th Order Approx.}      &              & & \multicolumn{4}{c}{\it $N$-th Order Approx.} \\
\cmidrule{5-8} \cmidrule{11-14}
  &         &  \it Europ.  &   \it Bench-       &        &         &         &     & \it Europ.       &  \it Bench-           &         &    &       \\
	& $S_{0}$ &   \it Price &  \it mark  & $N=0$  &  $N=1$  &  $N=2$  & $N=3$  &  \it Price   & \it mark & $N=0$  &  $N=1$  &  $N=2$ &   $N=3$  \\
	\midrule
	\midrule
 & $40.5$ & $0.142$ & $0.142$ & $0.142$ & $0.142$ & $0.142$ & $0.142$ & $0.307$ & $0.307$ & $0.307$ & $0.307$ & $0.307$ & $0.307$  \\
 (1) & $42.5$ & $0.771$ & $0.771$ & $0.771$ & $0.771$ & $0.771$ & $0.771$ & $1.543$ & $1.543$ & $1.543$ & $1.542$ & $1.543$ & $1.543$ \\
$r = 4.88\%$ & $45$ & $1.900$ & $1.900$ & $1.900$ & $1.900$ & $1.900$ & $1.900$ & $3.151$ & $3.151$ & $3.151$ & $3.150$ & $3.151$ & $3.151$ \\
 $\mathcal{T}=0.25$ & $47.5$ & $3.519$ & $3.519$ & $3.519$ & $3.519$ & $3.519$ & $3.519$ & $4.883$ & $4.883$ & $4.883$ & $4.882$ & $4.883$ & $4.883$ \\
 & $50$ & $5.548$ & $5.548$ & $5.548$ & $5.548$ & $5.548$ & $5.547$ & $6.760$ & $6.760$ & $6.760$ & $6.759$ & $6.761$ & $6.760$ \\
\midrule
\midrule
 & $40.5$ & $0.307$ & $0.307$ & $0.307$ & $0.307$ & $0.307$ & $0.307$ & $0.411$ & $0.411$ & $0.412$ & $0.411$ & $0.411$ & $0.411$  \\
 (2) & $42.5$ & $1.522$ & $1.522$ & $1.522$ & $1.521$ & $1.523$ & $1.521$ & $2.045$ & $2.046$ & $2.048$ & $2.044$ & $2.046$ & $2.046$ \\
$r = 4.88\%$ & $45$ & $3.100$ & $3.100$ & $3.100$ & $3.099$ & $3.102$ & $3.099$ & $4.080$ & $4.081$ & $4.085$ & $4.078$ & $4.081$ & $4.081$ \\
 $\mathcal{T}=0.75$ & $47.5$ & $4.828$ & $4.828$ & $4.829$ & $4.826$ & $4.831$ & $4.827$ & $6.125$ & $6.126$ & $6.132$ & $6.122$ & $6.126$ & $6.126$ \\
 & $50$ & $6.732$ & $6.732$ & $6.733$ & $6.729$ & $6.737$ & $6.731$ & $8.193$ & $8.195$ & $8.203$ & $8.190$ & $8.195$ & $8.195$ \\
\midrule
\midrule
 & $40.5$ & $0.404$ & $0.404$ & $0.404$ & $0.403$ & $0.404$ & $0.404$ & $0.456$ & $0.457$ & $0.458$ & $0.456$ & $0.457$ & $0.457$  \\
 (3) & $42.5$ & $1.976$ & $1.976$ & $1.977$ & $1.972$ & $1.978$ & $1.977$ & $2.264$ & $2.269$ & $2.276$ & $2.266$ & $2.269$ & $2.269$ \\
$r = 4.88\%$ & $45$ & $3.900$ & $3.900$ & $3.904$ & $3.893$ & $3.905$ & $3.903$ & $4.501$ & $4.510$ & $4.524$ & $4.506$ & $4.510$ & $4.510$ \\
 $\mathcal{T}=1.50$ & $47.5$ & $5.839$ & $5.840$ & $5.845$ & $5.829$ & $5.846$ & $5.843$ & $6.723$ & $6.736$ & $6.757$ & $6.730$ & $6.737$ & $6.737$ \\
 & $50$ & $7.829$ & $7.829$ & $7.837$ & $7.815$ & $7.837$ & $7.834$ & $8.937$ & $8.957$ & $8.983$ & $8.948$ & $8.957$ & $8.957$ \\

\bottomrule
{\bf RMSE ($\mathbf{\times 10^{-1}}$)} &   &   & \bf --  & \bf 0.028  & \bf 0.052  & \bf 0.033  & \bf 0.018  &   & \bf --  & \bf 0.099  &  \bf 0.036 & \bf 0.004  & \bf 0.003  \\
{\bf CPU (sec.)} &   &   &  \bf 1502.14 &  \bf 0.011  & \bf 0.036 & \bf 0.083  & \bf 0.171  &   &  \bf 1501.06 &  \bf 0.014 & \bf 0.053  & \bf 0.134  &  \bf 0.283 \\
\bottomrule
\end{tabular}
}
\end{center}
$\mbox{}$ \vspace{2em} \\
\noindent We start by considering regular down-and-out call options and regular up-and-out put options. To allow for a direct comparability of our results with the existing literature, we mainly rely on the parameters used in \cite{gh00} and \cite{ch07}, i.e.~we take $\sigma \in \{ 0.2, 0.4 \}$, $r=0.0488$, $\delta = 0.025$ and $K = 45$. For down-and-out call options we choose additionally $S_{0} \in \{40.5,42.5,45,47.5,50 \}$ and barrier level $L = 40$ while these parameters are ``reversed'' in the case of up-and-out put options, i.e.~we then consider $S_{0} \in \{40,42.5,45,47.5,49.5 \}$ and $L=50$. Finally, we fix times to maturity according to our previous scheme, i.e.~we consider the maturities $\mathcal{T} \in \{0.25, 0.75, 1.5 \}$. The results are summarized in Table~\ref{table 4} and Table~\ref{table 5}. 

\begin{figure}
\begin{subfigure}{.5\linewidth}
\centering
\includegraphics[scale=.16]{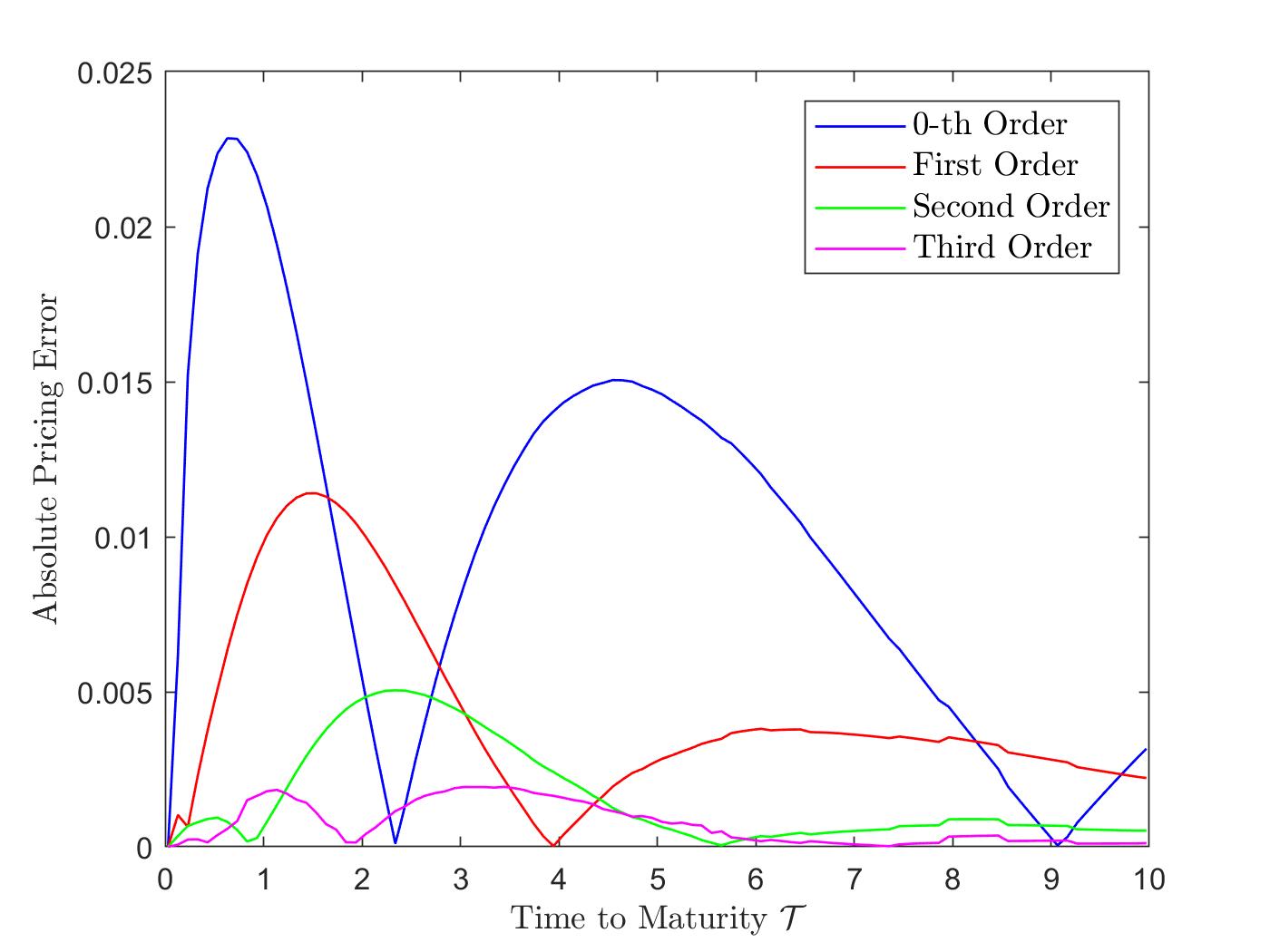}
\caption{Time to Maturity Graph: $\mathcal{T} \in (0,10)$.}
\label{fig:sub41}
\end{subfigure}%
\begin{subfigure}{.5\linewidth}
\centering
\includegraphics[scale=.16]{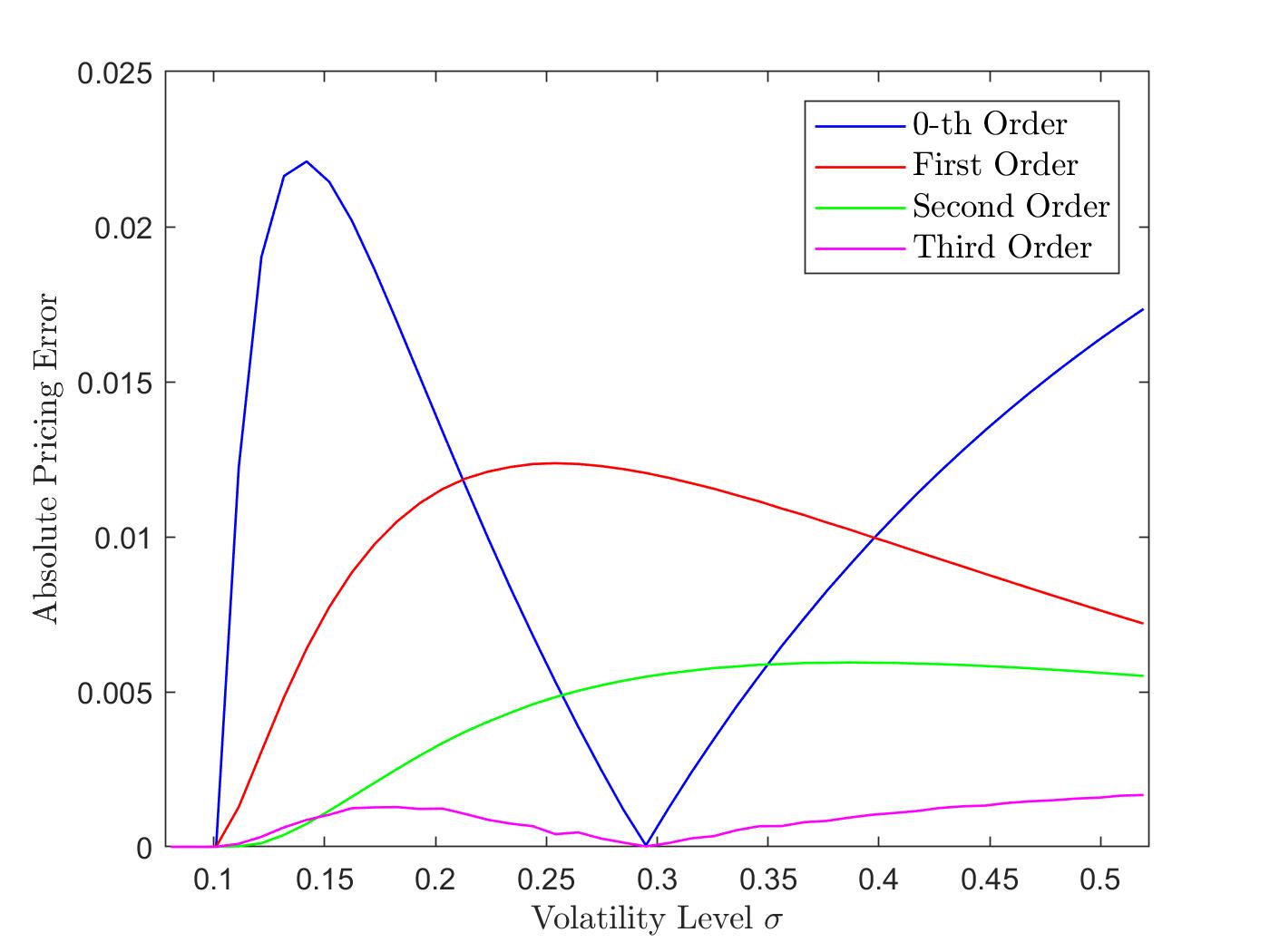}
\caption{Volatility Graph: $\sigma \in (0.075,0.525)$.}
\label{fig:sub42}
\end{subfigure}
\caption{Absolute up-and-out put option pricing errors as functions of the time to maturity $\mathcal{T} \in (0,10)$, and the volatility $\sigma \in (0.075,0.525)$. The remaining parameters are chosen as: $\sigma = 0.2$, $r=0.0488$, $\delta = 0.025$, $S_{0} = 40$, $K = 45$, $L = 50$ and $\mathcal{T} = 0.75$.}
\label{fig:test4}
\end{figure}

\begin{center}
\captionof{table}{Theoretical up-and-out put values for $K=45$, $\delta=0.025$ and barrier level $L=50$.}
\label{table 5}
\scalebox{0.75}{
\begin{tabular}{lrrrrrrrrrrrrr}  
\toprule
\multicolumn{14}{c}{\bf Up-and-Out Put Option Prices} \\
\bottomrule
  &         &      \multicolumn{6}{c}{\it Volatility Param.~$\sigma=0.2$}   & \multicolumn{6}{c}{\it Volatility Param.~$\sigma=0.4$} \\
\cmidrule(r){3-8} \cmidrule(l){9-14}
 \multicolumn{2}{c}{\it Parameters}   & \it European & \multicolumn{5}{c}{\it American} & \it European & \multicolumn{5}{c}{\it American} \\
\cmidrule(r){1-2} \cmidrule(r){3-3} \cmidrule(r){4-8} \cmidrule(l){9-9} \cmidrule(l){10-14} 
  &          &                 &     &    \multicolumn{4}{c}{\it $N$-th Order Approx.}      &              & & \multicolumn{4}{c}{\it $N$-th Order Approx.} \\
\cmidrule{5-8} \cmidrule{11-14}
  &         &  \it Europ.  &   \it Bench-       &        &         &         &     & \it Europ.       &  \it Bench-           &         &    &       \\
	& $S_{0}$ &   \it Price &  \it mark  & $N=0$  &  $N=1$  &  $N=2$  & $N=3$  &  \it Price   & \it mark & $N=0$  &  $N=1$  &  $N=2$ &   $N=3$  \\
	\midrule
	\midrule
 & $40$ & $4.981$ & $5.105$ & $5.089$ & $5.104$ & $5.106$ & $5.105$ & $6.039$ & $6.096$ & $6.084$ & $6.092$ & $6.098$ & $6.097$  \\
 (1) & $42.5$ & $3.055$ & $3.110$ & $3.100$ & $3.108$ & $3.110$ & $3.110$ & $4.319$ & $4.355$ & $4.347$ & $4.351$ & $4.355$ & $4.355$ \\
$r = 4.88\%$ & $45$ & $1.621$ & $1.644$ & $1.641$ & $1.642$ & $1.644$ & $1.644$ & $2.770$ & $2.791$ & $2.787$ & $2.789$ & $2.791$ & $2.791$ \\
 $\mathcal{T}=0.25$ & $47.5$ & $0.666$ & $0.673$ & $0.673$ & $0.673$ & $0.673$ & $0.673$ & $1.349$ & $1.358$ & $1.356$ & $1.357$ & $1.358$ & $1.358$ \\
 & $49.5$ & $0.122$ & $0.123$ & $0.123$ & $0.123$ & $0.123$ & $0.123$ & $0.267$ & $0.268$ & $0.268$ & $0.268$ & $0.268$ & $0.268$ \\
\midrule
\midrule
 & $40$ & $5.296$ & $5.552$ & $5.529$ & $5.544$ & $5.552$ & $5.553$ & $6.716$ & $6.877$ & $6.868$ & $6.866$ & $6.875$ & $6.879$  \\
 (2) & $42.5$ & $3.663$ & $3.811$ & $3.798$ & $3.804$ & $3.811$ & $3.812$ & $4.961$ & $5.072$ & $5.067$ & $5.063$ & $5.071$ & $5.074$ \\
$r = 4.88\%$ & $45$ & $2.272$ & $2.351$ & $2.346$ & $2.347$ & $2.352$ & $2.352$ & $3.265$ & $3.335$ & $3.332$ & $3.329$ & $3.334$ & $3.336$ \\
 $\mathcal{T}=0.75$ & $47.5$ & $1.073$ & $1.107$ & $1.105$ & $1.105$ & $1.107$ & $1.108$ & $1.615$ & $1.649$ & $1.648$ & $1.646$ & $1.649$ & $1.650$ \\
 & $49.5$ & $0.208$ & $0.214$ & $0.214$ & $0.214$ & $0.214$ & $0.214$ & $0.321$ & $0.328$ & $0.327$ & $0.327$ & $0.328$ & $0.328$ \\
\midrule
\midrule
 & $40$ & $5.396$ & $5.856$ & $5.842$ & $5.845$ & $5.853$ & $5.857$ & $6.789$ & $7.131$ & $7.142$ & $7.122$ & $7.126$ & $7.130$  \\
 (3) & $42.5$ & $3.860$ & $4.152$ & $4.146$ & $4.142$ & $4.149$ & $4.154$ & $5.040$ & $5.285$ & $5.294$ & $5.277$ & $5.280$ & $5.284$ \\
$r = 4.88\%$ & $45$ & $2.466$ & $2.637$ & $2.635$ & $2.630$ & $2.635$ & $2.638$ & $3.329$ & $3.487$ & $3.493$ & $3.481$ & $3.483$ & $3.486$ \\
 $\mathcal{T}=1.50$ & $47.5$ & $1.187$ & $1.266$ & $1.265$ & $1.262$ & $1.264$ & $1.266$ & $1.650$ & $1.727$ & $1.731$ & $1.725$ & $1.726$ & $1.727$ \\
 & $49.5$ & $0.231$ & $0.246$ & $0.246$ & $0.246$ & $0.246$ & $0.246$ & $0.328$ & $0.343$ & $0.344$ & $0.343$ & $0.343$ & $0.343$ \\

\bottomrule
{\bf RMSE ($\mathbf{\times 10^{-1}}$)} &   &   & \bf -- & \bf 0.095  & \bf 0.054  & \bf 0.014  & \bf 0.007  &   &  \bf -- & \bf 0.062  &  \bf 0.057 & \bf 0.024  & \bf 0.008  \\
{\bf CPU (sec.)} &   &   & \bf 1484.25  &  \bf 0.012  & \bf 0.041 & \bf 0.091  & \bf 0.185  &   &  \bf 1483.96 &  \bf 0.012 & \bf 0.041  & \bf 0.101  &  \bf 0.204 \\
\bottomrule
\end{tabular}
}
\end{center}
$\mbox{}$ \vspace{1em} \\
\noindent The simulation results show that our approximations for regular American barrier options have very similar properties to the ones obtained when analyzing our approximations for standard American options. As earlier, our higher order approximations outperform the $0$-th order method in any cases where the early exercise premium does not become meaningless and increasing in theses cases the order of our approximations substantially reduces the pricing error made by our method. Additionally, we note that a high pricing accuracy can be obtained by relying on higher order approximations. All these findings are confirmed by Figure~\ref{fig:sub41} and Figure~\ref{fig:sub42} where we have plotted for $r=0.0488$, $\delta = 0.025$, $S_{0}=40$, $K=45$ and barrier level $L=50$ the absolute up-and-out put pricing errors as functions of the time to maturity $\mathcal{T} \in (0,10)$ and of the volatility level $\sigma \in (0.075,0.525)$. Here, it is worth mentioning that our third order approximation exhibits a remarkable accuracy on the whole domains $\mathcal{T} \in (0,10)$ and $\sigma \in (0.075,0.525)$. Finally, we mention as earlier that increasing the order of our approximations leads to higher computational costs when executing the algorithm. However, we note that the costs of all our approximations -- especially of our higher order approximations -- is significantly lower than the costs of the respective versions for standard American options. This result is a direct consequence of the fact that, even for barrier options, European prices under the Black~\&~Scholes model can be computed using simple formulae, while in Merton's model already standard European prices are expressed in terms of (infinite) series. \vspace{1em} \\
\noindent To additionally illustrate the quality of our algorithm, we next provide in Table~\ref{table 6} a comparison of numerical results between our approximations and comparable methods. Although the Barone-Adesi~\&~Whaley extension of \cite{ai03} provides an important reference point for our approximations, we first note that it is already discussed throughout all our simulation studies since it corresponds to our $0$-th order version for American barrier options. Therefore, we focus on a comparison of results obtained with the modified quadratic approximation of \cite{ch07} and with our approximations of order up to three. Here, we rely once again on the parameter choices of \cite{ch07}, i.e.~we take $\sigma = 0.2$, $r= 0.0488$, $\delta = 0.025$, $K=45$, $L=50$, and initial values $S_{0} \in \{40,42.5,45,47.5,49.5 \}$. Nevertheless, we note that considering other parameters does not substantially change the results. This is in line with the analysis presented in Figure~\ref{fig:test4}. \vspace{1em} \\
\noindent The results in Table~\ref{table 6} show a clear dominance of all our higher order approximations over the modified quadratic scheme of \cite{ch07}. In fact, while the latter method provides a marginal increase in accuracy compared to the Barone-Adesi \& Whaley extension of \cite{ai03} (i.e.~compared to our $0$-th order version), our higher order approximations substantially decrease the pricing errors. This is clearly reflected in the resulting RMSEs. In terms of efficiency (CPU), the modified quadratic approximation of \cite{ch07} has the advantage to be very much comparable to the Barone-Adesi~\&~Whaley scheme. This is however not surprising, as this method essentially replicates the Barone-Adesi~\&~Whaley ansatz of \cite{ai03} while including an additional parameter. \vspace{1em} \\
\noindent We lastly 
\noindent turn to reverse up-and-out put options, i.e.~we look at up-and-out put options in the case where the barrier level $L$ and strike price $K$ have the following relation: $L < K$. This situation is characterized by the fact that an American up-and-out put option holder will always exercise his option at the time the price process touches the barrier level, since this allows him a recovery of $K-L$. Hence, the American up-and-out put option turns in this case into an option with rebate as defined in Section~\ref{pwr1}.\ref{pwr2} and dealing with this situation can be done accordingly.\footnote{We recall few central results for the pricing of options with rebates in Appendix C.} \vspace{1em} \\
\begin{center}
\captionof{table}{Theoretical up-and-out put values for $K=45$, $\delta=0.025$ and barrier level $L=50$.}
\label{table 6}
\scalebox{0.75}{
\begin{tabular}{lrrrcrrrr}  
\toprule
\multicolumn{9}{c}{\bf Up-and-Out Put Option Prices} \\
\bottomrule
 \multicolumn{2}{c}{\it Parameters}   & \it European & \multicolumn{6}{c}{\it American} \\
\cmidrule(r){1-2} \cmidrule(r){3-3} \cmidrule(r){4-9} \cmidrule(l){9-9}  
  &          &                 &     &   &  \multicolumn{4}{c}{\it $N$-th Order Approx.}   \\
\cmidrule{6-9} 
  &         &  \it Europ.  &   \it Bench-       &   Mod.~Quad.  &    &         &         &     \\
	& $S_{0}$ &   \it Price &  \it mark  &  Approx. & $N=0$  &  $N=1$  &  $N=2$  & $N=3$  \\
	\midrule
	\midrule
 (1) & $40$ & $4.981$ & $5.105$ & $5.090$ & $5.089$ & $5.104$ & $5.106$ & $5.105$  \\
 $r = 4.88\%$ & $42.5$ & $3.055$ & $3.110$ & $3.101$ & $3.100$ & $3.108$ & $3.110$ & $3.110$ \\
 $\sigma = 0.2$ & $45$ & $1.621$ & $1.644$ & $1.641$ & $1.641$ & $1.642$ & $1.644$ & $1.644$ \\
 $\mathcal{T}=0.25$ & $47.5$ & $0.666$ & $0.673$ & $0.673$ & $0.673$ & $0.673$ & $0.673$ & $0.673$  \\
 & $49.5$ & $0.122$ & $0.123$ & $0.123$ & $0.123$ & $0.123$ & $0.123$ & $0.123$  \\
\midrule
\midrule
(2) & $40$ & $5.296$ & $5.552$ & $5.537$ & $5.529$ & $5.544$ & $5.552$ & $5.553$  \\
$r = 4.88\%$ & $42.5$ & $3.663$ & $3.811$ & $3.805$ & $3.798$ & $3.804$ & $3.811$ & $3.812$ \\
$r = 4.88\%$ & $45$ & $2.272$ & $2.351$ & $2.351$ & $2.346$ & $2.347$ & $2.352$ & $2.352$ \\
 $\mathcal{T}=0.75$ & $47.5$ & $1.073$ & $1.107$ & $1.108$ & $1.105$ & $1.105$ & $1.107$ & $1.108$ \\
 & $49.5$ & $0.208$ & $0.214$ & $0.214$ & $0.214$ & $0.214$ & $0.214$ & $0.214$ \\
\midrule
\midrule
 (3) & $40$ & $5.396$ & $5.856$ & $5.870$ & $5.842$ & $5.845$ & $5.853$ & $5.857$ \\
 $r = 4.88\%$ & $42.5$ & $3.860$ & $4.152$ & $4.169$ & $4.146$ & $4.142$ & $4.149$ & $4.154$ \\
$\sigma = 0.2$ & $45$ & $2.466$ & $2.637$ & $2.651$ & $2.635$ & $2.630$ & $2.635$ & $2.638$ \\
 $\mathcal{T}=1.50$ & $47.5$ & $1.187$ & $1.266$ & $1.274$ & $1.265$ & $1.262$ & $1.264$ & $1.266$ \\
 & $49.5$ & $0.231$ & $0.246$ & $0.248$ & $0.246$ & $0.246$ & $0.246$ & $0.246$ \\
\bottomrule
{\bf RMSE ($\mathbf{\times 10^{-1}}$)} &   &   &  \bf -- &  \bf 0.093 &  \bf 0.095  & \bf 0.054  & \bf 0.014  & \bf 0.007  \\
{\bf CPU (sec.)} &   &   & \bf 1484.25  &  \bf 0.012 & \bf 0.012  & \bf 0.041 & \bf 0.091  & \bf 0.185   \\
\bottomrule
\end{tabular}
}
\end{center}
$\mbox{}$ \vspace{1.2em} \\
\begin{center}
\captionof{table}{Theoretical up-and-out put values for $K=50$, $\delta=0.06$ and barrier level $L=49$.}
\label{table 7}
\scalebox{0.75}{
\begin{tabular}{lrrrrrrrrrrrrr}  
\toprule
\multicolumn{14}{c}{\bf Up-and-Out Put Option Prices} \\
\bottomrule
  &         &      \multicolumn{6}{c}{\it Volatility Param.~$\sigma=0.2$}   & \multicolumn{6}{c}{\it Volatility Param.~$\sigma=0.4$} \\
\cmidrule(r){3-8} \cmidrule(l){9-14}
 \multicolumn{2}{c}{\it Parameters}   & \it European & \multicolumn{5}{c}{\it American} & \it European & \multicolumn{5}{c}{\it American} \\
\cmidrule(r){1-2} \cmidrule(r){3-3} \cmidrule(r){4-8} \cmidrule(l){9-9} \cmidrule(l){10-14} 
  &          &                 &     &    \multicolumn{4}{c}{\it $N$-th Order Approx.}      &              & & \multicolumn{4}{c}{\it $N$-th Order Approx.} \\
\cmidrule{5-8} \cmidrule{11-14}
  &         &  \it Europ.  &   \it Bench-       &        &         &         &     & \it Europ.       &  \it Bench-           &         &    &       \\
	& $S_{0}$ &   \it Price &  \it mark  & $N=0$  &  $N=1$  &  $N=2$  & $N=3$  &  \it Price   & \it mark & $N=0$  &  $N=1$  &  $N=2$ &   $N=3$  \\
	\midrule
	\midrule
 & $35$ & $14.829$ & $15.000$ & $15.000$ & $15.000$ & $15.000$ & $15.000$ & $14.810$ & $15.000$ & $15.000$ & $15.000$ & $15.000$ & $15.000$  \\
 (1) & $40$ & $9.966$ & $10.013$ & $10.020$ & $10.012$ & $10.017$ & $10.013$ & $9.918$ & $10.006$ & $10.012$ & $10.006$ & $10.006$ & $10.006$ \\
$r = 4.88\%$ & $45$ & $5.046$ & $5.055$ & $5.062$ & $5.054$ & $5.058$ & $5.055$ & $4.985$ & $5.017$ & $5.022$ & $5.018$ & $5.017$ & $5.017$ \\
 $\mathcal{T}=0.50$ & $48$ & $2.025$ & $2.027$ & $2.029$ & $2.027$ & $2.028$ & $2.027$ & $2.000$ & $2.008$ & $2.009$ & $2.008$ & $2.008$ & $2.008$ \\
 & $48.5$ & $1.514$ & $1.515$ & $1.516$ & $1.515$ & $1.515$ & $1.515$ & $1.500$ & $1.504$ & $1.505$ & $1.504$ & $1.504$ & $1.504$ \\
\midrule
\midrule
 & $35$ & $14.647$ & $15.000$ & $15.000$ & $15.000$ & $15.000$ & $15.000$ & $14.575$ & $15.000$ & $15.000$ & $15.000$ & $15.000$ & $15.000$  \\
 (3) & $40$ & $9.889$ & $10.020$ & $10.036$ & $10.020$ & $10.028$ & $10.020$ & $9.775$ & $10.006$ & $10.014$ & $10.006$ & $10.006$ & $10.006$ \\
$r = 4.88\%$ & $45$ & $5.027$ & $5.064$ & $5.078$ & $5.065$ & $5.071$ & $5.065$ & $4.924$ & $5.017$ & $5.023$ & $5.018$ & $5.017$ & $5.017$ \\
 $\mathcal{T}=1.00$ & $48$ & $2.021$ & $2.030$ & $2.033$ & $2.030$ & $2.031$ & $2.030$ & $1.985$ & $2.008$ & $2.009$ & $2.008$ & $2.008$ & $2.008$ \\
 & $48.5$ & $1.512$ & $1.516$ & $1.518$ & $1.516$ & $1.517$ & $1.516$ & $1.493$ & $1.504$ & $1.505$ & $1.504$ & $1.504$ & $1.504$ \\
\midrule
\midrule
 & $35$ & $14.450$ & $15.000$ & $15.000$ & $15.000$ & $15.000$ & $15.000$ & $14.319$ & $15.000$ & $15.000$ & $15.000$ & $15.000$ & $15.000$  \\
 (4) & $40$ & $9.786$ & $10.021$ & $10.044$ & $10.023$ & $10.021$ & $10.021$ & $9.614$ & $10.006$ & $10.014$ & $10.006$ & $10.006$ & $10.006$ \\
$r = 4.88\%$ & $45$ & $4.987$ & $5.067$ & $5.085$ & $5.068$ & $5.066$ & $5.066$ & $4.852$ & $5.017$ & $5.023$ & $5.018$ & $5.017$ & $5.017$ \\
 $\mathcal{T}=1.50$ & $48$ & $2.011$ & $2.030$ & $2.035$ & $2.031$ & $2.030$ & $2.030$ & $1.967$ & $2.008$ & $2.009$ & $2.008$ & $2.008$ & $2.008$ \\
 & $48.5$ & $1.507$ & $1.516$ & $1.519$ & $1.517$ & $1.516$ & $1.516$ & $1.484$ & $1.504$ & $1.505$ & $1.504$ & $1.504$ & $1.504$ \\

\bottomrule
{\bf RMSE ($\mathbf{\times 10^{-2}}$)} &   &   & \bf --  & \bf 0.978  & \bf 0.056  & \bf 0.304  & \bf 0.021  &   &  \bf -- & \bf 0.428  &  \bf 0.031 & \bf 0.003  & \bf 0.002  \\
{\bf CPU (sec.)} &   &   & \bf 1493.02  &  \bf 0.012  & \bf 0.039 & \bf 0.083  & \bf 0.167  &   & \bf 1515.77  &  \bf 0.012 & \bf 0.041  & \bf 0.088  &  \bf 0.173 \\
\bottomrule
\end{tabular}
}
\end{center}
$\mbox{}$ \vspace{1em} \\
\noindent To allow for a better comparability of our results, we rely on the parameter choice made in \cite{ch07}, i.e.~we take $\sigma \in \{0.2, 0.4 \}$, $r=0.0488$, $\delta = 0.06$, $K=50$, $S_{0} \in \{35, 40, 45, 48, 48.5\}$ and barrier level $L=49$. Our approximations are implemented based on the ansatz offered in Section \ref{pwr1}.\ref{pwr2} In particular, this means that we first compute the price of the relevant European up-and-out put options with rebate $\mathcal{R}(K,L) := (K-L)^{+}$ and subsequently compute the corresponding early exercise premium via our approximations. Consequently, when referring to the European price we always think of rebate-type options. The results are summarized in Table~\ref{table 7}.  \vspace{1em} \\
\noindent As earlier, our simulation results show a clear dominance of the higher order approximations over the $0$-th order algorithm. However, compared to the case of regular options, our higher order approximations seem to provide even more accuracy. This is easily deduced by comparing the RMSEs and noting that we have used different scaling parameters. Additionally, the results are consistent with the observations made so far for both standard American options as well as regular American barrier options: When using higher order approximations American-type options are priced with a high accuracy and increasing the order of our method generally leads to substantially more precision. Finally, we note that all these findings as well as the consistency obtained among the results suggest that applying the same method to other types of derivatives -- for instance to lookback options, as done in \cite{ch07} -- is expected to deliver similar conclusions. However, since the main techniques would not differ much from the ones presented here, we do not detail these extensions. 

\section{Conclusion}
\label{SEC6}
\noindent The present article extended the current literature on pricing American-type options in two directions. First, we have considered the problem of pricing standard American options in jump-diffusion models. Here, we have extended the ansatz introduced under the Black \& Scholes framework in \cite{fa15} to a model of constant jumps as well as to Merton's jump-diffusion model. The resulting approximations offer a generalization of the method proposed in \cite{ba91} and allow for a considerable increase in accuracy, when compared with the latter method. Secondly, we have considered the pricing of American barrier options under the model of Black \& Scholes. Here, we have offered a generalization of the methods proposed in \cite{ai03} and \cite{ch07} that is based on the techniques developed in the context of standard American options. We have tested all our approximations of up to order three using numerical simulations. Our numerical analysis showed a clear dominance of higher order approximations over their respective $0$-th order version and revealed that significantly more pricing accuracy is obtained when relying on approximations of the first few orders. Additionally, they suggested that increasing the order of any approximation by one generally refines the pricing precision, however that this happens at the expense of greater computational complexity. \vspace{2em} \\
\noindent \acknow{I would like to thank Walter Farkas and Giovanni Barone-Adesi for their advice as well as the two anonymous reviewers and Christoph Reisinger (the Editor) for their constructive comments. I also thank Jérôme Detemple, Sander Willems, Alexander Smirnow, Urban Ulrych, Jakub Rojcek, Matthias Feiler and the participants of the Gerzensee Research Days 2018 for their valuable suggestions.} \vspace{1em} \\
\section*{Appendix}
\renewcommand{\theequation}{A.\arabic{equation}}
\subsection*{Appendix A: Constant Jump Model}
\noindent Let us review few well-known results on European call options that are crucially needed in the implementation of our $N$-th order approximations under Model~(\ref{mixed}). Being close to the Black \& Scholes model, Model~(\ref{mixed}) is particularly manageable and many properties can be derived by slightly adapting their counterparts in the Black \& Scholes framework. Using standard methods, one derives in particular that $\mathcal{C}_{E}(\mathcal{T},x;K)$, the price of a European call option on $(S_{t})_{t \geq 0}$ having maturity $\mathcal{T} \geq 0$, initial value $S_{0} = x \geq 0$ and strike price $K \geq 0$, equals
\begin{equation}
\label{memert1}
\mathcal{C}_{E}(\mathcal{T},x;K)  =  \sum \limits_{n=0}^{\infty} e^{-( \lambda+ r) \mathcal{T}} \frac{(\lambda \mathcal{T})^{n}}{n!} \; \mathcal{BS}\left(xe^{(r-\delta-\lambda(e^{\varphi}-1) + \frac{n \varphi}{\mathcal{T}})\mathcal{T}}, \sigma, \mathcal{T}; K \right) ,
\end{equation}
\noindent where
\begin{equation}
\mathcal{BS} \left(X,\Sigma,T;K \right):= X  \mathcal{N} \left(d_{1} \bigg(\frac{X}{K}, \Sigma, T \bigg) \right) - K \mathcal{N} \left(d_{2} \bigg(\frac{X}{K}, \Sigma, T \bigg) \right),
\label{Notation1}
\end{equation}
\noindent $\mathcal{N}(\cdot)$ denotes the standard normal CDF and 
\begin{equation}
d_{1} \left(y,\varsigma,s \right) := \frac{1}{\sqrt{\varsigma^{2}s}} \log(y) + \frac{1}{2} \sqrt{\varsigma^{2}s}, \hspace{2em} d_{2}(y,\varsigma,s) := d_{1}(y,\varsigma,s) - \sqrt{\varsigma^{2} s} .
\label{Notation2}
\end{equation}
\noindent For $\partial_{x}\mathcal{C}_{E}(\cdot)$, we first obtain from the Black \& Scholes/Garman \& Kohlhagen model (cf.~\cite{gk83}) that
$$ \partial_{x} \left[ e^{-r\mathcal{T}} \mathcal{BS}\left(xe^{(r-\delta-\lambda(e^{\varphi}-1) + \frac{n \varphi}{\mathcal{T}})\mathcal{T}}, \sigma, \mathcal{T}; K \right) \right] = e^{-(\delta +\lambda(e^{\varphi}-1)-\frac{n \varphi}{\mathcal{T}}) \mathcal{T}} \; \mathcal{N} \left(d_{1} \bigg(\frac{xe^{(r-\delta-\lambda(e^{\varphi}-1) + \frac{n \varphi}{\mathcal{T}})\mathcal{T}}}{K}, \sigma, \mathcal{T} \bigg) \right) $$
\noindent and see that, for any $\mathcal{T} \in [0,T]$, we have
$$ \left \lvert \partial_{x} \left[ e^{-r\mathcal{T}} \mathcal{BS}\left(xe^{(r-\delta-\lambda(e^{\varphi}-1) + \frac{n \varphi}{\mathcal{T}})\mathcal{T}}, \sigma, \mathcal{T}; K \right) \right] \right \rvert \leq  e^{-(\delta+\lambda(e^{\varphi}-1)-\frac{n \varphi}{\mathcal{T}}) \mathcal{T}} .$$
\noindent The latter condition allows us to interchange differentiation and summation in the above series representation by means of the dominated convergence theorem and gives us finally that
\begin{equation}
\label{memert2}
\partial_{x} \mathcal{C}_{E}(\mathcal{T},x;K)  =  \sum \limits_{n=0}^{\infty} e^{-( \delta + \lambda e^{\varphi}) \mathcal{T}} \; \frac{(\lambda \mathcal{T} e^{\varphi})^{n}}{n!} \; \mathcal{N} \left(d_{1} \bigg(\frac{xe^{(r-\delta-\lambda(e^{\varphi}-1) + \frac{n \varphi}{\mathcal{T}})\mathcal{T}}}{K}, \sigma, \mathcal{T} \bigg) \right).
\end{equation}
\noindent Using the same approach, higher order Greeks can be also derived from the corresponding Black \& Scholes properties. While both $\mathcal{C}_{E}(\cdot)$ and $\partial_{x}\mathcal{C}_{E}(\cdot)$ are explicitly needed in the derivation of our approximations, higher order Greeks can help improving the stability of the higher order algorithms (cf.~Remark~1).

\subsection*{Appendix B: Merton's Jump-Diffusion Model}
\noindent Following the line of Appendix A, we now briefly recall central results on European options under Merton's jump-diffusion model (cf.~\cite{me76}). First, one obtains that the price of a European call option under Merton's Model (\ref{MertMod}) having maturity $\mathcal{T} \geq 0$, initial value $S_{0} = x \geq 0$ and strike price $K \geq 0$, equals
\begin{equation}
\label{MeMERT}
\mathcal{C}_{E}^{\mathcal{M}}(\mathcal{T},x;K)  =  \sum \limits_{n=0}^{\infty} e^{-( \lambda+ r) \mathcal{T}} \frac{(\lambda \mathcal{T})^{n}}{n!} \; \mathcal{BS}\left(xe^{(r-\delta-\lambda \zeta + \frac{n \log(1+\zeta)}{\mathcal{T}})\mathcal{T}}, \Sigma_{n}, \mathcal{T}; K \right) ,
\end{equation}
\noindent where $\Sigma_{n} := \sqrt{\sigma^{2} + \frac{n \sigma_{\mathcal{M}}^{2}}{\mathcal{T}}}$ and we have used Notation (\ref{Notation1}) and (\ref{Notation2}).
\noindent Secondly, one readily computes the delta $\partial_{x} \mathcal{C}_{E}^{\mathcal{M}}(\cdot)$ and obtain that it equals
\begin{equation}
\partial_{x} \mathcal{C}_{E}^{\mathcal{M}}(\mathcal{T},x;K) = \sum \limits_{n=0}^{\infty} e^{-(\delta + \lambda (1+\zeta) ) \mathcal{T}} \; \frac{\big(\lambda \mathcal{T} (1+\zeta)\big)^{n}}{n!} \; \mathcal{N} \left(d_{1} \bigg(\frac{xe^{(r-\delta-\lambda \zeta + \frac{n \log(1+\zeta)}{\mathcal{T}})\mathcal{T}}}{K}, \Sigma_{n}, \mathcal{T} \bigg) \right).
\end{equation}
\noindent As in the model of constant jumps, we note that both $\mathcal{C}_{E}^{\mathcal{M}}(\cdot)$ and $\partial_{x}\mathcal{C}_{E}^{\mathcal{M}}(\cdot)$ are explicitly needed in the derivation of our approximations while further, higher order Greeks can be derived to help improving the stability of the higher order algorithms.
\subsection*{Appendix C: Barrier Options with Rebate}
\noindent In this Appendix, we briefly review some well-known results to arrive at a valuation formula for 
\begin{equation}
\mathcal{R}(K,L) \mathbb{E}^{\mathbb{Q}}_{x} \left[ B_{\tau_{L}}(r)^{-1} \, \mathds{1}_{ \{\tau_{L} \leq \mathcal{T} \} } \right],
\label{RAbaBa}
\end{equation} 
\noindent the rebate term in (\ref{BarrEURO}). Further details can be found in the well-written book \cite{jy06}. \vspace{1em} \\
\noindent First, we note that, for $S_{0} > L$,  
\begin{align}
\tau_{L} &: = \inf \{ t >0 : \; S_{t} \leq L \} \\
& \; = \inf \{ t >0 : \; \nu t + W_{t} \leq y \},
\end{align}
\noindent where $(W_{t})_{t \geq 0}$ is a Brownian motion and $\nu$, $y$ are given by $\nu : = \frac{1}{\sigma} \big(r-\delta - \frac{1}{2}\sigma^2 \big)$ and $y := \frac{1}{\sigma} \log\left(\frac{L}{S_0} \right) <0$.
\noindent Hence, computing the rebate term reduces to valuing a particular Laplace transform for the hitting time of a drifted Brownian motion. These results are known in closed form. Indeed, we have that, for $y < 0$,
\begin{equation}
\mathbb{E}^{\mathbb{Q}}_{S_{0}} \left[ e^{-r \tau_{L}} \mathds{1}_{ \{\tau_{L} \leq \mathcal{T} \} } \right] = e^{(\nu - \gamma) y} \mathcal{N} \left( \frac{-\gamma \mathcal{T} + y}{\sqrt{\mathcal{T}}}\right) + e^{(\nu + \gamma) y} \mathcal{N} \left( \frac{\gamma \mathcal{T} + y}{\sqrt{\mathcal{T}}}\right),
\label{formulaA1}
\end{equation}
\noindent where $\gamma$ is chosen to satisfy
\begin{equation}
\gamma = \pm \sqrt{2r + \nu^2 }.
\end{equation}
\noindent Therefore, Formula (\ref{formulaA1}) provides us with a closed form expression for the term in (\ref{RAbaBa}) and similar results can be obtained in the case of an up-barrier (cf.~\cite{jy06}). \vspace{1em} \\
\noindent As before, we note that these closed form results are crucially needed for the computation of European barrier-type options with rebate in the implementation of our $N$-th order approximations.

\end{document}